\documentclass[acmtog,nonacm]{acmart}

\acmJournal{TOG}

\usepackage{overpic}
\usepackage{arydshln}
\usepackage{amsmath}
\usepackage{graphics}
\usepackage{placeins} %?
\usepackage{makecell} %?
\usepackage{wrapfig}
\usepackage{amssymb}
\usepackage{multirow}
\usepackage{enumerate}
\usepackage{color}

\newcommand{\ZY}[1]{{#1}}
\newcommand{\XR}[1]{{#1}}
\newcommand{\NW}[1]{{#1}}

\usepackage{graphicx}
\usepackage{bbding}
\usepackage{pifont} %?
\usepackage{wasysym} %?
\usepackage{utfsym} %?
\usepackage{fontawesome} %?

\begin{document}
% \begin{CJK}{UTF8}{gbsn}
%%
%% The "title" command has an optional parameter,
%% allowing the author to define a "short title" to be used in page headers.
\title{Globally Consistent Normal Orientation for Point Clouds by Regularizing the Winding-Number Field}
%% The "author" command and its associated commands are used to define
%% the authors and their affiliations.
%% Of note is the shared affiliation of the first two authors, and the
%% "authornote" and "authornotemark" commands
%% used to denote shared contribution to the research.

\author{Rui Xu}
\affiliation{\institution{Shandong University} 
\country{China}}
\email{xrvitd@163.com}

\author{Zhiyang Dou}
\affiliation{  \institution{The University of Hong Kong}
\country{China}}\email{zhiyang0@connect.hku.hk}

\author{Ningna Wang}
\affiliation{  \institution{The University of Texas at Dallas}
\country{USA}}
\email{ningna.wang@utdallas.edu}

\author{Shiqing Xin}
\authornote{Co-corresponding authors: Shiqing Xin and Changhe Tu. }
\affiliation{  \institution{Shandong University}
\country{China}}\email{xinshiqing@sdu.edu.cn}

\author{Shuangmin Chen}
\affiliation{  \institution{Qingdao University of Science and Technology}
\country{China}}\email{csmqq@163.com}

\author{Mingyan Jiang}
\affiliation{  \institution{Shandong University}
\country{China}}\email{jiangmingyan@sdu.edu.cn}

\author{Xiaohu Guo}
\affiliation{  \institution{The University of Texas at Dallas}
\country{USA}}\email{xguo@utdallas.edu}

\author{Wenping Wang}
\affiliation{  \institution{Texas A\&M University}
\country{USA}}\email{wenping@tamu.edu}

\author{Changhe Tu} 
\authornotemark[1]
\affiliation{  \institution{Shandong University}
\country{China}}
\email{chtu@sdu.edu.cn}

% \author{Ben Trovato}
% \authornote{Both authors contributed equally to this research.}
% \email{trovato@corporation.com}
% \orcid{1234-5678-9012}
% \author{G.K.M. Tobin}
% \authornotemark[1]
% \email{webmaster@marysville-ohio.com}
% \affiliation{%
%   \institution{Institute for Clarity in Documentation}
%   \streetaddress{P.O. Box 1212}
%   \city{Dublin}
%   \state{Ohio}
%   \country{USA}
%   \postcode{43017-6221}
% }

%%
%% By default, the full list of authors will be used in the page
%% headers. Often, this list is too long, and will overlap
%% other information printed in the page headers. This command allows
%% the author to define a more concise list
%% of authors' names for this purpose.
% \renewcommand{\shortauthors}{Trovato and Tobin, et al.}

%%
%% The abstract is a short summary of the work to be presented in the
%% article.
\begin{abstract}
Estimating normals with globally consistent orientations for a raw point cloud has many downstream geometry processing applications.
Despite tremendous efforts in the past decades, it remains challenging to deal with an unoriented point cloud with various imperfections, particularly in the presence of data sparsity coupled with nearby gaps or thin-walled structures. 
In this paper, we propose a smooth objective function to characterize the requirements of an acceptable winding-number field, which allows one to find the globally consistent normal orientations starting from a set of completely random normals.
By taking the vertices of the Voronoi diagram of the point cloud as examination points, we consider the following three requirements:
(1)~the winding number is either~0 or~1, (2)~the occurrences of~1 and the occurrences of~0 are balanced around the point cloud, and (3)~the normals align with the outside Voronoi poles as much as possible. 
Extensive experimental results show that our method outperforms the existing approaches, especially in handling sparse and noisy point clouds, as well as shapes with complex geometry/topology.
\end{abstract}

%%
%% The code below is generated by the tool at http://dl.acm.org/ccs.cfm.
%% Please copy and paste the code instead of the example below.
%%
\begin{CCSXML}
<ccs2012>
   <concept>
       <concept_id>10010147.10010371.10010396.10010400</concept_id>
       <concept_desc>Computing methodologies~Point-based models</concept_desc>
       <concept_significance>500</concept_significance>
       </concept>
 </ccs2012>
\end{CCSXML}

\ccsdesc[500]{Computing methodologies~Point-based models}

%%
%% Keywords. The author(s) should pick words that accurately describe
%% the work being presented. Separate the keywords with commas.
\keywords{raw point cloud, normal orientation, winding number, Voronoi diagram, optimization}  

\begin{teaserfigure}
  \centering
  \vspace{-2mm}
  \includegraphics[width=\textwidth]{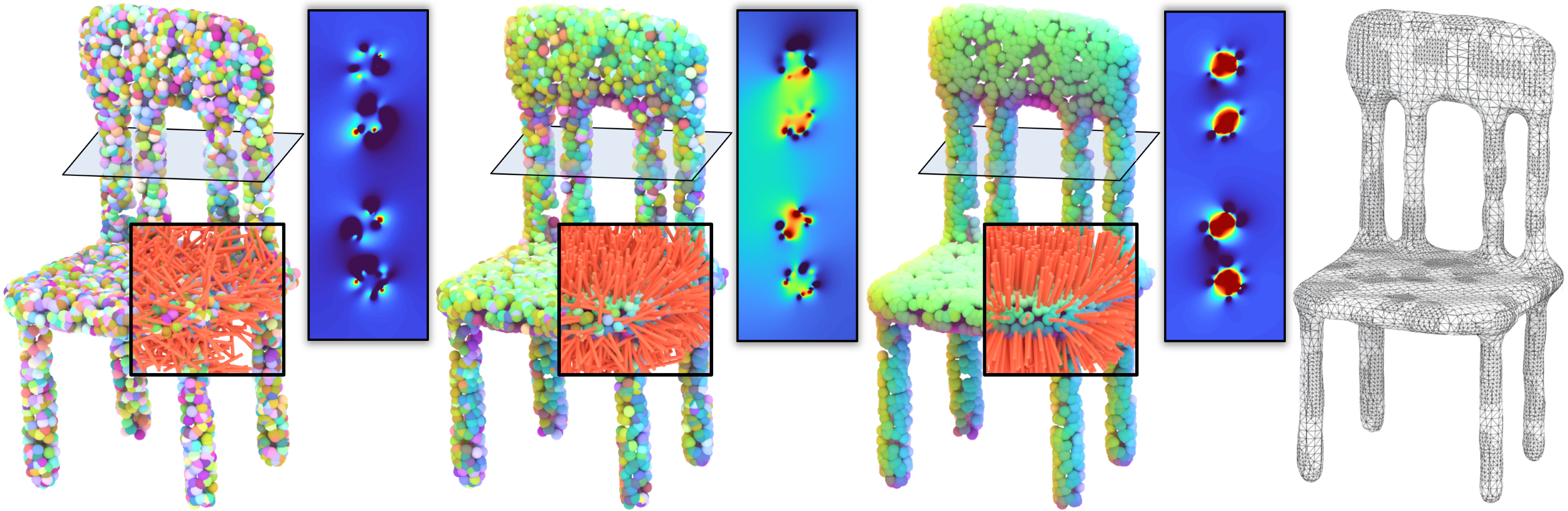}
  \vspace{-1mm}
  \makebox[\linewidth][l]{ \textbf{(a)~Initial normal vectors \hspace{1.9cm} (b)~20 iterations \hspace{2.5cm} (c)~40 iterations   \hspace{2.1cm} (d)~Reconstruction}}
  \vspace{-6mm}
  \caption{
  % In computer graphics, a closed and orientable surface separates $\mathbb{R}^3$ into the interior part and the exterior part, whose winding numbers are valued at $1$ and $0$, respectively. 
  % Inspired by this observation, 
  %Inspired by the winding-number field of a closed and orientable surface, which separates $\mathbb{R}^3$ into the interior and the exterior, we creatively solve the inverse problem: providing globally consistent orientation of an un-oriented point cloud by regularizing the winding-number distribution.
  For a closed and orientable surface, the winding number
  is 0 on the exterior and 1 on the interior.
  Inspired by this fact, in this paper we consider a reverse problem: given an un-oriented point cloud, is it possible to find the globally consistent normal orientations by regularizing the winding-number distribution?
  We propose a smooth objective function to characterize the requirements of an acceptable winding-number field. Starting from a set of completely random normals~(a), we repeatedly optimize their directions~(b,c) until the objective function cannot be reduced. With the computed normals, one can simply call the screened Poisson reconstruction~(SPR) solver to produce the final surface~(d). Note that
  we use RGB mapping to visualize the normals and provide a sectional view to visualize the change of the winding-number field, where ``blue'' and ``red'' indicate 0 and 1, respectively.% while ``blue'' represents 0.  
  % separates the space $\mathbb{R}^3$ into two regions of $0$ (exterior\XR{in blue}) and $1$ (interior\XR{in red}) values\XG{do we need to add a colorbar here to show color-value mapping?}\XR{I tried it and it's kind of ugly. So try to explain in text...}, as long as their normal orientations are globally consistent. Inspired by this fact, in this paper we solve for a reverse problem: given an un-oriented point cloud, we solve for its globally consistent normal orientations by regularizing its winding-number distribution. 
  % we creatively solve the problem of globally consistent orientation of an unoriented point cloud by regularizing the winding-number distribution. 
  % We propose a smooth objective function to characterize the requirements of an acceptable winding-number field. Starting from a set of completely random normals~(a), we repeatedly optimize their directions~(b,c) until the objective function converges. With the computed normals, one can simply call the classic Poisson surface reconstruction to produce the final surface~(d). Note that the normals are visualized using RGB mapping.
  }
  \label{fig:teaser}
\end{teaserfigure}

% \received{20 February 2007}
% \received[revised]{12 March 2009}
% \received[accepted]{5 June 2009}

%%
%% This command processes the author and affiliation and title
%% information and builds the first part of the formatted document.
\maketitle
\newcommand{\rspace}{\mathbb{R}}
\newcommand{\vorcell}{\Omega^{vor}}
\newcommand{\anypoint}{\mathbf{x}}
\newcommand{\sample}{\mathbf{p}}
\newcommand{\query}{\mathbf{q}}
\newcommand{\cross}{\mathcal{A}}
\newcommand{\area}{a}
\newcommand{\normal}{\mathbf{n}}
\newcommand{\numV}{M}
\newcommand{\numP}{N}
\newcommand{\ie}{\textit{i.e., }}
\newcommand{\eg}{\textit{e.g., }}

\newcommand{\vecdiagonal}{\mathbf{v}}

\section{Introduction}
An unoriented point cloud becomes more informative
if it is equipped with a set of normals with globally consistent orientations.
Predicting reliable normals serves as a crucial step for many downstream tasks, e.g., surface reconstruction~\cite{kazhdan2005reconstruction, kazhdan2006poisson, kazhdan2013screened, xu2022rfeps,wang2021neural}, shape registration~\cite{pomerleau2015review}, determining inside/outside information~\cite{jacobson2013robust, barill2018fast}, shape analysis~\cite{grilli2017review, dou2022coverage, zapata2019fast}.
{Despite significant progress~\cite{ hoppe1992surface, dey2004provable, dey2005normal, alliez2007voronoi, konig2009consistent, merigot2010voronoi, boltcheva2017surface,  guerrero2018pcpnet, li2022neaf,  hou2022iterative,
dipole_propagation} being made on this problem, it is still a stumbling task of discovering the globally consistent normals for an unoriented point cloud while allowing for various imperfections.}

%with a sound theoretical basis

% Globally consistent normal orientation for point clouds is an important problem, given its wide applications in computer graphics and computer vision communities. A consistent normal estimation is a crucial pre-requisite for many techniques,  Although the past decades have witnessed tremendous efforts made to ac hieve consistent and accurate normal orientation~\cite{ hoppe1992surface, amenta1998surface, dey2004provable, dey2005normal, ouyang2005normal, alliez2007voronoi, huang2009consolidation, konig2009consistent, merigot2010voronoi, wang2012variational, boltcheva2017surface,  guerrero2018pcpnet, li2022neaf, 
% dipole_propagation}, globally consistent normal estimation in the presence of imperfect remains an open problem. 

\begin{figure}[!t]
    \centering
    \begin{overpic}[width=0.99\linewidth]{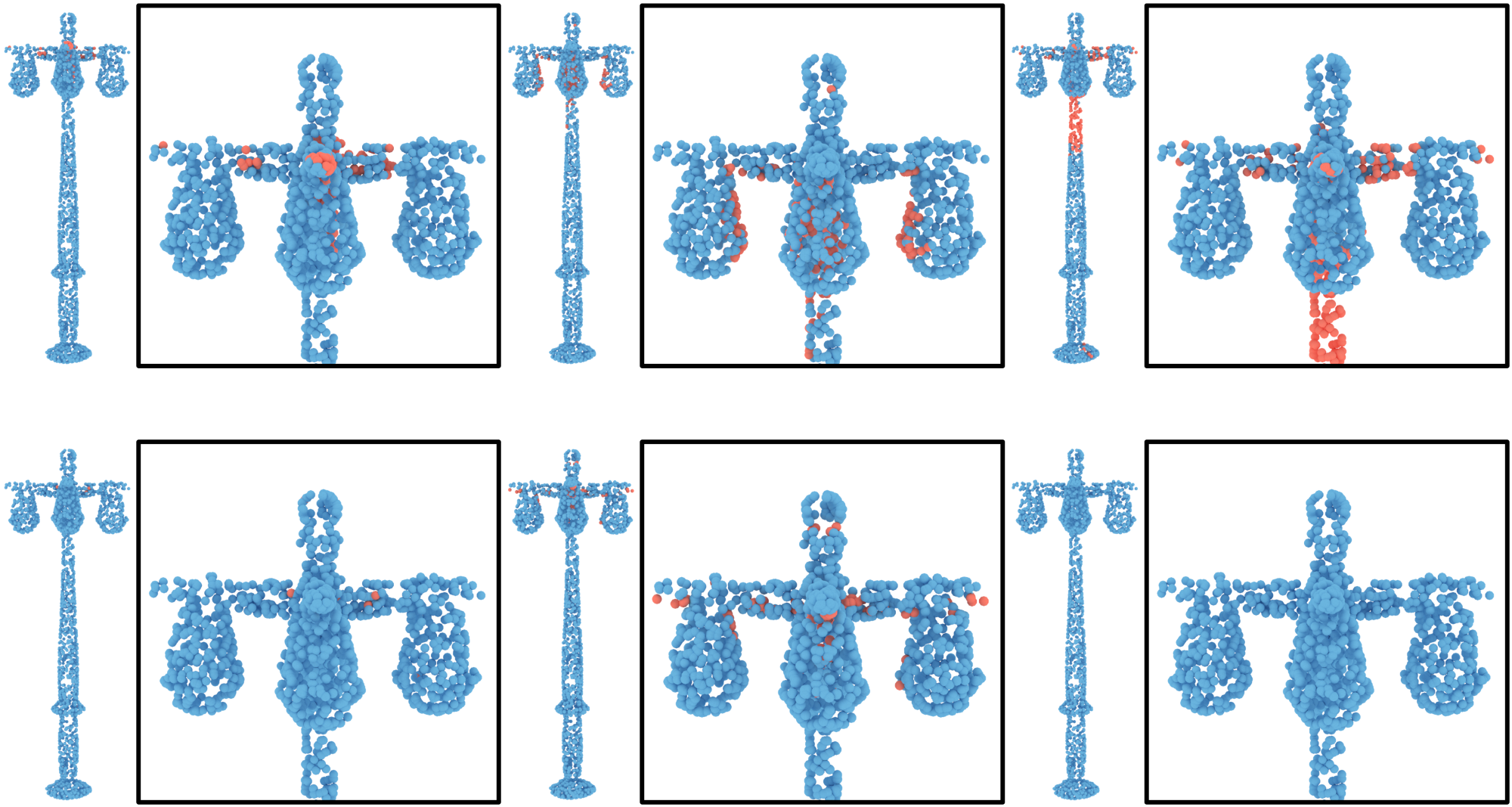}
    \put(11, 25.5){\textbf{K\"{o}nig} \shortcite{konig2009consistent}}
    \put(43, 25.5){\textbf{PCPNet} \shortcite{guerrero2018pcpnet}}
    \put(77, 25.5){\textbf{Dipole} \shortcite{dipole_propagation}}
    \put(12, -3){\textbf{iPSR} \shortcite{hou2022iterative}}
    \put(45, -3){\textbf{PGR} \shortcite{PGR2022Siyou}}
    \put(83, -3){\textbf{Ours}}
    \end{overpic}
    %\vspace{5pt}
    \caption{
    % \XR{Comparison of normal orientation with existing methods. The red point indicates the angle error between the oriented normal and the ground truth normal is larger than $90$ degree.}
    Existing normal orientation approaches are not able to deal with various imperfections such as noise, thin structures, 
    nearby surfaces and sharp features. Note that the red points indicate a false orientation, \ie the angle between the ground-truth normal and the predicted normal is larger than $90$ degrees. 
}
    \vspace{-15pt}
    \label{fig:intro_comp}
\end{figure}
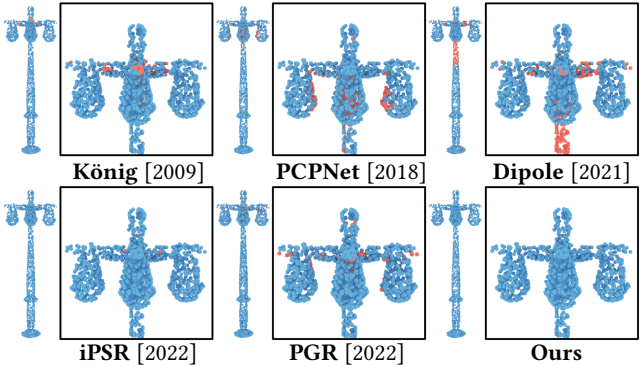

Most of the existing research works~\cite{hoppe1992surface,pauly2003shape,cazals2005estimating,alliez2007voronoi,levin1998approximation}
first compute a normal tensor for each point, regardless of orientation,
followed by spreading the orientation flags through propagation~\cite{dipole_propagation}.
They are not able to deal with various imperfections such as noise,
thin structures, nearby surfaces, and sharp features
since the normals do not rigorously satisfy the property of spatial coherence.
In contrast,
the recently proposed iPSR~\cite{hou2022iterative} and 
Parametric Gauss Reconstruction (PGR)~\cite{PGR2022Siyou} 
focus more on the global consistency of normal orientations,
and achieve better results. 
However, they still suffer from data sparsity coupled with nearby gaps, thin-walled structures, or highly complex geometry/topology.
Fig.~\ref{fig:intro_comp}
demonstrates the results of various approaches,
where the red points indicate a false orientation.

In recent years, the winding number,
as a powerful tool for inside-outside tests,
has gained increasing attention in digital geometry processing,
ranging from meshing~\cite{Hu2018TMW} to reconstruction~\cite{barill2018fast,wang2022restricted}.
% Suppose that the surface is closed and orientable, and 
Despite the ability to distinguish the interior part (the winding number is close to 1) from the exterior part (the winding number is close to 0),
it heavily depends on the support of reliable normals. 
% reveals the global information about normal orientation
% A closed and orientable surface separates~$R^3$ into the interior part and the exterior part,
% where the winding number is valued at $1$ and $0$, respectively.  
% Obviously, the winding number reveals the global information about normal orientation.
% However, they cannot be computed without the support of normals. 
Our hypothesis is that only when the normals are oriented with global consistency, 
the winding-number field could be approximately binary-valued with $1$ and $0$. This inspires us to optimize the normals such that the winding-number field becomes fully regularized. %\XG{Or should we discuss the PGR and its limitation here?}
Based on this hypothesis, we propose an all-in-one functionality to characterize the requirements of a winding-number field
from three aspects:
% \SQ{focus on the idea, delay the implementation details..}
% In the implementation, we have three considerations:
(a)~the winding number should be close to either $1$ or $0$ at any query point,
(b)~when the query points are scattered in the neighborhood of input samples $\sample_i$, the occurrences
of $1$ and the occurrences of $0$ should be approximately balanced,
and
(c)~the sample $\sample_i$'s normal vector should align well with the direction towards the outside Voronoi pole~\cite{amenta1998surface}.
Note that 
the first two requirements are used to regularize the distribution of the winding number 
while the last requirement 
enforces the computed normals 
to be as accurate as the Voronoi-based approaches~\cite{alliez2007voronoi}.
The three terms can be integrated into a smooth objective function
with regard to the normals such that the best configuration of normals can be found by solving an unconstrained optimization.

\begin{figure*}[!t]
	\centering
\vspace{-2mm}
\includegraphics[width=\textwidth]{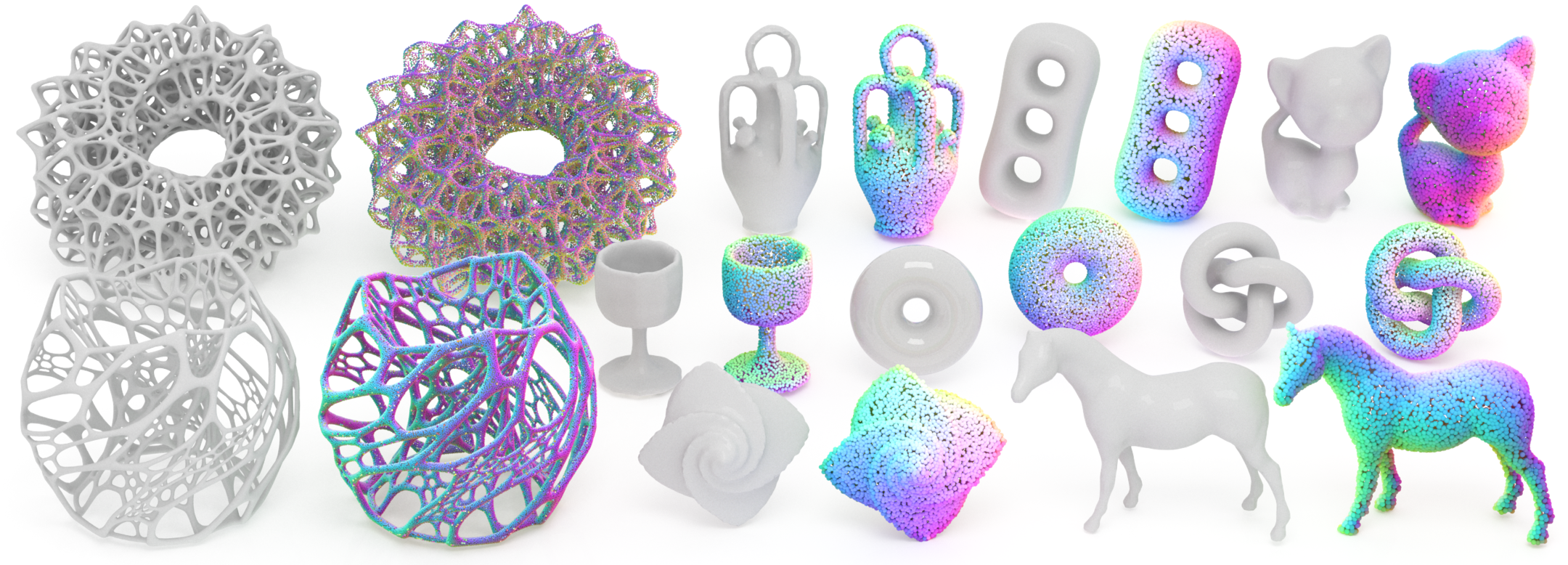}
\vspace{-8mm}
\caption{
We equip the input unoriented point clouds with our computed normals (rendered 
with RGB mapping). By feeding the points and the normals together into the screened Poisson reconstruction (SPR) solver, 
we get high-fidelity reconstruction results (in gray), which exhibit the high quality of the computed normals.
% predicted by our method.
% Note that we equip the input unoriented point clouds with normals (rendered in a color-coded style) predicted by our method. 
}
\label{FIG:gallery}
 \vspace{-3mm}
\end{figure*}

In the implementation, we use L-BFGS to solve the proposed optimization problem. Starting from a completely random normal setting, it generally requires about $30$-$50$ iterations to arrive at the termination.
We use the same set of parameters to test our method on various unoriented point clouds, including synthetic data and real scans. 
Both quantitative statistics and visual comparison show that our method has the advantage of normal accuracy and consistency. 
%Compared with previous SOTA methods, our method has some appealing features. 
It is not only robust to noise and data sparsity (see Fig.~\ref{fig:comp_noise} and Fig.~\ref{fig:comp_lowsampling}),
but also can handle challenging shapes with complex geometry/topology (see Fig.~\ref{fig:comp_diffcult}).
Furthermore, our method can be even applied to incomplete point clouds that encode an open surface (see Fig.~\ref{fig:comp_incomplete}).
In Fig.~\ref{FIG:gallery},
we provide a gallery of results produced by our approach.
% First, our method is robust to noise and data sparsity (see Fig.~\ref{fig:comp_noise} and Fig.~\ref{fig:comp_lowsampling}). Second, our method can handle challenging shapes with complex geometry/topology (see Fig.~\ref{fig:comp_diffcult}). Finally, our method can be even applied to incomplete point clouds that encode an open surface (see Fig.~\ref{fig:comp_incomplete}).

% The contributions of this paper can be summarized as follows:
% \begin{itemize}
%     \item First, we formulate an all-in-one energy optimization framework (in Sec.~\ref{sec:method}) for globally orientating normals of point clouds, via regularizing the winding-number field. Our formulation is mathematically simple yet powerful, and can robustly handle various imperfections, such as nearby features, thin or sharp features and noises.
    
%     \item Second, we creatively reshape and integrate the \textit{double well function} in our optimization (in Sec.~\ref{sec:method_polar}). This function perfectly matches the favorable binary features of the winding-number field, thus provides a compelling tool for winding number regularization. 
    
%     \item Third, we apply the Voronoi-based strategies (in Sec.~\ref{sec:rw_voronoi}) during implementation which effectively improved our robustness to noise and efficiencies. \NW{Not sure, shall we move the ablation study back?} \XG{I think we should move its ablation back to main paper if it is claimed a contribution.}
% \end{itemize}

% As a result, our method significantly outperforms previous SOTA methods while superior robustness to noise. Another appealing feature of our method is its capability to process input point clouds with much fewer points and challenging shapes with thin-plane or highly complex structures, whereas previous methods typically fail for the normal orientation. We conduct extensive experiments to demonstrate the remarkable
% effectiveness of our method for orientating normal vectors of the point clouds across various shapes.

\section{Related Work}
\subsection{Estimating Normal Orientations for Point Clouds}
The problem of point cloud orientation has been extensively researched in the past decades.
%~\cite{hoppe1992surface, pauly2003shape, levin2004mesh, mitra2003estimating, yoon2007surface, amenta1998surface, dey2004provable, dey2005normal, ouyang2005normal, alliez2007voronoi, merigot2010voronoi, wang2012variational, boltcheva2017surface, boulch2016deep, li2022neaf, guerrero2018pcpnet, zhou2020normal, dipole_propagation, zhu2021adafit, xiang2021walk, zhou2022refine, ben2019nesti, lenssen2020deep, ben2020deepfit}. 
In general, attention must be paid to orientation and accuracy for achieving normal consistency. 
% The difference lies in that some approaches divide it into two sub-problems while some approaches take into account the two aspects as a whole. 
Existing methods can be divided into two categories: optimization methods and learning techniques.
The latter can be further divided into regression-based approaches
and surface fitting-based approaches.

% \XR{TODO: add PGR~\cite{PGR2022Siyou}.
% For reference(from Prof. Bin Wang's paper): 
% Among them, PGR~\cite{PGR2022Siyou} treats the normals and surface elements in the Gauss Lemma as unknowns and solves for a globally optimal indicator function. However, the dense linear system of PGR incurs a quadratic complexity of computational time and memory footprint, which limits its practical usage especially on large models \XG{Is our method faster than theirs? If not, we may not want to highlight this fact. In fact, we should highlight those facts shown in our comparison experiments}\XR{No. they are faster. We're better off criticizing him for not being able to handle complex pipes than for needing video memory.}. iPSR~\cite{hou2022iterative} iteratively runs Poisson surface reconstruction~\cite{kazhdan2006poisson} and updates the normals using surface generated in the previous iteration. Although this approach exhibits good performance, there is currently no theoretical guarantee for its convergence. The non-convergence phenomenon may occur in thin structures \XG{as shown in Fig.\ref{fig:comp_diffcult} of this paper}.
% }

\paragraph{Optimization-based Approaches}
Hoppe et al.~\shortcite{hoppe1992surface} pioneered
on normal orientation.
Their approach first uses Principal Component Analysis (PCA)
to initialize the normal tensors,
and then makes their orientations consistent by a minimum spanning tree (MST) based  propagation. 
% by propagation via a minimum spanning tree (MST) where the weight of each edge is assigned as the similarity between the normal of each endpoint. The local estimation of the normal is obtained by Principal Component Analysis (PCA). 
Besides the MST-based propagation, more propagation strategies
include multi-seed~\cite{xie2004surface}, Hermite curve~\cite{konig2009consistent}, and edge collapse~\cite{jakob2019parallel}. The dipole propagation~\cite{dipole_propagation} is also a competing algorithm for propagating the orientation flags. 
In terms of accuracy improvement, 
many techniques are proposed, e.g.,
exponentially decaying function~\cite{levin2004mesh},
local least square fitting~\cite{mitra2003estimating},
truncated Taylor expansion~\cite{cazals2005estimating}, moving least squares~\cite{levin1998approximation}, multi-scale kernel~\cite{aroudj2017visibility}, ensemble framework~\cite{yoon2007surface}.
Wang et al.~\shortcite{wang2012variational}
proposed to minimize a combination of the Dirichlet energy and the coupled-orthogonality deviation such that the normals are perpendicular to the surface of the underlying shape. \ZY{In terms of handling sparse data, VIPSS~\cite{huang2019variational}, as a variational method, reconstructs an implicit surface from an un-oriented point set.}

% \ZY{Dirichlet energy together with coupled-orthogonality deviation~\cite{wang2012variational}} and Voronoi diagram~\cite{amenta1998surface, dey2004provable, dey2005normal, ouyang2005normal, 
%  alliez2007voronoi, merigot2010voronoi, wang2012variational, boltcheva2017surface}.

There are also many research works
on orienting normal vectors for shapes with corners or geometry edges.
For example, L0 norm~\cite{sun2015denoising} or L1 norm~\cite{sun2015denoising, avron2010l1}
is based on the observation that a general surface is smooth
almost everywhere except at some small number of sharp features.
% since these points define a normal cone,
% rather than a single normal vector.
% The literature~\cite {sun2015denoising, avron2010l1, zhang2013point, liu2015quality} then began to take sharp features into consideration. 
% ~\cite{sun2015denoising, avron2010l1} proposed to minimize the L0 or L1 norm to achieve \SQ{???}.
% ~\cite{li2010robust, zhang2013point, liu2015quality} oriented normal vectors based on robust statistics and point clustering locally. 
As each feature point is allowed to own a range of normal vectors, 
Zhang et al.~\shortcite{zhang2018multi}
employed the pair consistency voting strategy to compute multiple normals for feature points.
Xu et al.~\shortcite{xu2022rfeps} used optimal transport to regularize normal vectors for the points nearby geometry edges.
Besides, statistics and subspace segmentation~\cite{li2010robust, zhang2013point, liu2015quality} are used to estimate normals for point clouds with sharp features.

In recent years, much attention has been paid to the global consistency of normal orientations, such as \ZY{Stochastic Poisson Surface Reconstruction (SPSR)~\cite{sellan2022stochastic}}, iterative Poisson Surface Reconstruction (iPSR)~\cite{hou2022iterative} and Parametric Gauss Reconstruction (PGR)~\cite{PGR2022Siyou}. 
For example,
iPSR repeatedly refines
 the surface by feeding the normals computed in the preceding iteration
 into the Poisson surface reconstruction solver.
PGR treats surface normals and surface element areas as unknown parameters, 
facilitating the Gauss formula to interpret the indicator as a member of some parametric function space. 
% viewing surface normals and surface element areas as unknown parameters, the Gauss formula interprets the indicator as a member of some parametric function space.  
Global methods achieve better results than local methods. 
% focus more on the global consistency of normal orientations,
% and achieve better results. 
However, they still suffer from data sparsity coupled with nearby gaps, thin-walled structures
or highly complex geometry/topology. 
For example, iPSR may disconnect thin structures 
%\SQ{???}\XR{The non-convergence phenomenon may occur in thin structures?},
while PGR may generate bulges for tubular shapes. 
Furthermore, PGR's application on large models is constrained by the super-linear growth of GPU memory.
% a quadratic complexity of computational time and memory footprint, which limits its practical usage especially on large models
% For reference(from Prof. Bin Wang's paper): 
% Among them, PGR~\cite{PGR2022Siyou} treats the normals and surface elements in the Gauss Lemma as unknowns and solves for a globally optimal indicator function. However, the dense linear system of PGR incurs a quadratic complexity of computational time and memory footprint, which limits its practical usage especially on large models \XG{Is our method faster than theirs? If not, we may not want to highlight this fact. In fact, we should highlight those facts shown in our comparison experiments}\XR{No. they are faster. We're better off criticizing him for not being able to handle complex pipes than for needing video memory.}. iPSR~\cite{hou2022iterative} iteratively runs Poisson surface reconstruction~\cite{kazhdan2006poisson} and updates the normals using surface generated in the previous iteration. Although this approach exhibits good performance, there is currently no theoretical guarantee for its convergence. The non-convergence phenomenon may occur in thin structures \XG{as shown in Fig.\ref{fig:comp_diffcult} of this paper}.
\paragraph{Regression-based Approaches} {Regression-based methods model normal estimation as a regression or classification task where the surface normals are directly regressed from the feature extracted from the local patches.} Specifically, PCPNet~\cite{guerrero2018pcpnet} 
encodes the multiple-scale features of local patches in a structured manner,
which enables one to estimate local shape properties such as normals and curvature. 
% proposed a multiscale point cloud normal estimation architecture. 
Nesti-Net~\cite{ben2019nesti}
estimates the multi-scale property of a point on a local coarse Gaussian grid, which defines a suitable representation for the CNN architecture and enables accurate normal estimation.
% \cite{ben2019nesti} predicted the optimal scale rather than directly concatenating multiple scales.
Zhou et al.~\shortcite{zhou2020normal} 
proposed a multi-scale selection strategy to select the most suitable scale for each point through a joint analysis of multiscale features.
% For single-scale normal estimation, Local Plane Features Constraint (LPFC) is used in our networks to ensure the robustness of the normal estimation network. Besides, the binary classifier used in our LPFC can well obtain the main part of the patch, especially when the sampling scale is large.
% •
% A scale selection strategy for scale prediction is employed in our method. In this paper, we propose a novel scale estimation network, which is used to select the most suitable scale of each point through a joint analysis of multiscale features extracted from single-scale networks.
% extended the scale selection strategy and imposed extra constraints on the features.
% However, learning multiple neighborhood scales jointly leads to the phenomenon of over-smoothing. 
Hashimoto and Saito~\shortcite{hashimoto2019normal}
used a point network and a voxel network to estimate normal vectors without sacrificing
the inference speed.
% To efficiently incorporate local and spatial structures, 
% propose a joint model that combines both local and global features. 
Although the regression-based methods typically outperform traditional data-independent methods, the regression-based methods rely on a large amount of training data for network training and 
are limited by the generalization capability because the brute-force training course may cause the network to overfit the normal vectors from the training data.

% Although the aforementioned normal estimation methods
% often achieve better results than traditional data-independent methods,
% they have to depend on a large amount
% of training data to obtain a normal prediction neural model.
% {Although the regression-based methods often achieve better results than traditional data-independent methods, they are weak in generalization capability and sensitive to noisy inputs because the brute-force regression manner forces the network to memorize normal vectors. }
%To tackle the issue, Surface fitting-based methods are proposed.}
% \SQ{is true?} 

\paragraph{Surface fitting-based approaches}  Different from those regression-based methods, surface fitting-based approaches estimate a fitting surface by taking advantage of its neighboring points.
%\XG{should iPSR and PGR both surface-fitting approaches? They are not local fitting, but global fitting instead?}. 
In particular, 
% Observing that the normal estimation problem 
% can be formulated as an iteratively re-weighting least squares (IRLS) scheme,
Lenssen et al.~\shortcite{lenssen2020deep}
% formulated the normal estimation problem  as an iteratively re-weighting least squares (IRLS) scheme
% and 
presented a light-weight graph neural network 
that parameterizes a local quaternion transformer and a deep kernel function to iteratively re-weight graph edges in a large-scale point neighborhood graph.
DeepFit~\cite{ben2020deepfit} achieves scale-free normal estimation by
per-point weight estimation for weighted least squares.
Zhu et al.~\shortcite{zhu2021adafit} predicted an additional offset to improve the quality of normal estimation.
% adjust the distribution of clouds and introduced a cascaded scale aggregation layer adapting to the scale of the local neighborhood. 
% adds an additional offset prediction to improve the quality of normal estimation
% the fitting technique  are proposed~\cite{lenssen2020deep, ben2020deepfit, dipole_propagation, zhu2021adafit}. Specifically, \cite{lenssen2020deep} and ~\cite{ben2020deepfit} first
% utilize the network to predict point-wise weights for the selection of the neighboring points, then estimate the surface normal by the differentiable and weighted least squares plane or polynomial surface fitting. 
% ~\cite{zhu2021adafit} additionally predicted an offset to adjust the distribution of clouds and introduced a cascaded scale aggregation layer adapting to the scale of the local neighborhood. 
Recently, the dipole propagation~\cite{dipole_propagation}
establishes a consistent normal orientation in a local phase and a global phase. 
However, tests show that dipole cannot deal with the point sparsity or tubular structures.
% \ZY{dipole typically fails when the input point cloud is with sparse points or encodes high genus shapes. It also performs poorly on the shapes with slice plane or tubular structures.}

% \SQ{what's the difference between this paragraph and the previous graph?} \ZY{ updated. The first paragraph covers regression based methods (directly predict the normal for each point on the point cloud.) The second paragraph represents another solution (better than those in the first paragraph) called the "surface fitting based method" (first fitting a local surface at the point then estimate the normal using the local surface). The rationale for proposing the second style is to overcome the weak generalization capability and sensitivity to noisy inputs of the first one. }

Although deep learning approaches
show great potential in normal estimation,
it is still notoriously hard for both point-based regression approaches and 
surface fitting-based approaches
to robustly deal with 
different noise levels, outliers, thin-plate structures, and varying levels of detail.

% challenging task due to difficulties associated with sampling density, noise, outliers, and detail level.
% It so the local and global components into two different sub-problems.
% proposed to achieve normal orientation for point clouds with dipole propagation. The key idea is to solve the local and global sub-problems one by one by first training a neural network to learn a coherent normal direction at the patch level and then to propagate the orientation across all coherent patches using dipole propagation.
% However, learning-based techniques typically rely on sufficient training data and suffer from the generalization problem.
% Our method, as an optimization framework, is able to handle various shapes robustly, yielding pleasing normal orientation results.

\subsection{Voronoi-based Normal Orientation}
\label{sec:rw_voronoi}
Voronoi diagrams, as a powerful tool to encode spatial proximity, are extensively used to estimate normal vectors~\cite{amenta1998surface, dey2004provable, dey2005normal, ouyang2005normal, 
 alliez2007voronoi, merigot2010voronoi, wang2012variational, boltcheva2017surface, kolluri2004spectral, grimm2011shape}. 
% Adopting Voronoi Diagram for point cloud normal estimation has a long history~\cite{amenta1998surface, dey2004provable, dey2005normal, ouyang2005normal, 
%  alliez2007voronoi, merigot2010voronoi, wang2012variational, boltcheva2017surface}. 
Amenta and Bern~\shortcite{amenta1998surface} 
proved that when the point density satisfies the standard 
of local feature size,
one can roughly recover the real normals 
and even construct a discrete interpolation-type surface that is conformal to the base surface. 
The central idea is to identify inside poles and outside poles 
from the Voronoi diagram of the input point cloud,
and use the poles to help orient the point cloud and assign their normals. 
Observing the Voronoi diagram 
can locally represent the most likely direction of the normal to the surface, Alliez et al.~\cite{alliez2007voronoi} 
proposed to compute an implicit function by solving a generalized eigenvalue problem. It can be seen from the existing approaches that inside poles and outside poles are robust to noise,
which helps find the dominant Delaunay balls in a noise-resistant manner~\cite{dey2004provable}. 
Generally speaking, Voronoi diagrams 
% are a useful tool to orient a point cloud,
% but the limitation lies in that Voronoi Diagrams are computed based on straight-line distances,
can produce faithful results for dense point clouds
but are weak in dealing with thin-plate structures or sharp features. 
In this paper, we thoroughly investigate the winding number by analyzing all the Voronoi vertices of the input point cloud. Our approach utilizes the winding-number requirements to ensure global normal consistency while relying on the Voronoi diagram to accurately predict the normals.

\subsection{Winding Number}
The winding number was first introduced by~\cite{meister1769generalia}.
For a smooth manifold surface,
it can be computed using a contour integral in complex analysis.
As a powerful tool for inside-outside tests,
it has been widely used in many higher-level geometry processing operations including tetrahedral meshing~\cite{hu2020fast}, reconstruction~\cite{barill2018fast,wang2022restricted}, normal orientation for point clouds~\cite{dipole_propagation}, shape analysis~\cite{wang2022computing}, shape modeling~\cite{sellan2021swept}, animation~\cite{nuvoli2022skinmixer}.
For example,
Jacobson et al.~\shortcite{jacobson2013robust} introduced a winding-number-based function to guide an inside-outside segmentation of a polygonal surface. %flawed input mesh.

% segmentation of 
% that is able to handle the mesh inputs with self-intersections, open boundaries, and non-manifold pieces by utilizing the harmonic nature of the winding number. 
% We only require reasonably consistent orientation of the input triangle mesh. By generalizing the winding number for arbitrary triangle meshes, we define a function that is a perfect segmentation for watertight input and is well-behaved otherwise. This function guides a graphcut segmentation of a constrained Delaunay tessellation (CDT), providing a minimal description that meets the boundary exactly 
Barill et al.~\shortcite{barill2018fast} derived a differential form of the winding number function and gave a tree-based fast algorithm to reduce the asymptotic complexity of generalized winding number computation,
and also demonstrated a variety of new applications.
% \ZY{including tetrahedral meshing~\cite{hu2020fast}, reconstruction~\cite{barill2018fast,wang2022restricted}, normal orientation for point clouds~\cite{dipole_propagation}, shape analysis~\cite{wang2022computing}, shape modeling~\cite{sellan2021swept}, animation~\cite{nuvoli2022skinmixer}, just to name a few.} 
% proposed to further generalized the winding number to point clouds and proposed a hierarchical algorithm for fast evaluation, which significantly accelerated the speed. Both classic and generalized winding numbers require orientation. 
% \SQ{I guess there are other related works on winding number...
% for example, tet mesh in the wild? } \ZY{done.}

It's known that if the input point cloud is equipped with a meaningful normal setting,
the winding number can robustly distinguish the inside from the outside in a global manner and is valued at 1~(inside) and 0~(outside). 
This observation motivates us to regularize the winding-number field by repeatedly tuning the normals so that they become consistent. 

%Note that the winding numbers are examined at only inside poles and outside poles to guarantee the noise resistance. 

% we consider from a different angle by revealing and demonstrating that the generalized winding number could serve for the globally consistent normal orientation of the point clouds.

% \subsection{Normal..}
% \ZY{?}

\begin{figure}[!tp]
\centering
\begin{overpic}
[width=0.99\linewidth]{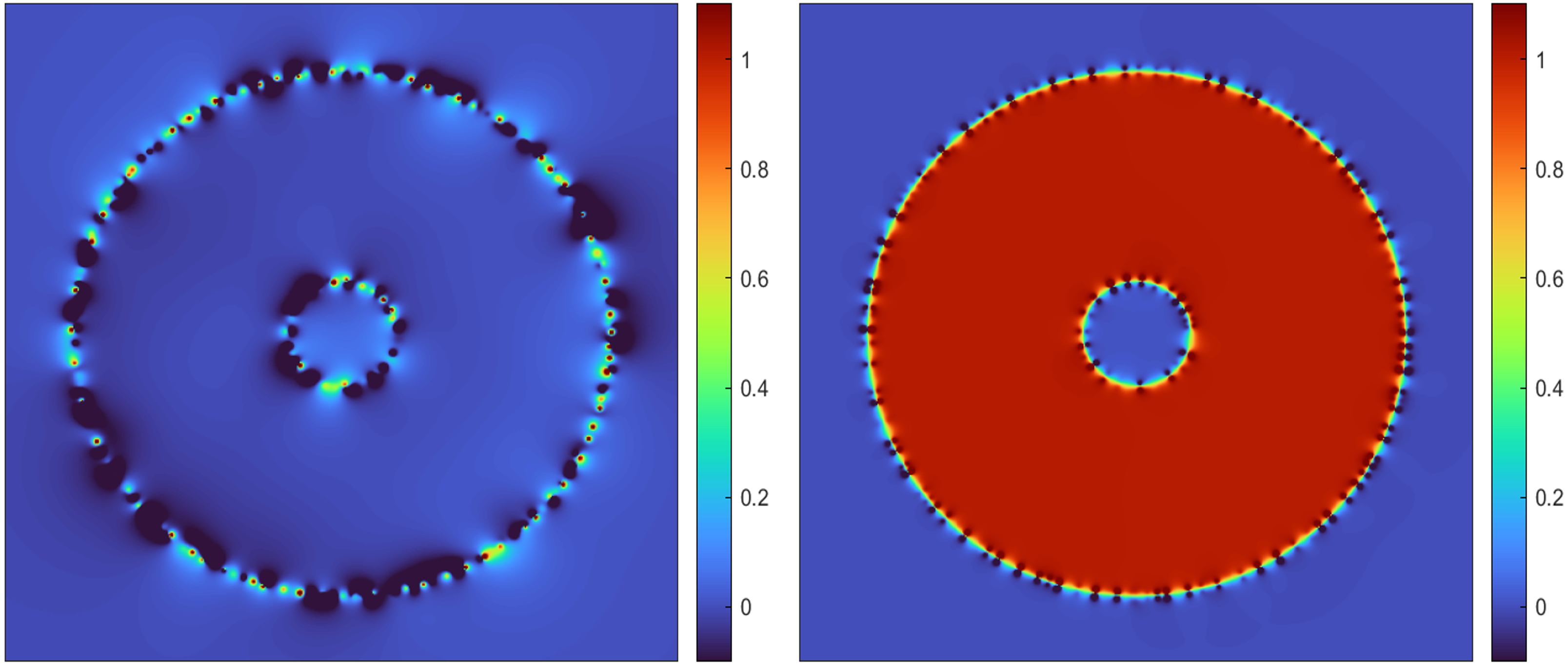}
\put(6,-3.5){\textbf{(a)~Random Normals}}
\put(52,-3.5){\textbf{(b)~Ground Truth Normals}}
\end{overpic}
\vspace{3pt}
\caption{(a)~If the normals are random,
the winding number tends to be $0$ everywhere. 
(b)~If the normals can encode a closed and orientable shape, 
the winding number is valued at $1$ (interior) and $0$ (exterior).
}
\vspace{-4mm}
\label{fig:2dwinding}
\end{figure}

% \begin{figure}[!htp]
%     \centering
%     \begin{overpic}[width=0.99\linewidth]{figures/2_pre/dw_voronoi.pdf}
%     \put(15, -5){(a)}
%     \put(70, -5){(b)}
%     \end{overpic}
%     \vspace{5pt}
%     \caption{(a)~A simple form of \textit{double well function} in Eq.~\ref{eq:doublewell}. (b)~Voronoi diagram in 2D. The Voronoi cell $\vorcell_i$ of point $\sample_i$ is highlight in blue with Voronoi vertices in orange.\XR{Both are no used.}}
%     \label{fig:dw_voronoi}
% \end{figure}

\section{Preliminaries}
% \SQ{to continue...}
\subsection{Generalized Winding Number}
\label{sec:pre_winding}
The theory of winding number can be generalized
to polygonal meshes~\cite{jacobson2013robust}, triangle soup and point clouds~\cite{barill2018fast}.
% The \textit{generalized winding number} is a robust method for determining whether a point is inside or outside of a point cloud ~\cite{barill2018fast}.\XR{Point cloud? Generalized should be mesh? Do we need to mention the mesh winding number in the preliminaries?} 
Suppose $\{\sample_i\}_{i=1}^N$ are samples from a continuous surface with normals $\{\normal_i\}_{i=1}^N$.
% Given a point cloud~$\{\sample_i\}_{i=1}^N$ with normals $\{\normal_i\}_{i=1}^N$, 
The generalized winding number~$w$ at the query point $\query$ can be expressed as an area-weighted sum of the overall contribution of the point set~\cite{barill2018fast}:
\begin{equation}
w(\query) = \sum_{i=1}^N \area_i \frac{(\sample_i - \query)\cdot \normal_i}{4\pi\left \| (\sample_i - \query) \right \|^3},
\label{eq:windingnum_ori}
\end{equation}
where $\area_i$ is the dominating area of the point $\sample_i$. 
Obviously, the normals are central to the computation of winding numbers. 
When the normals are random, see Fig.~\ref{fig:2dwinding}~(a),
the winding number tends to be $0$ everywhere. 
If the normals can encode a closed and orientable shape, instead, see Fig.~\ref{fig:2dwinding}~(b), 
the winding number is about $1$ for the interior points
and $0$ for the exterior points.
% we show a 2D example of winding number. 
% For the noise-free situation~(see Fig.~\ref{fig:2dwinding}(a)),\XR{Need to change.}
% the winding number is valued at 0 and 1 if $\area_i$ is estimated accurately. 
% \SQ{to revise..}
% If we add extra noise to the input point set, 
% the winding number may have a small perturbation but the overall distribution remains almost unchanged~(see Fig.~\ref{fig:2dwinding}(b)).\XR{Move to Fig.4?}

\noindent{\bf Remark: }
How to estimate $\area_i$ is a problem when the base surface is not available.
A commonly used technique~\cite{barill2018fast} is to project  
$\sample_i$'s $k$-nearest neighbors onto the tangent plane of $\sample_i$. Thus $\area_i$ is approximated by the area of the $\sample_i$'s cell of the 2D Voronoi diagram.
However, this strategy depends on the choice of~$k$.
% However, it has to depend on the choice of~$k$.
Since the estimation of~$\area_i$ is essential to the computation of the winding number, we adopt a parameter-free strategy in Sec.~\ref{sec:Implementation}.
% \ZY{projecting each point $p_i$ and its K-nearest neighbors ($k=20$) onto a plane and computing area of the 2D Voronoi cell of $p_i$ on the plane.}
% ...\SQ{@zhiyang, please complete this paragraph } \ZY{done. Actually, there can be many ways to estimate $a_i$, I choose the one in Fast Winding number.}
% The estimation of $\area_i$ will be discussed in Sec.~\NW{TODO}.
% This bounded polarization feature empower us to apply normal orientation for point clouds.

\subsection{Voronoi Vertices for Examining Winding Number}
\label{sec:pre_voro}

\begin{figure}[!tp]
    \centering
    \begin{overpic}[width=\linewidth]{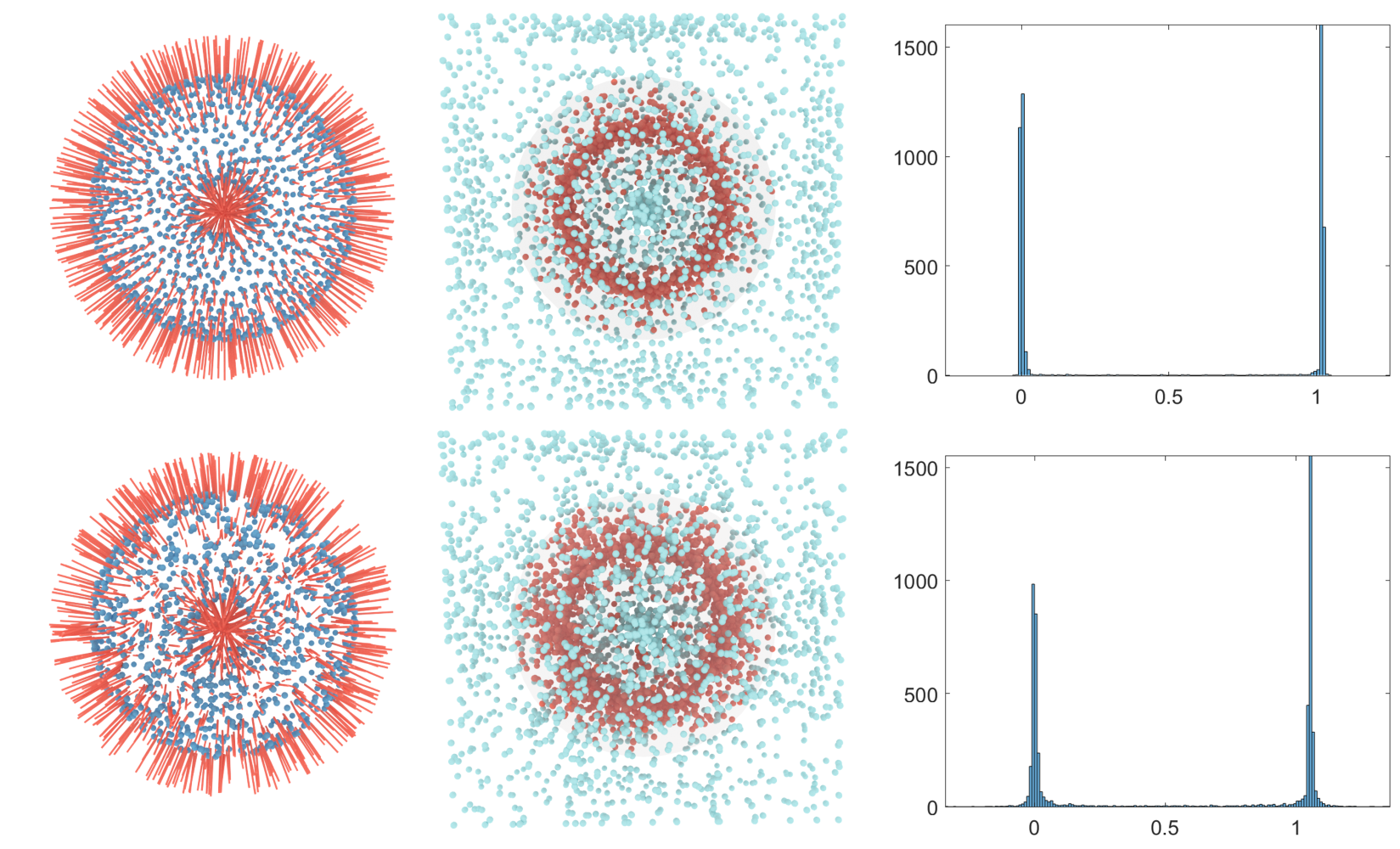}
    \put(3, -3){(a)~Point cloud}
    \put(30, -3){(b)~Voronoi vertices}
    \put(67, -3){(c)~WN distribution}
    \end{overpic}
    %\vspace{3pt}
    \caption{
    Assuming the point cloud is equipped with a ground-truth normal setting, we examine the winding number at the Voronoi vertices of the point cloud.
    Top row:~For a noise-free point cloud, the winding number is close to either $0$ or $1$ for Voronoi vertices. 
    Bottom row:~When noises are added to point positions, the histogram of the winding number 
    has a slight change, but still demonstrates two peaks close to 0 and 1.
    % \XR{Top row: uniform and noise free. Bottom row: not uniform and noisy.}
    % Given a point cloud equipped with normals~(a,b),
    % we examine the winding number at the Voronoi vertices of the point cloud.
    % The Voronoi vertices whose winding number is close to $1$
    % are colored in red, while the Voronoi vertices whose winding number is close to $0$ are colored in cyan.
    % we pre-compute the inside poles (red) and the outside poles (cyan) of the point set 
    % for examining the winding number (b).
    Note that the intersection points between the Voronoi edges and the 1.3x bounding box are also included for the winding number query. 
    }
    \label{fig:3dvoronoidis}
    \vspace{-10pt}
\end{figure}

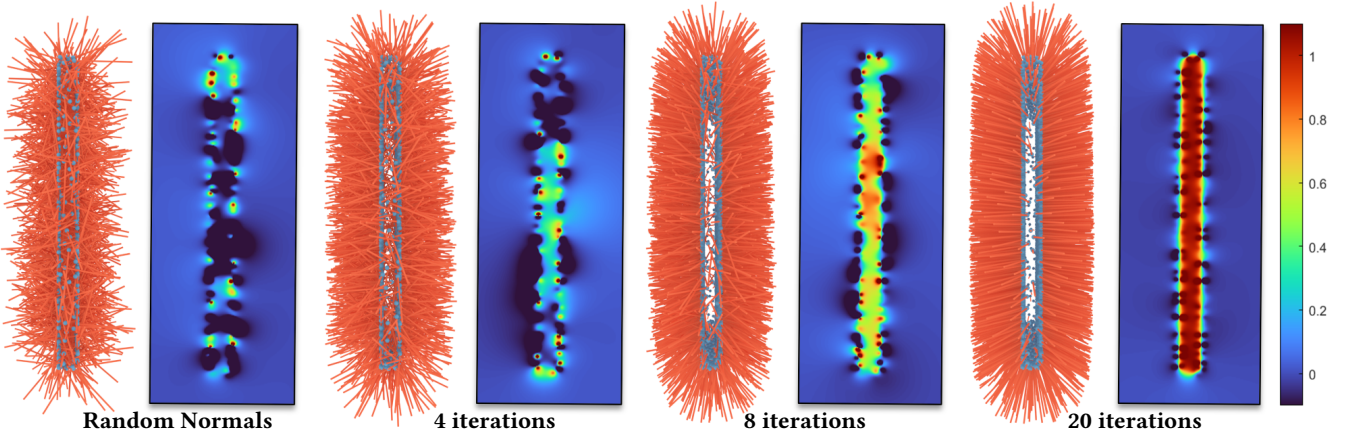
\begin{figure*}[t]
\centering
\begin{overpic}
[width=\textwidth]{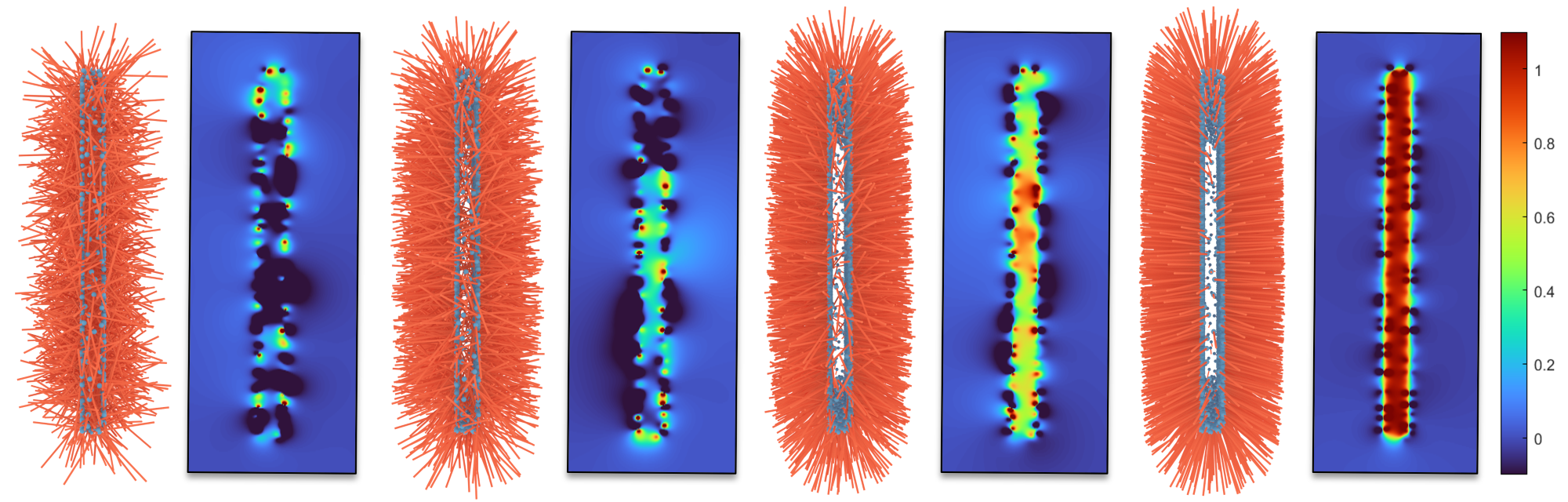}
\vspace{3pt}
\put(7,0){\textbf{Random Normals}}
\put(33,0){\textbf{4 iterations}}
\put(56,0){\textbf{8 iterations}}
\put(80,0){\textbf{20 iterations}}
\end{overpic}
\vspace{-4mm}
\caption{
The optimization progress using our method. The input model is a thin board with randomly initialized normals (in orange) of the point cloud. 
We cut the board in the middle with a plane 
to show the sectional view of the winding-number field (visualized in a color-coded style).
% During the iteration, we also show the winding number distribution in the color map of all query points in a slice plane. 
% \ZY{"in" a slice plane? maybe "for"?} \XR{Maybe not mention "query points", here should be a field?}
}
\label{fig:iteration}
\end{figure*}

% \begin{wrapfigure}{r}{0.28\textwidth}
% \vspace{-3.5mm}
%   %\hspace{-15mm}
%   \begin{center}
%     \includegraphics[width=50mm]{figures/2_pre/fig_voronoi.pdf}
%   \end{center}
%   %\hspace{-5mm}
%   \vspace*{-4mm}
%   \vspace{-5mm}
% \end{wrapfigure}

\begin{wrapfigure}{r}{4cm}
\vspace{-3.5mm}
  \hspace*{-4mm}
  \centerline{
  \includegraphics[width=45mm]{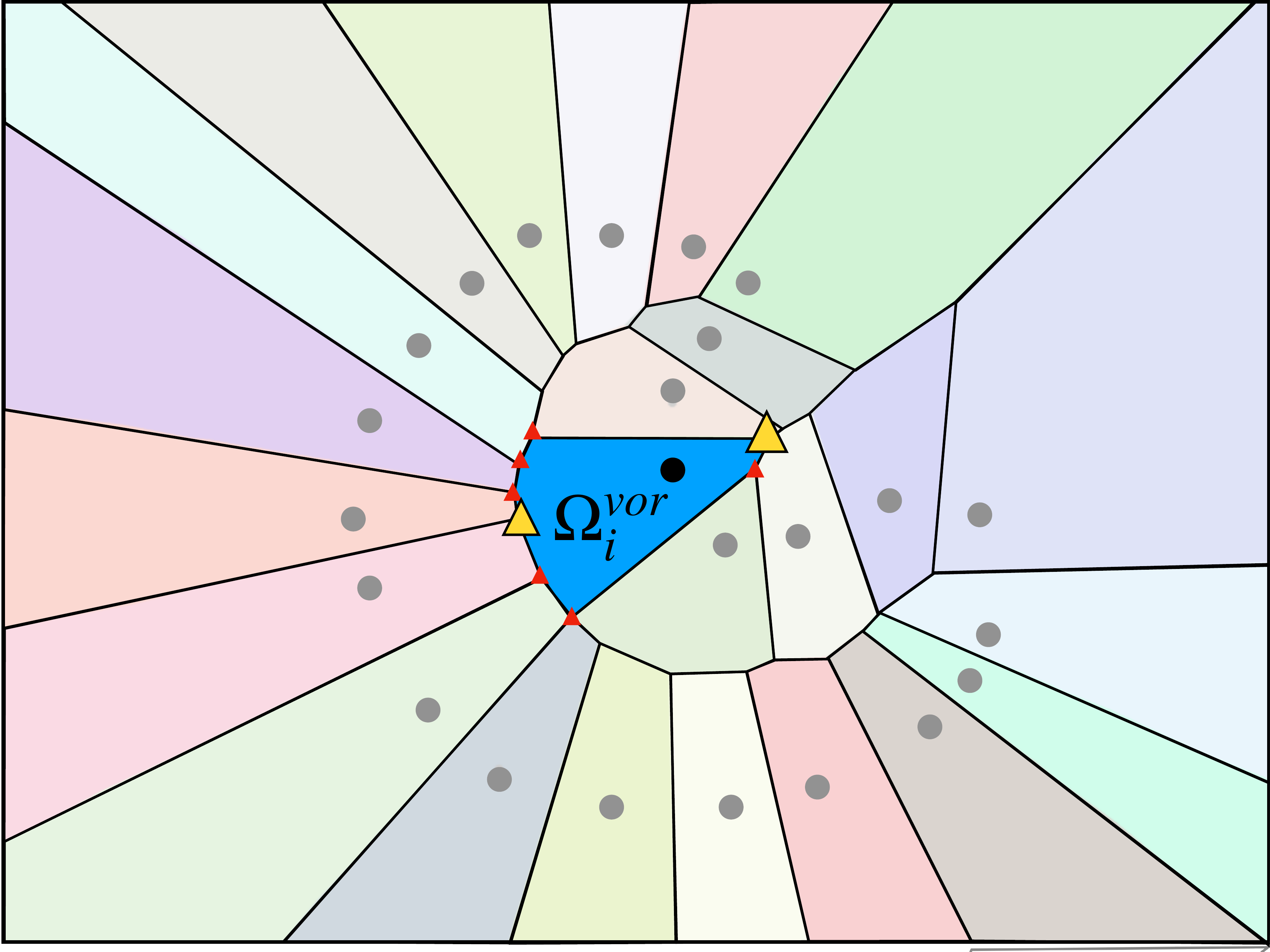}
  }
  \vspace*{-4mm}
\end{wrapfigure}

The \textit{Voronoi diagram} (VD) of a set of points $\{ \sample_i \}_{i=1}^{N}$ 
partitions the entire space into $N$ cells based on spatial proximity. 
In 3D, it includes Voronoi vertices, Voronoi edges, Voro- noi faces, and Voronoi cells as the atomic elements. 
Let $\vorcell_i$ be the cell of $\sample_i$;
See the 2D example in the inset figure.
The two farthest vertices of~$\vorcell_i$, located on both sides of the surface, are defined as poles~\cite{amenta2001power}, which are helpful for orienting normals.
Note that the inside poles and the outside poles are hard to be distinguished before the normals are determined.
Therefore, in this paper, we use all the Voronoi vertices, a superset of the Voronoi poles, for examining the winding number given by a point cloud. 

%weigh ? by the winding-number score,
%rather than insist on identifying inside poles and outside poles.
% \NW{We have observed that the distributions of the winding number are less affected by noises when equipped with faithful normals.}

As shown in
the top row of Fig.~\ref{fig:3dvoronoidis}(c), the winding number is close to either $0$ or $1$ at the Voronoi vertices for a noise-free point cloud. If we add noises to point positions at a level
of 0.5\%, the histogram just changes slightly (see
the top row of Fig.~\ref{fig:3dvoronoidis}(c)). 
Note that in Fig.~\ref{fig:3dvoronoidis}~(b), the Voronoi vertices are colored in red (resp. cyan) if the winding number is close to $1$ (resp. $0$).
% whose winding number is close to $1$ are colored in red, while the Voronoi vertices whose winding number is close to $0$ are colored in cyan.
% \XR{Do we need to talk about the distribution of the WNF?}
Besides, we use a 1.3x bounding box to enclose the point cloud and add the intersection points between the Voronoi edges and the box as examination points.
% Note that the intersection points on the box are forced to own a winding number of~$0$.
% \XR{Do we really need this? }
One may consider a different strategy for generating the examination points, e.g., adding Gaussian noise to the input point cloud. 
Based on our tests, most of the Voronoi vertices are 
remote from the surface and noise-insensitive,
which accounts for why we take the Voronoi vertices as examination points.
We conduct the ablation study in Supplementary Material.

\section{Method}
\label{sec:method}

% \subsection{Winding Number}

% Given an input point cloud with normals, the generalized winding number~\cite{barill2018fast} is given by

% \ZY{add continuous definition.}
% \begin{equation}
% w = \sum_{i=1}^Na_i \frac{(p_i - q)\cdot \normal_i}{4\pi\left \| (p_i - q) \right \|^3}.
% \label{eq:windingnum_ori}
% \end{equation}
% \ZY{TBC.}
% The result of the winding number can then be used for inside-outside tests; See Fig.~\ref{} for a demonstration.

% Fig.~\ref{}.
% \ZY{some features of Winding number.}
% \paragraph{feature1}

% \paragraph{feature2}

% \subsection{Optimization Framework}
% Driven by the aforementioned features, we xx.
The winding number,
in its nature,
can reflect global inside-outside information,
which motivates us to compute the normals by regularizing the winding-number field. 
In the implementation, we examine the winding number at the Voronoi vertices of the point cloud. 
% (We take all the Voronoi vertices as examination points, which is a little different from the definition of poles.) 
We hope that
the computed normals can not only lead to a reasonable 
winding-number field
but also accurately align with Voronoi poles.
The requirements can be summarized into the following three aspects.

\paragraph{$w(\query)$ is valued at 0 or 1.}
Although one can construct a surface such that
the winding number is valued at any integer, 
we only consider the common case where the winding number is either $0$ or $1$. 
% When the normals are correctly set, the winding number is close to either 0 or 1.
%which is the main requirement enforced on the winding number. 
We shall include a term~$f_{01}(\normal)$ to characterize the basic requirement of a valid winding-number field.

\paragraph{The winding-number values are balanced for $\sample_i$'s Voronoi vertices.} 
Sample $\sample_i$ dominates a cell in the Voronoi diagram.
In general cases, it is unlikely that all the Voronoi vertices of $\sample_i$'s cell are located inside or outside. 
Therefore, we hope the number of $1$'s and the number of 0's are balanced when we consider the winding number of $\sample_i$'s Voronoi vertices. This observation leads to a balance term~$f_{B}(\normal)$.

\paragraph{Normals align with Voronoi poles.}
Like the power crust techniques~\cite{amenta2001power},
Voronoi poles are very helpful in predicting the normals.
Let $\query_{k}^i$ be the $k$-th Voronoi vertex of $\sample_i$'s Voronoi cell. 
We hope the vector $\query_{k}^i- \sample_i$ has similar orientation with $\normal_i$ if $w(\query_{k}^i)\approx 0$
but reverse orientation with $\normal_i$ if $w(\query_{k}^i)\approx 1$.
The alignment requirement leads to a term~$f_{A}(\normal)$.

By summarizing them together,
we get a functional w.r.t. the normals,
\begin{equation}
f(\normal) =  \frac{f_{01}(\normal) + \lambda_{B} f_{B}(\normal) + \lambda_{A}f_{A}(\normal)}{N},
\label{eq:all}
\end{equation}
where $\lambda_{B}$ and $\lambda_{A}$ 
are two parameters to tune the influence of 
$f_{B}$ and $f_{A}$, respectively. 
We establish the details of the separate terms in the following subsections,
while delaying the ablation study of $\lambda_{B}$ and $\lambda_{A}$ 
in supplementary material.
Fig.~\ref{fig:iteration} gives an example of how the normals 
change with the decreasing of the value of~$f$.

\begin{figure}[t]
    \centering
\begin{overpic}[width=\linewidth]{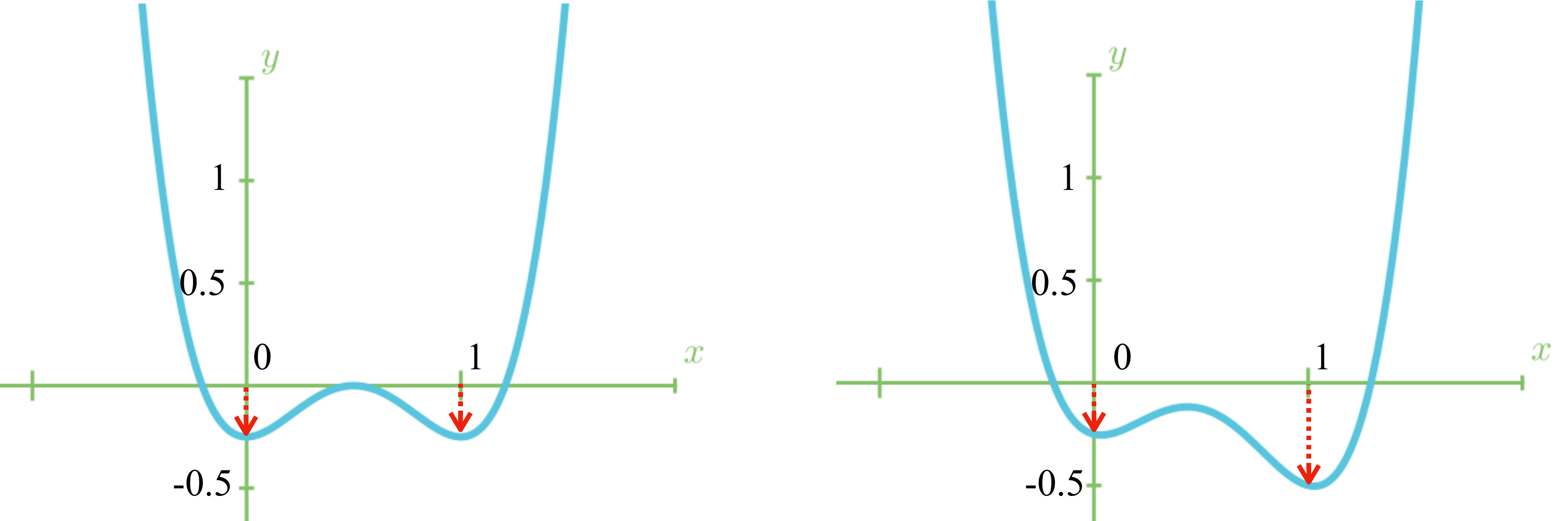}
    \put(5, 0){(a)}
    \put(59, 0){(b)}
\end{overpic}
    \caption{(a) A standard double well function. (b) A new version with a shear correction for suppressing the randomness of normals. }
    % Two variances of simple double well functions (Sec.~\ref{sec:pre_double_well}): (a) Eq.~\ref{eq:doublewell} w/o last term $\frac{w_j}{D}$, (b) Eq.~\ref{eq:doublewell}. The detail discussion is given in the ablation study in Sec.~\ref{sec:ablation}.
    \label{fig:dw_variance}
\end{figure}

\subsection{The 0-1 Term~$f_{01}$}
\label{sec:method_polar}
\paragraph{Double well function}
In the continuous setting, 
the winding number is valued at 0 or 1 
if the input surface is closed and topologically equivalent to 
a single-layer orientable surface. 
Therefore,
we need to define an energy function to pull the winding number to the binary states as far as possible. 
For this purpose, we introduce the {double well function} 
inspired by one of the most important functions in the field of quantum mechanics~\cite{jelic2012double}. 
A simple form of the double well function can be written as:
\begin{equation}
    f_{DW}(x) = 4\left(x-0.5\right)^4 - 2\left(x-0.5\right)^2,
\end{equation}
with two valleys at $x=0$ and $x=1$, respectively,
as Fig.~\ref{fig:dw_variance}(a) shows.
% The parameter $\sigma$ is used to tune the rate of change,
% and chosen to 0.5 in this paper.
% \XR{Maybe 'Scale' not 'Rate of change'?}

\paragraph{A new double well function with a shear correction}
If we equip a point cloud with a set of random normals,
the resulting winding number tends to be 0 for an arbitrary query point; See Fig.~\ref{fig:2dwinding}(a).
In order to encourage the occurrence of $1$'s for the winding number of examination points,
we need to tune the double well function with a shear correction, as Fig.~\ref{fig:dw_variance}(b) shows.
In this way, the 0-1 term~$f_{01}$ can be defined by  
the overall contribution of the winding number $w_{j}=w_{j}(\normal)$ at $\query_j,j=1,2,\cdots,M.$
\begin{equation}
f_{01}(\normal)= \sum_j^{\numV}\left(f_{DW}(w_{j})-\frac{w_{j}}{D}\right),
% (\frac{w_{j}-0.5}{\sqrt{\sigma}})^4-(\frac{w_{j}-0.5}{\sqrt{\sigma}})^2-\frac{w_{j}}{D},
\label{eq:doublewell}
\end{equation}
where the parameter $D$ is used to tune the degree of shear correction.
We make the ablation study about $D$
in the supplementary material
and empirically set $D=4$ for all the experiments. 

\subsection{The Balance Term $f_{B}$}
\label{sec:method_variance}

\begin{wrapfigure}{r}{2cm}
\vspace{-3.5mm}
  \hspace*{-5mm}
  \centerline{
  \includegraphics[width=26mm]{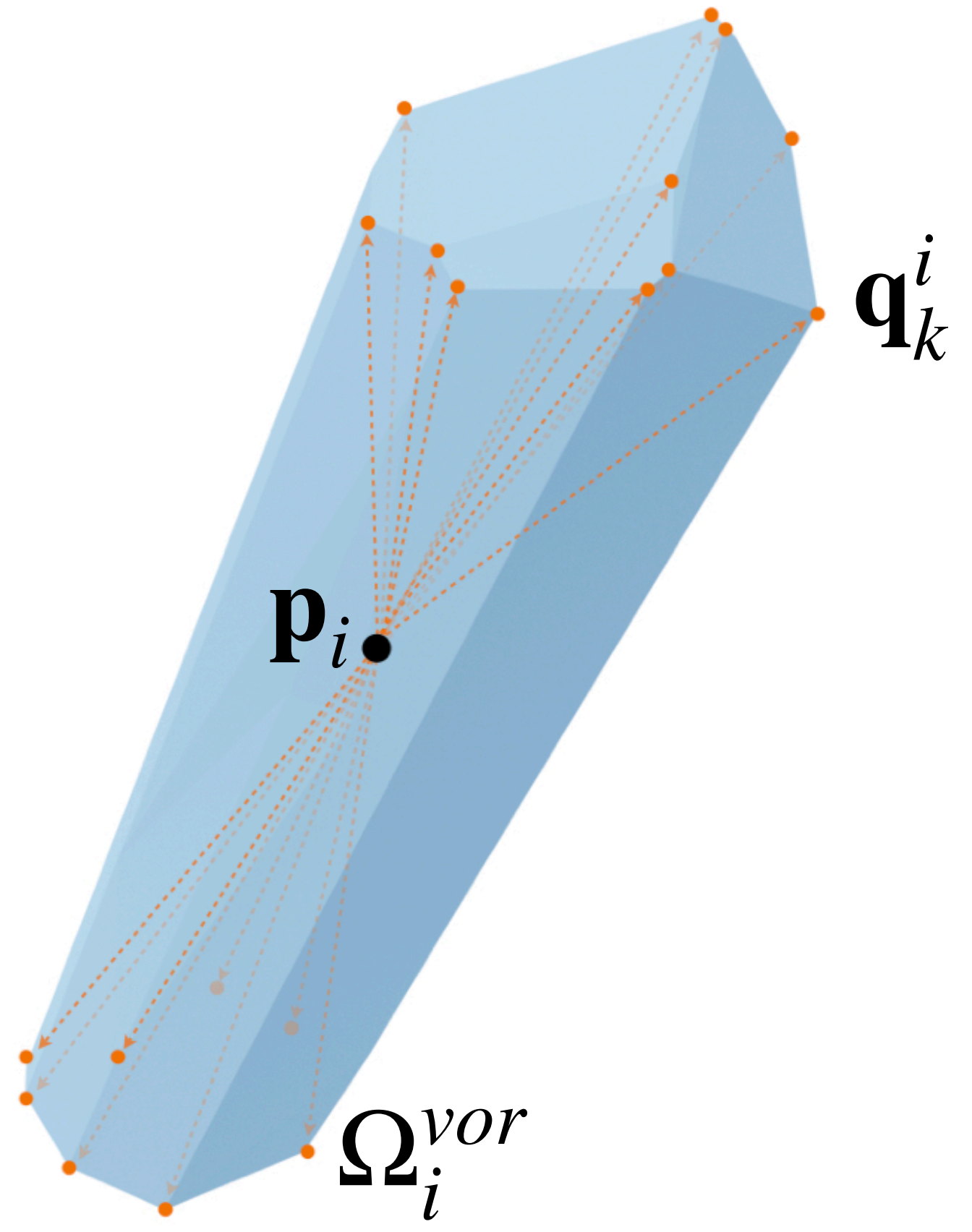}
  }
  \vspace*{-4mm}
\end{wrapfigure}

Let $\vorcell_i$ be the Voronoi cell dominated by the point 
$\sample_i$ of the point cloud.
If the point density meets the local feature size standard~\cite{amenta1998surface},
one half of $\vorcell_i$ is located inside the surface,
and the other half is located outside. 
Therefore,
it is reasonable to suppress the occurrence of the situation
that all vertices of $\vorcell_i$ are inside the shape or outside the shape. 
In other words,
the winding-number scores at the vertices of~$\vorcell_i$
should be balanced, which can be achieved
by maximizing the variance of the winding-number scores.
Let $\overline w^i$ be the average score for $\vorcell_i$.
The variance can be measured by $\sum_k^{M_i}(w_k^i- \overline w^i)^2$,
where $M_i$ is the total number of vertices of $\vorcell_i$,
and $w_k^i=w(\mathbf{q}_k^i)$ is the winding-number score for the $k$-th vertex $\mathbf{q}_k^i$.
The balance term can be defined by the overall winding-number variance.
% \XR{Old caption: $\sample_i$'s Voronoi cell $\vorcell_i$ has $M_i$ vertices, i.e., $\query_k^i,k\in[0, M_i]$.}
% then 
% , the ideal Voronoi vertices of a given Voronoi cell $\vorcell_i$ are located either outside or inside. Therefore, the distribution of winding number computed within the cell $\vorcell_i$ should exhibit a relatively large variance. We take the advantage of this feature to encourage larger variance among winding numbers of a Voronoi cell, which further promotes the normal orientation. To measure the variance of the winding number for each point $p_i$, we consider the winding number $w_{k}^i$ of the $k$-th vertices within its Voronoi cell. And we define the deviation as $||w_{k}^i - \overline w^i||^2$ where $\overline w_i = \sum_{j=1}^{M_i}w_{k}^i$ and $M_i$ is the number of vertices of the given Voronoi cell $\vorcell_i$.
\begin{equation}
f_{B}(\normal) = -\sum_i^N\left(\frac{1}{M_i}\sum_k^{M_i}(w_k^i- \overline w^i)^2\right).
\label{eq:Variance}
\end{equation}

\subsection{The Alignment Term $f_{A}$}
\label{sec:method_normal_contraint}

% \begin{figure}[h]
%     \centering
%     \begin{overpic}[width=0.8\linewidth]{figures/3_method/area.pdf}
%     \put(28, 2){\textbf{(a)}}
%     \put(81, 2){\textbf{(b)}}
%     \end{overpic}
%     \caption{ We give a simple yet effective way to weigh the influence of $\sample_i$
%     for computing the winding number.
%     (a) $\sample_i$'s Voronoi cell $\vorcell_i$ has $M_i$ vertices, i.e., $\query_k^i,k\in[0, M_i]$. (b) Let $\mathbf{q}_{max}^i$ be the farthest Voronoi vertex.
%     We cut $\vorcell_i$ by the plane orthogonal to $\mathbf{q}_{max}^i-\mathbf{p}_i$
%     and use the area of the cut polygon to define the influence of $\sample_i$. \XR{split.}
%     }
%     \label{fig:area_cell}
% \end{figure}
As pointed out in~\cite{amenta1998surface},
Voronoi poles are useful for orienting the normals.
Let $\sample_i$ be a point in the
given point cloud, 
$\sample_i$'s Voronoi cell $\vorcell_i$ has $M_i$ vertices, i.e., $\query_k^i,k=1,2,\cdots,M_i$.
If $\query_{k}^i$ is the inside (resp. outside) pole of $\sample_i$,
$\sample_i-\query_{k}^i$ (resp. $\query_{k}^i-\sample_i$) approximately aligns with the normal vector of~$\normal_i$.
In this paper, we turn the observation 
into an alignment requirement by enforcing 
the two sequences
% the
% positive correlation between $\normal_i\cdot (\query_{k}^i- \sample_i)$
% and $w_{k}^{i}$.
% In other words, we hope the sequence
$$
\normal_i\cdot (\query_{k}^i- \sample_i),\quad k=1,2,\cdots,M_i
$$
and 
$$
w_{k}^{i},\quad k=1,2,\cdots,M_i
$$
to have exactly the reverse ordering. 
According to the rearrangement inequality~\cite{hardy1952inequalities},
we hope $\sum_k^{\numV_i}{w_{k}^{i}}\normal_i\cdot (\query_{k}^i- \sample_i)$ to get minimized. 
Therefore, we can define the alignment term as follows.
\begin{equation}
f_{A}(\normal) = \sum_i^N\left(\frac{1}{\numV_i}\sum_k^{\numV_i}{w_{k}^{i}}\normal_i\cdot (\query_{k}^i- \sample_i)\right).
\label{eq:normal}
\end{equation}

% $M_i$ denotes the number of the Voronoi vertices of cell $\vorcell_i$ given the seed $\sample_i$, and $\query_k^i$ is the $k$-th Voronoi vertices of $\vorcell_i$. \NW{change $\query_k^i$ if not easy to understand.}
% \XR{Why $w_k^i$, need to discuss here. I think we need help of Prof.~Xin.... } \XG{For smooth surface and clean point samples, their Voronoi cells tend to be long and thin, with Voronoi vertices away from the seed. In this case the vector from seed to Voronoi vertex is close to the normal direction. However, if there are noises in the point samples, their Voronoi cells are of much diverse shape and Voronoi vertices could be located close to the surface (see Fig 3 of [Alliez et al. 2007]), making a strong justification here needed.}

\begin{figure}[t]
    \centering
    \begin{overpic}[width=1.0\linewidth]{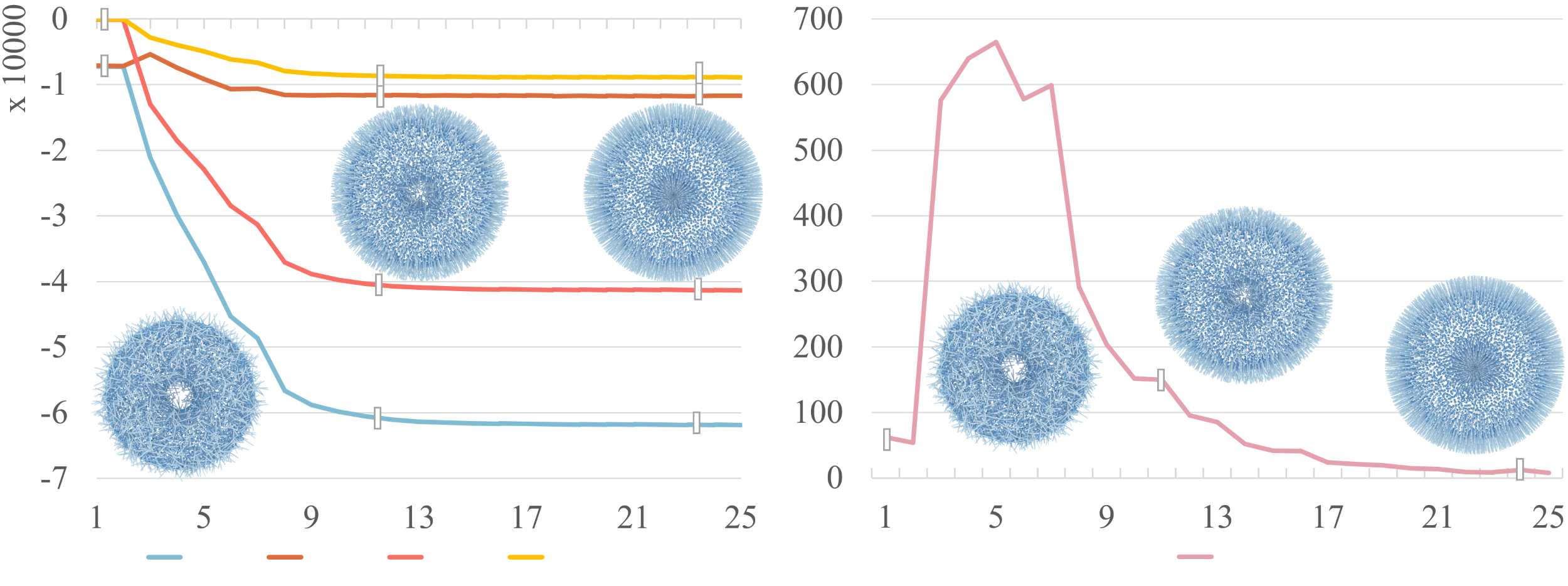}
    \put(13, 0.7){$f$}
    \put(20, 0.7){$f_{01}$}
    \put(27, 0.7){$f_B$}
    \put(35, 0.7){$f_A$}
    \put(78, 0.7){$\| g \|$}
    \end{overpic}
    \vspace{-6mm}
    \caption{Plot on the decreasing of the functional value and the gradient norm.
    The experiment is made on the torus model with 4K points. 
    }
    \label{fig:dropline}
\end{figure}

\begin{figure}[t]
    \centering
    \vspace{-3mm}
    \begin{overpic}[width=1.0\linewidth]{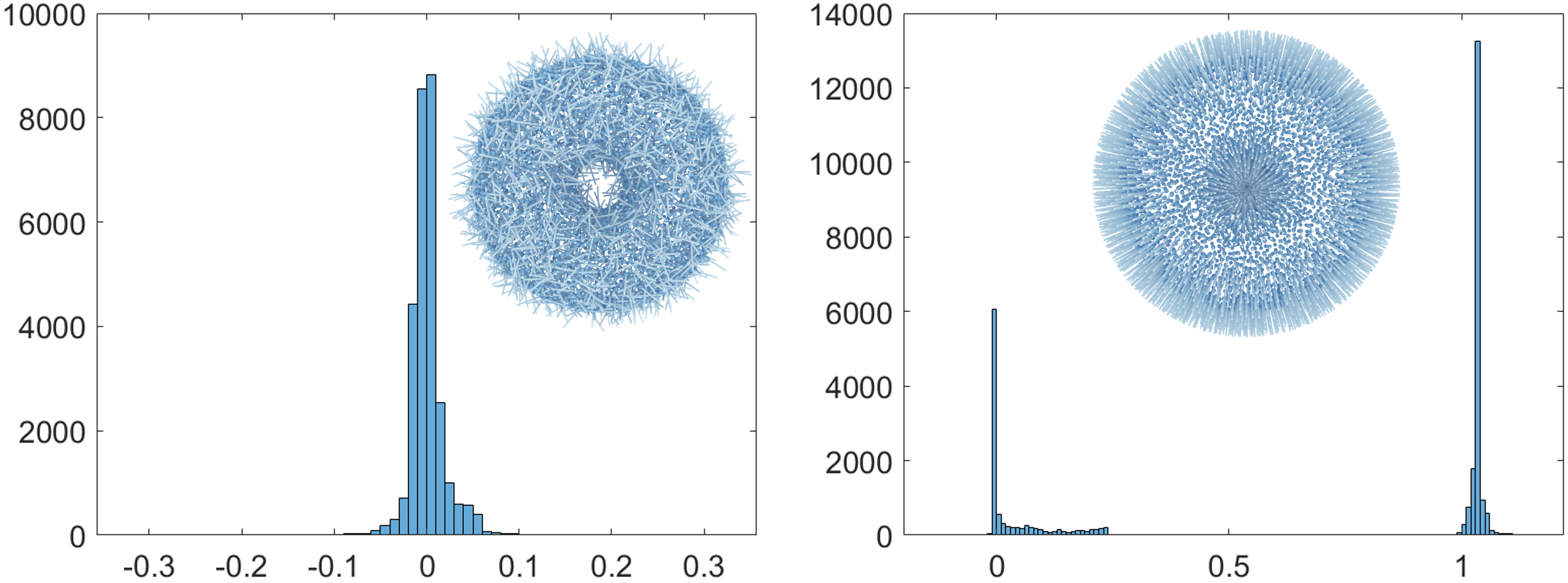}
    \put(21, -3){(a)~Start}
    \put(72, -3){(b)~End}
    \end{overpic}
    \caption{Histograms of winding number distribution for the Voronoi vertices of the torus point clouds, before and after optimization. It can be seen that the distribution of the winding number
    is regularized by optimization. 
    }
    \label{fig:distribution}
\end{figure}

\begin{table*}[!h]
\centering

\caption{\XR{Statistics on 
the truth percentages under different sampling conditions and
noise levels. }
% Comparison of normal orientation with SOTA methods with different sampling conditions.
}
\vspace{-3mm}
\label{tab:normalconsis}
\resizebox{1.0\linewidth}{!}{
\begin{tabular}{l|ccccccc|ccccccc|ccccccc}
\toprule
Sampling    & \multicolumn{7}{c|}{White Noise Sampling without noise}                                                                                                                                 & \multicolumn{7}{c|}{White Noise Sampling With 0.25\% noise}                                                                                                                                         & \multicolumn{7}{c}{White Noise Sampling With 0.5\% noise}                                                                                                                \\ \cmidrule{1-22}

Models      & Hoppe                             & K\"{o}nig        & PCPNet & Dipole & PGR                               & iPSR    & Ours                              & Hoppe                             & K\"{o}nig        & PCPNet                           & Dipole & PGR                               & iPSR    & Ours                              & Hoppe                             & K\"{o}nig        & PCPNet & Dipole & PGR                              & iPSR    & Ours                              \\ \cmidrule{1-22}
82-block    & 99.730                            & \textbf{99.980}  & 89.300 & 97.080 & 99.275                            & 99.175  & \textbf{99.980}  & 84.980                            & 99.900                            & 89.750                           & 98.030 & 98.800                            & 98.150  & \textbf{99.930}  & 98.950                            & 99.730                            & 89.030 & 98.230 & 97.350                           & 98.000  & \textbf{99.880}  \\
bunny       & 99.700                            & 97.080                            & 93.000 & 94.350 & \textbf{100.000} & 99.625  & 99.750                            & 98.550                            & 96.730                            & 92.050                           & 94.930 & 99.875                            & 99.700  & \textbf{99.980}  & 97.600                            & 96.980                            & 92.630 & 96.180 & 99.125                           & 99.375  & \textbf{99.580}  \\
chair       & 69.580                            & 86.730                            & 86.030 & 80.580 & \textbf{100.000} & 99.975  & \textbf{100.000} & 88.030                            & 85.880                            & 86.480                           & 73.550 & 99.975                            & \textbf{100.000} & \textbf{100.000} & 72.830                            & 65.05                             & 86.430 & 77.450 & 99.275                           & 99.400  & \textbf{99.400}  \\
cup-22      & 61.180                            & 60.780                            & 68.450 & 55.980 & 99.950                            & 99.400  & \textbf{99.950}  & 93.400                            & 59.380                            & 68.350                           & 56.230 & 99.925                            & 99.400  & \textbf{99.950}  & 55.930                            & 60.100                            & 67.700 & 57.900 & 99.350                           & 98.725  & \textbf{99.850}  \\
cup-35      & 99.800                            & 59.880                            & 83.400 & 54.400 & \textbf{100.000} & \textbf{100.000} & \textbf{100.000} & 99.530                            & 60.780                            & 83.00                            & 52.430 & 99.900                            & 99.950  & \textbf{100.000} & 99.000                            & 60.830                            & 82.100 & 53.030 & 98.725                           & 99.950  & \textbf{100.000} \\
fandisk     & 98.880                            & 99.850                            & 96.880 & 86.750 & 99.725                            & 99.275  & \textbf{100.000} & 99.480                            & 99.900                            & 96.800                           & 86.700 & 99.650                            & 99.050  & \textbf{99.950}  & 97.530                            & 99.580                            & 96.850 & 95.800 & 98.850                           & 97.325  & \textbf{99.750}  \\
holes       & \textbf{100.000} & \textbf{100.000} & 94.830 & 90.100 & \textbf{100.000} & \textbf{100.000} & \textbf{100.000} & \textbf{100.000} & \textbf{100.000} & 94.600                           & 90.900 & \textbf{100.000} & \textbf{100.000} & \textbf{100.000} & 99.850                            & \textbf{100.000} & 93.750 & 91.000 & 99.075                           & 99.975  & \textbf{100.000} \\
horse       & 95.700                            & 89.700                            & 95.930 & 90.780 & 99.425                            & 99.500  & \textbf{99.800}  & 92.680                            & 90.700                            & 95.550                           & 93.080 & 99.250                            & 99.325  & \textbf{99.750}  & 93.630                            & 89.880                            & 95.530 & 89.630 & 96.850                           & \textbf{98.425}  & 97.500  \\
kitten      & 99.680                            & \textbf{99.980}  & 94.100 & 98.230 & 99.950                            & 99.950  & \textbf{99.980}  & 99.750                            & \textbf{100.000} & 94.050                           & 98.380 & 99.825                            & 99.950  & \textbf{100.000} & 99.630                            & \textbf{99.980}  & 94.550 & 97.930 & 98.150                           & 99.825  & \textbf{99.980}  \\
knot        & 99.850                            & 99.980                            & 80.780 & 56.230 & 99.925                            & \textbf{100.000} & \textbf{100.000} & 99.980                            & 99.980                            & 81.380                           & 69.980 & 98.725                            & \textbf{100.000} & \textbf{100.000} & 99.780                            & 93.730                            & 80.600 & 70.530 & 95.300                           & \textbf{100.000} & 99.980  \\
lion        & 96.380                            & 92.300                            & 94.830 & 89.980 & 94.975                            & 97.750  & \textbf{99.700}  & 94.580                            & 93.130                            & 94.880                           & 93.400 & 93.500                            & 96.325  & \textbf{99.550}  & 88.450                            & 93.330                            & 94.250 & 88.880 & 89.325                           & 93.325  & \textbf{94.830}  \\
mobius      & \textbf{100.000} & 55.150                            & 89.800 & 53.950 & \textbf{100.000} & 87.250  & \textbf{100.000} & 68.600                            & 55.130                            & 86.980                           & 54.200 & 99.175                            & 80.525  & \textbf{99.380}  & 55.980                            & 55.130                            & 82.030 & 53.650 & \textbf{94.900} & 68.225  & 85.780                            \\
mug         & 98.450                            & 66.980                            & 77.350 & 68.250 & 99.925                            & 99.875  & \textbf{100.000} & 98.480                            & 67.480                            & 77.330                           & 67.000 & 99.925                            & 99.900  & \textbf{100.000} & 68.230                            & 68.000                            & 76.930 & 66.150 & 99.050                           & 99.700  & \textbf{100.000} \\
octa-flower & 50.900                            & 88.030                            & 99.280 & 95.300 & 98.600                            & 95.350  & \textbf{99.330}  & 53.800                            & 58.180                            & \textbf{99.000} & 95.180 & 97.725                            & 95.525  & 98.800                            & 89.330                            & 86.930                            & 98.200 & 95.500 & 96.225                           & 93.775  & \textbf{98.550}  \\
sheet       & 51.200                            & 51.130                            & 83.980 & 52.450 & \textbf{100.000} & 99.075  & \textbf{100.000} & 99.480                            & 51.300                            & 83.400                           & 52.700 & 99.950                            & 98.950  & \textbf{100.000} & 51.100                            & 51.050                            & 81.130 & 59.550 & 99.875                           & 97.225  & \textbf{99.980}  \\
torus       & \textbf{100.000} & \textbf{100.000} & 96.880 & 99.950 & \textbf{100.000} & \textbf{100.000} & \textbf{100.000} & \textbf{100.000} & \textbf{100.000} & 96.750                           & 99.980 & \textbf{100.000} & \textbf{100.000} & \textbf{100.000} & \textbf{100.000} & \textbf{100.000} & 96.780 & 99.800 & 99.975                           & \textbf{100.000} & \textbf{100.000} \\
trimstar    & 97.200                            & \textbf{100.000} & 91.050 & 96.700 & 98.600                            & 99.650  & \textbf{100.000} & 99.050                            & \textbf{100.000} & 90.930                           & 94.350 & 98.150                            & 99.325  & \textbf{100.000} & 98.880                            & \textbf{100.000} & 91.080 & 94.830 & 96.225                           & 98.500  & \textbf{100.000} \\
vase        & 95.680                            & 90.900                            & 83.330 & 75.980 & 99.100                            & 99.825  & \textbf{100.000} & 94.650                            & 90.130                            & 83.130                           & 82.480 & 99.000                            & 99.475  & \textbf{100.000} & 86.430                            & 88.830                            & 82.700 & 90.700 & 97.875                           & 98.550  & \textbf{99.650}  \\ \bottomrule
\end{tabular}
}
\end{table*}

\subsection{Implementation Details}\label{sec:Implementation}
\begin{wrapfigure}{r}{2cm}
\vspace{-3.5mm}
  \hspace*{-4mm}
  \centerline{
  \includegraphics[width=30mm]{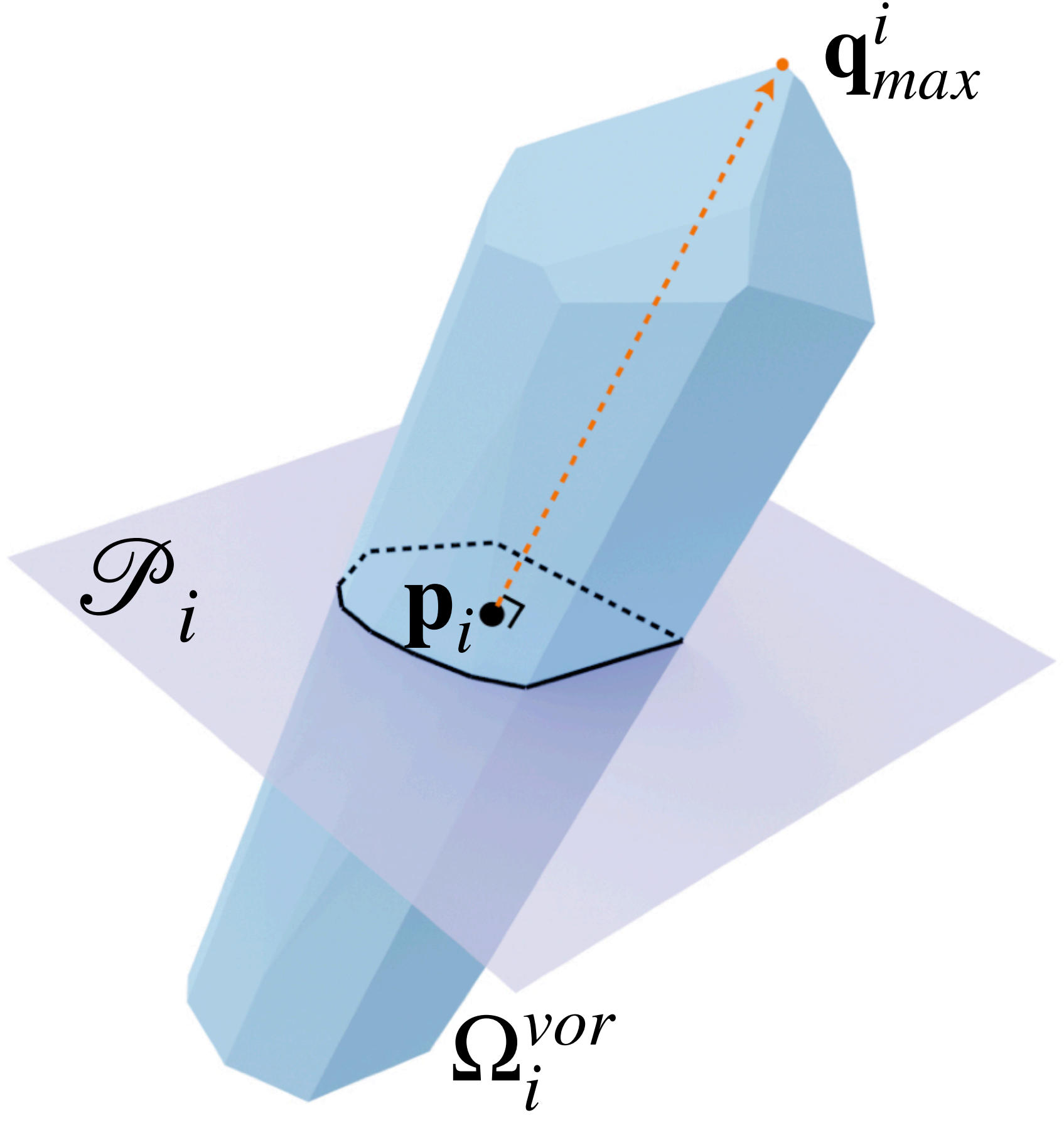}
  }
  \vspace*{-4mm}
\end{wrapfigure}

\paragraph{Area weight of $\sample_i$}
Let $\sample_i$ be a point in the given point cloud. 
The estimation of the winding number at an arbitrary point
has to input the weighting area of $\sample_i$; See Eq.~\eqref{eq:windingnum_ori}.
A typical way for defining the weighting area $a_i$ is based on KNN~\cite{barill2018fast}.
However, it has to include a parameter $k$ to resist the irregular distribution of points (typically $k=20$). 
% the technique for estimating $\sample_i$ includes 
% a parameter \SQ{radius?k?}.
In this paper, we use a parameter-free strategy for estimating $a_i$.
As the inset figure shows,
$\mathbf{q}_{max}^i\in\vorcell_i$ is the farthest Voronoi vertex to $\sample_i$.
We build a plane orthogonal to $\mathbf{q}_{max}^i-\mathbf{p}_i$
and use it to  cut $\vorcell_i$ into two halves,
resulting in a convex cut polygon.
We use the area of the cut polygon to define~$a_i$.
% \XR{Old caption:  (b) Let $\mathbf{q}_{max}^i$ be the farthest Voronoi vertex.
% We cut $\vorcell_i$ by the plane orthogonal to $\mathbf{q}_{max}^i-\mathbf{p}_i$
% and use the area of the cut polygon to define the influence of $\sample_i$.}

% To estimate the facet area $a_i$ at the point $\sample_i$ in Eq.~\ref{eq:windingnum_ori}, we make full use of the structure of $\sample_i$'s corresponding Voronoi cell $\vorcell_i$. As shown in Fig~\ref{fig:area_cell} (b), we find the Voronoi vertex $\query_{max}^i$ of the $\vorcell_i$ with the furthest distance to $p_i$. The pair of $\sample$ and vector $\overrightarrow{\sample_i \query_{max}^i}$ defines a cutting-plane $\mathcal{P}_i$. Therefore, the area $\area_i$ of point $\sample_i$ can be defined as $\area_i = \mathcal{P}_i \cap \vorcell_i$.
% \XR{We are using an Oriented Bounding Box(OBB) and re-scale it to 1.2x size as the bounding box for the Voronoi structure.}

\paragraph{Optimization details}
The overall objective function 
takes the normals $\normal_i,i=1,2,\cdots,N,$ as variables. 
As $\normal_i$ is required to be a unit vector,
we parameterize a normal vector as 
\begin{equation}
\normal_i =  \left(\sin(u_i)\cos(v_i),\sin(u_i)\sin(v_i),\cos(u_i)\right).
\label{eq:ni}
\end{equation}
In this way, we turn the problem of minimizing~$f(\normal)$ into an unconstrained optimization problem. 

The function $f=f(\normal)$ can be viewed as a composite function of
$$f=f(w_1,w_2,\cdots,w_M)$$
and
$$w_j=w_j(\normal_1,\normal_2,\cdots,\normal_N).$$
At the same time, $\normal_i$ is a composite function of $u_i$ and $v_i$. 
Therefore,
the gradients of the overall function can be quickly computed by the chain rule.
We omit the form of the detailed gradients for brevity. 
Fig.~\ref{fig:dropline} plots how the objective function and the gradient norm 
are decreased during the optimization. 
It can be seen from Fig.~\ref{fig:distribution} that the normals become globally consistent
upon the regularization of the winding number.
%\SQ{@xu rui, a plot figure?}\XR{Done.}
% \XR{Need to ref fig.~\ref{fig:distribution}?}

\noindent{\bf Remark.}
In order to show that the minimization of~$f$ can arrive at the termination,
we need to prove the fact that the objective function has a lower bound. 
%First of all, we need t show that $w_j$ cannot be infinitely large.
Observing that $f_{01}$ is quartic about $w_j$ (with a positive leading coefficient)
but the other two terms have a lower degree,
it is easy to know that $f$ approaches $+\infty$ if one of the winding numbers is sufficiently large, which naturally constrains every~$w_j$ in a limited range, e.g.,~$[W_1, W_2]$. 
%Next, we shall show that the function value must have a low bound. 
Therefore, the boundedness of $f$ follows immediately from the boundedness of $w_j$. 
See more rigorous proof in the supplemental material.

\section{Experimental Results}

\subsection{Experimental Setting}

\paragraph{Platform}
Our experiments are conducted  on a computer with an AMD Ryzen 9 5950X CPU and 32 GB memory. We run the GPU-based approaches 
\cite{guerrero2018pcpnet,zhu2021adafit,li2022neaf,dipole_propagation,PGR2022Siyou} on an NVIDIA GeForce RTX 3090 card.

\paragraph{Point clouds and Normalization} 
% All the point clouds are normalized to a range of $[-0.5,0.5]^3$.
% We test our method with various sampling conditions by taking both point density and noise into consideration; See Sec~\ref{sec:normal_ori_recon} for more details. 
We make the tests on a total of 18 models of various shapes (see Table~\ref{tab:normalconsis}).
All the point clouds are normalized to a range of $[-0.5,0.5]^3$.
We use two types of sampling strategies, i.e., white noise sampling and blue noise sampling~\cite{gptoolbox}.
Besides the noise-free point clouds,
% we further add Gaussian noise to each point cloud
% at two levels, i.e., $0.25\%$ and $0.5\%$ of the longest edge of the bounding box. 
\XR{We scale all models to $[-0.5,0.5]^3$ so that the longest edge of the bounding box is always 1.0. For noise generation, we use the standard Gaussian distribution with $\mu = 0.0$ and $\sigma^2 = 1.0$ to produce noise displacement. Each point 
is given a random displacement that is added to the original position. The noise level is controlled by a scale factor of $0.25\%$ and $0.5\%$, respectively. }

\paragraph{Parameters}
In all the experiments, we adopt the same parameter setting: $\lambda_A = 10.0$, $\lambda_B = 50.0$, and $D = 4.0$.
% \SQ{@Rui, $\lambda_V$?}
We use the L-BFGS algorithm implemented in C++ for solving the optimization. 
The termination condition is set by requiring the difference between the objective function values at two consecutive steps not to exceed a threshold of $1.0$.
%In this paper, we set $\tau=1.0$ for all experiments. %with a maximum step set to be $50$ steps.

\paragraph{Approaches}
We include five state-of-the-art (SOTA) methods~\cite{hoppe1992surface,konig2009consistent,dipole_propagation,PGR2022Siyou,guerrero2018pcpnet} for comparison.
PGR~\cite{PGR2022Siyou} 
receives an un-oriented point cloud as the input and
outputs a polygonal surface,
but we focus more on the quality of its computed normals.
% Although PGR~\shortcite{PGR2022Siyou} is a reconstruction method, it also iteratively optimizes the normals, so we compare the final normals obtained by PGR as its output.
Note that~\cite{hoppe1992surface}
and~\cite{konig2009consistent}
% use the Euclidean minimal spanning tree (EMST)
% to propagate orientations, which 
need to pre-compute a Riemannian graph to 
encode the proximity between points.
In our experiments, 
% we set the radius of KNN to be $0.05$. 
we take two closely spaced points as neighbors 
if the distance between them is less than 0.05.
Besides, PCPNet~\cite{guerrero2018pcpnet} has multiple pre-trained models, we use \textit{multi\_scale\_oriented\_normal} in all experiments. 
For PGR~\cite{PGR2022Siyou} and Dipole~\cite{dipole_propagation}, we
follow the default setting.

\begin{figure*}[!h]
%\vspace{5px}
\centering
\begin{overpic}
[width=0.99\linewidth]{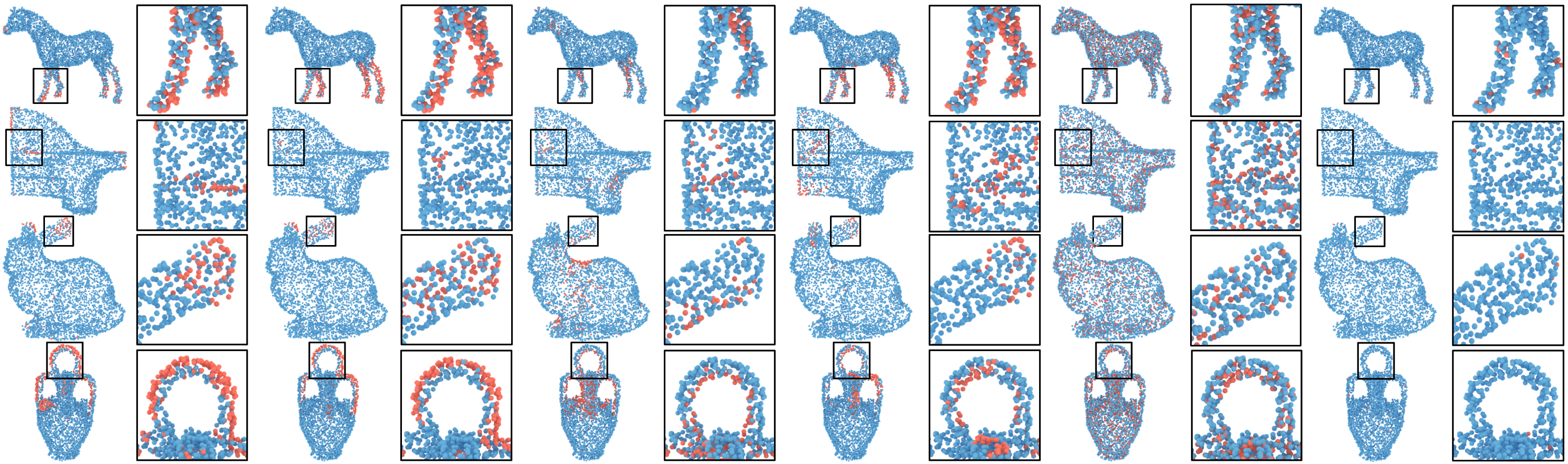}
\put(6, -1.5){\textbf{Hoppe}}
\put(23, -1.5){\textbf{K{\"o}nig}}
\put(39, -1.5){\textbf{PCPNet}}
\put(57, -1.5){\textbf{Dipole}}
\put(74, -1.5){\textbf{PGR}}
\put(90, -1.5){\textbf{Ours}}
\end{overpic}
 \caption{Comparison on the ratio of true normals
 between our approach and the existing five methods: Hoppe~\cite{hoppe1992surface}, K{\"o}nig~\cite{konig2009consistent}, PCPNet~\cite{guerrero2018pcpnet}, Dipole~\cite{dipole_propagation} and
 PGR~\cite{PGR2022Siyou}.
 The true predictions and false predictions 
 are colored in blue and red, respectively. 
 % w.r.t. the ground-truth (GT).
 % Here the true predictions in blue while the false predictions are colored in red. 
 % normal orientation with existing methods. The red point indicates the angle error between the oriented normal and the ground truth normal is larger than $90$ degree. 
 Note that the level of Gaussian noise is $0.5\%$.
 % \NW{Updated}
 %We use white noise sampling with 0.5\% noise.
 }
 \vspace{-2mm}
 % \XR{Need change, red points is bigger than 90. All uniform and noisy(WS and 0.5\% noise).} }
\label{fig:comp_normals}
\end{figure*}

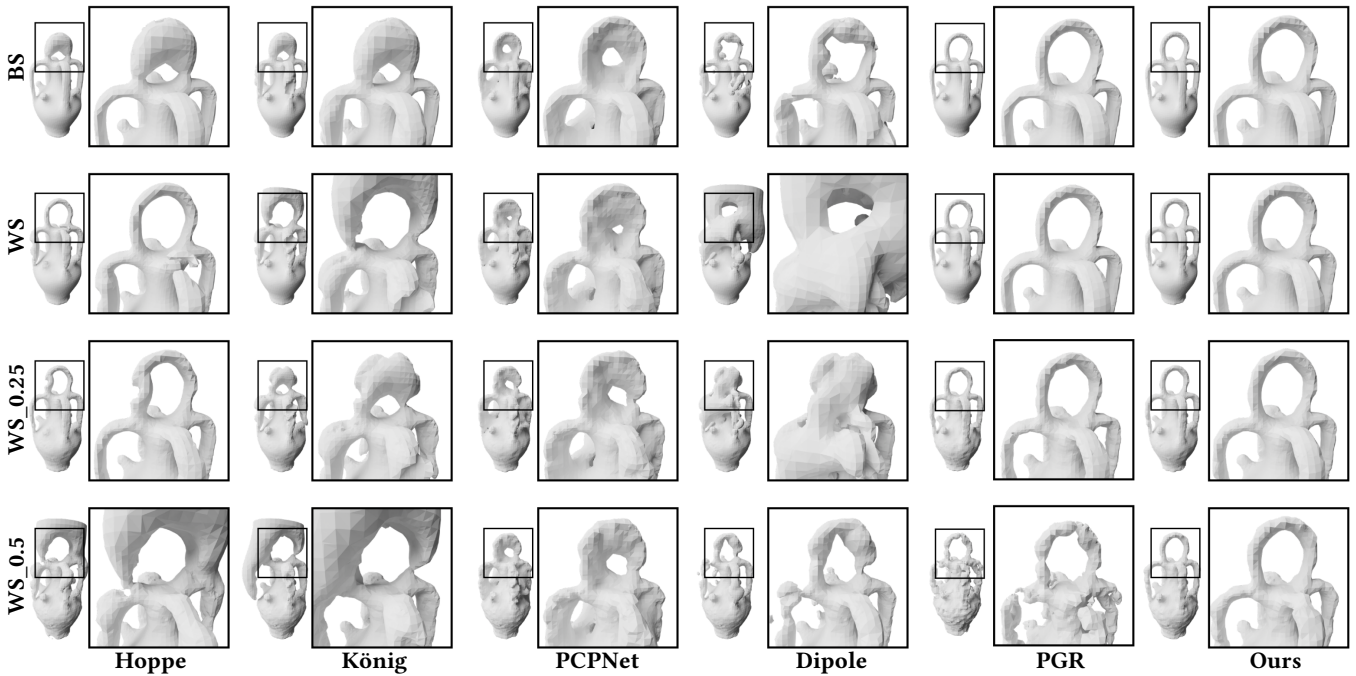
\begin{figure*}[!h]
%\vspace{7px}
\centering
\begin{overpic}
[width=0.99\linewidth]{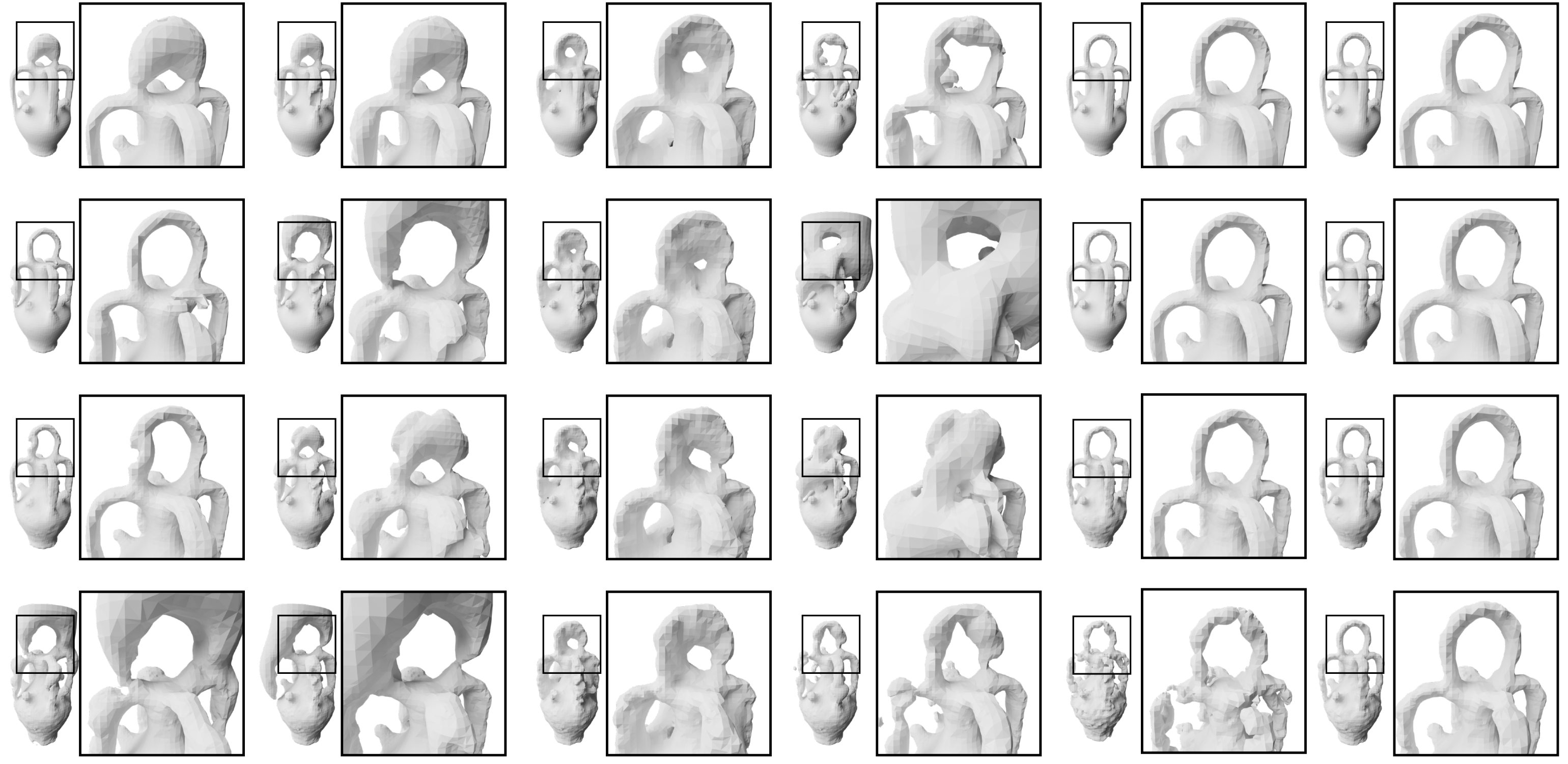}
\put(7, -1){\textbf{Hoppe}}
\put(24, -1){\textbf{K{\"o}nig}}
\put(40, -1){\textbf{PCPNet}}
\put(58, -1){\textbf{Dipole}}
\put(76, -1){\textbf{PGR}}
\put(92, -1){\textbf{Ours}}
\put(-1,43){\rotatebox{90}{\textbf{BS}}}
\put(-1,30){\rotatebox{90}{\textbf{WS}}}
\put(-1,15){\rotatebox{90}{\textbf{WS\_0.25}}}
\put(-1,3){\rotatebox{90}{\textbf{WS\_0.5}}}
\end{overpic}
% \vspace{-5mm}
\caption{Visual comparison of the reconstructed surfaces
at different sampling conditions and different levels of noise.
We show results of four different sampling conditions: BS (blue noise sampling), WS (white noise sampling), WS\_0.25 (white noise sampling with 0.25\% noise) and WS\_0.5 (white noise sampling with 0.5\% noise).
% among our method and four other methods: Hoppe~\cite{hoppe1992surface}, K{\"o}nig~\cite{konig2009consistent}, PCPNet~\cite{guerrero2018pcpnet} and Dipole~\cite{dipole_propagation}. We show results of four different sampling conditions: BS (blue noise sampling), WS (white noise sampling), WN\_0.25 (white noise sampling with 0.25\% noise) and WS\_0.5 (white noise sampling with 0.5\% noise).\XR{Change to thin tube.}
}
\vspace{-2mm}
\label{fig:comp_recon}
\end{figure*}

\subsection{Comparisons}
\label{sec:comparison}

\paragraph{Indicators}
We evaluate the performance from two aspects.
On the one hand, we keep track of the percentage of correctly oriented normals. For an input point~$\sample_i$,
the predicted orientation is true
if the angle between the computed normal and the ground-truth normal is less than $90$ degrees. 
On the other hand, 
we make statistics about the reconstruction quality 
by feeding the point clouds and the normals together into the 
SPR solver~\cite{kazhdan2006poisson}.
Specially, we use the \textit{Chamfer Distance}~(CD)
to measure the error between the ground-truth surface
and the reconstructed surface (with the support of predicted normals).
%\NW{~\cite{kazhdan2006poisson}.}
% \NW{We use \textit{Chamfer Distance}~(CD) as the metric to compute the error between the surface reconstructed from predicted normals and the ground truth normals.}

%bounding box 的最长轴 * 0.25\% * 正态分布的noise 
%bounding box 的最长轴 * 0.5\% * 正态分布的noise 

% \XR{Introduce other methods, hoppe and knoig need set knn or radius... In our experiments, we set all radius as $0.05$. }
%可以在这里拉踩一下其他方法

\paragraph{Quality of predicted normals}
% \subsubsection{Normal Comparisons}
% \label{sec:comp_normal}
% To provide an in-depth comparison, we conduct a comparison on normal orientation with point cloud at different levels of noises: 1) uniform sampled points without noise; 2) unevenly sampled points cloud without noise; 3) uneven sampled points cloud with $0.25\%$ noises; 4) uneven sampled points cloud with $0.5\%$ noises. Same as~\cite{dipole_propagation}, we calculate the percentage of correctly oriented normals. A normal is identified as correctly oriented if the angle between this normal and the ground truth normal is less than $90$ degree. 
% The results are summarized in Table~\ref{tab:normalconsis}. 
In Table~\ref{tab:normalconsis},
we show the statistics of 
the \textit{truth percentages} of normals
over the $18$ models, under different sampling conditions and
noise levels.  
The statistics show that 
our approach has a higher truth percentage than the SOTA methods.
For example, for the blue noise sampling point clouds,
our approach can achieve a percentage of $100\%$ 
for $88.9\%$ of the tested models,
much higher than the SOTA methods. 
Furthermore, we give a visual comparison in 
Fig.~\ref{fig:comp_normals}
where the points are colored differently depending on 
whether the normal orientation is correctly predicted.  
It can be clearly seen that 
% our approach outperforms the SOTA methods in terms of the truth percentage. 
% Specially, 
our approach has an advantage in
predicting the normals for points in the tubular regions
and thin regions with sharp features and corners; See the highlighted regions.

Besides, we also make statistics about 
the Root Mean Square Error (RMSE) of angles between 
the estimated normal and the ground-truth normal,
which also shows that our algorithm 
has advantage in prediction accuracy. 
The detailed statistics are included in the supplementary material.
% show the comparison with SOTA methods in terms of normal estimation, for which we use the angle Root Mean Square Error~(RMSE) between the estimated normal and the ground truth normal as the metric. The detail of normal estimation is given in the Supplementary Material.

% Next, we show the comparison with SOTA methods in terms of normal estimation, for which we use the angle Root Mean Square Error~(RMSE) between the estimated normal and the ground truth normal as the metric. As shown in Table\XR{in appendix...}, our method achieves competitive results compared with SOTA methods.

% Notably, although our method is designed for normal orientation, there are some other works focusing on normal estimation, which could be considered for orientating normals. We compare with representative methods PCA~\cite{Rusu_ICRA2011_PCL}, AdaFit~\cite{zhu2021adafit} and NeAF~\cite{li2022neaf} and report the result in Table~\ref{tab:estimation4orientation}\XR{in appendix...}. Due to the lack of consideration of globally consistent normals orientation during the estimation, these methods perform poorly compared with our method.

\begin{table*}[t]
\centering

\caption{
%\XR{Comparison of the surface reconstruction qualityby feeding the predicted normals into the Poisson surface reconstruction solver~\cite{kazhdan2006poisson}. Note that the Chamfer Distance between the reconstructed surfaceand the ground-truth surface is scaled by $100$ times for a better presentation.}
\ZY{Reconstruction quality comparison using Poisson surface reconstruction solver~\cite{kazhdan2006poisson} with predicted normals. The Chamfer Distance between the reconstructed surface and the ground-truth surface is presented scaled by a factor of $100$ for a better presentation.}}
\vspace{-3mm}
\label{tab:recon_poisson}
\resizebox{1.0\linewidth}{!}{
\begin{tabular}{l|ccccccc|ccccccc|ccccccc}
\toprule
\multicolumn{1}{l|}{Sampling}    & \multicolumn{7}{c|}{White Noise Sampling without noise}                                                                                                                                                                                                  & \multicolumn{7}{c|}{White Noise Sampling With 0.25\% noise}                                                                                                                                                                                                      & \multicolumn{7}{c}{White Noise Sampling With 0.5\% noise}                                                                                                                                                                                   \\\cmidrule{1-22}
\multicolumn{1}{l|}{Models}      & Hoppe                           & K\"{o}nig      & PCPNet               & Dipole               & PGR                             & iPSR                            & \multicolumn{1}{c|}{Ours}                            & Hoppe                           & K\"{o}nig      & PCPNet                          & Dipole                          & PGR                             & iPSR                            & \multicolumn{1}{c|}{Ours}                            & Hoppe                           & K\"{o}nig      & PCPNet                          & Dipole                          & PGR                             & iPSR                            & Ours                            \\
\cmidrule{1-22}
\multicolumn{1}{l|}{82-block}    & 0.149                           & 0.149                           & 0.489                & 0.195                & 0.158                           & 0.165                           & \multicolumn{1}{c|}{\textbf{0.130}} & 0.593                           & 0.155                           & 0.521                           & 0.184                           & 0.202                           & 0.195                           & \multicolumn{1}{c|}{\textbf{0.134}} & 0.175                           & \textbf{0.172} & 0.578                           & 0.203                           & 0.462                           & 0.244                           & 0.184                           \\
\multicolumn{1}{l|}{bunny}       & 0.143                           & 0.230                           & 0.326                & 0.358                & \textbf{0.092} & 0.103                           & \multicolumn{1}{c|}{0.112}                           & 0.130                           & 0.285                           & 0.355                           & 0.354                           & 0.173                           & 0.156                           & \multicolumn{1}{c|}{\textbf{0.126}} & 0.198                           & 0.285                           & 0.379                           & 0.280                           & 0.322                           & 0.203                           & \textbf{0.157} \\
\multicolumn{1}{l|}{chair}       & 1.758                           & 0.739                           & 0.425                & 0.583                & 0.118                           & \textbf{0.074} & \multicolumn{1}{c|}{0.076}                           & 0.599                           & 0.761                           & 0.461                           & 0.912                           & 0.166                           & 0.116                           & \multicolumn{1}{c|}{\textbf{0.080}} & 1.545                           & 2.311                           & 0.516                           & 0.795                           & 0.397                           & \textbf{0.165} & 0.172                           \\
\multicolumn{1}{l|}{cup-22}      & 1.673                           & 1.631                           & 1.523                & 1.787                & 0.121                           & 0.131                           & \multicolumn{1}{c|}{\textbf{0.112}} & 0.338                           & 1.680                           & 1.539                           & 1.778                           & 0.203                           & 0.191                           & \multicolumn{1}{c|}{\textbf{0.139}} & 1.898                           & 1.715                           & 1.568                           & 1.764                           & 0.347                           & 0.267                           & \textbf{0.193} \\
\multicolumn{1}{l|}{cup-35}      & 0.133                           & 1.221                           & 0.903                & 1.613                & \textbf{0.095} & 0.115                           & \multicolumn{1}{c|}{0.098}                           & 0.119                           & 1.215                           & 0.884                           & 1.771                           & 0.171                           & 0.170                           & \multicolumn{1}{c|}{\textbf{0.094}} & 0.147                           & 1.278                           & 0.979                           & 1.721                           & 0.297                           & 0.214                           & \textbf{0.134} \\
\multicolumn{1}{l|}{fandisk}     & 0.128                           & 0.123                           & 0.195                & 0.637                & 0.081                           & 0.090                           & \multicolumn{1}{c|}{\textbf{0.073}} & 0.127                           & 0.126                           & 0.211                           & 0.643                           & 0.149                           & 0.120                           & \multicolumn{1}{c|}{\textbf{0.103}} & 0.191                           & 0.171                           & 0.254                           & 0.219                           & 0.307                           & 0.181                           & \textbf{0.169} \\
\multicolumn{1}{l|}{holes}       & \textbf{0.036} & \textbf{0.036} & 0.252                & 0.286                & 0.072                           & 0.075                           & \multicolumn{1}{c|}{0.073}                           & \textbf{0.039} & \textbf{0.039} & 0.267                           & 0.200                           & 0.191                           & 0.144                           & \multicolumn{1}{c|}{0.070}                           & \textbf{0.051} & \textbf{0.051} & 0.328                           & 0.254                           & 0.576                           & 0.193                           & 0.149                           \\
\multicolumn{1}{l|}{horse}       & 0.291                           & 0.482                           & 0.165                & 0.307                & 0.085                           & 0.082                           & \multicolumn{1}{c|}{\textbf{0.075}} & 0.281                           & 0.465                           & 0.192                           & 0.224                           & 0.167                           & \textbf{0.088} & \multicolumn{1}{c|}{0.090}                           & 0.305                           & 0.522                           & 0.238                           & 0.358                           & 0.454                           & 0.244                           & \textbf{0.178} \\
\multicolumn{1}{l|}{kitten}      & 0.064                           & \textbf{0.061} & 0.268                & 0.076                & 0.061                           & 0.099                           & \multicolumn{1}{c|}{0.088}                           & 0.066                           & \textbf{0.065} & 0.265                           & 0.079                           & 0.177                           & 0.109                           & \multicolumn{1}{c|}{0.084}                           & 0.073                           & \textbf{0.072} & 0.299                           & 0.092                           & 0.319                           & 0.161                           & 0.144                           \\
\multicolumn{1}{l|}{knot}        & \textbf{0.040} & \textbf{0.040} & 0.817                & 1.634                & 0.166                           & 0.105                           & \multicolumn{1}{c|}{0.064}                           & \textbf{0.045} & 0.046                           & 0.822                           & 0.998                           & 0.338                           & 0.140                           & \multicolumn{1}{c|}{0.068}                           & \textbf{0.063} & 0.311                           & 0.875                           & 0.931                           & 0.652                           & 0.191                           & 0.123                           \\
\multicolumn{1}{l|}{lion}        & 0.229                           & 0.351                           & 0.211                & 0.397                & 0.112                           & 0.117                           & \multicolumn{1}{c|}{\textbf{0.087}} & 0.268                           & 0.349                           & 0.224                           & 0.248                           & 0.268                           & 0.198                           & \multicolumn{1}{c|}{\textbf{0.112}} & 0.388                           & 0.383                           & 0.274                           & 0.365                           & 0.619                           & 0.251                           & \textbf{0.228} \\
\multicolumn{1}{l|}{mobius}      & \textbf{0.117} & 1.551                           & 0.563                & 1.507                & 0.126                           & 0.156                           & \multicolumn{1}{c|}{0.237}                           & 2.161                           & 1.549                           & 0.649                           & 1.529                           & \textbf{0.238} & 0.287                           & \multicolumn{1}{c|}{0.417}                           & 2.819                           & 1.743                           & 0.755                           & 1.607                           & \textbf{0.410} & 0.435                           & 0.634                           \\
\multicolumn{1}{l|}{mug}         & 0.135                           & 1.228                           & 1.208                & 1.506                & 0.135                           & 0.906                           & \multicolumn{1}{c|}{\textbf{0.125}} & 0.146                           & 1.238                           & 1.214                           & 1.514                           & 0.214                           & 0.929                           & \multicolumn{1}{c|}{\textbf{0.118}} & 1.276                           & 1.244                           & 1.253                           & 1.624                           & 0.322                           & 1.240                           & \textbf{0.146} \\
\multicolumn{1}{l|}{octa-flower} & 2.424                           & 0.601                           & 0.130                & 0.245                & \textbf{0.093} & 0.139                           & \multicolumn{1}{c|}{0.164}                           & 2.445                           & 1.505                           & \textbf{0.146} & 0.232                           & 0.193                           & 0.165                           & \multicolumn{1}{c|}{0.177}                           & 0.521                           & 0.703                           & \textbf{0.181} & 0.238                           & 0.372                           & 0.293                           & 0.247                           \\
\multicolumn{1}{l|}{sheet}       & 1.607                           & 1.613                           & 3.327                & 1.563                & 0.098                           & 0.650                           & \multicolumn{1}{c|}{\textbf{0.091}} & 0.123                           & 1.607                           & 2.909                           & 1.511                           & 0.167                           & 0.717                           & \multicolumn{1}{c|}{\textbf{0.116}} & 1.630                           & 1.635                           & 3.046                           & 1.545                           & 0.380                           & 0.881                           & \textbf{0.219} \\
\multicolumn{1}{l|}{torus}       & \textbf{0.019} & \textbf{0.019} & 0.215                & 0.019                & 0.054                           & 0.163                           & \multicolumn{1}{c|}{0.043}                           & 0.026                           & 0.026                           & 0.232                           & \textbf{0.021} & 0.135                           & 0.190                           & \multicolumn{1}{c|}{0.064}                           & 0.044                           & 0.044                           & 0.250                           & \textbf{0.033} & 0.264                           & 0.240                           & 0.125                           \\
\multicolumn{1}{l|}{trimstar}    & 0.214                           & 0.137                           & 0.417                & 0.175                & 0.158                           & 0.147                           & \multicolumn{1}{c|}{\textbf{0.112}} & 0.154                           & 0.142                           & 0.422                           & 0.208                           & 0.340                           & 0.153                           & \multicolumn{1}{c|}{\textbf{0.110}} & 0.158                           & \textbf{0.153} & 0.454                           & 0.238                           & 0.670                           & 0.225                           & 0.170                           \\
\multicolumn{1}{l|}{vase}        & 0.217                           & 0.440                           & 0.613                & 0.950                & 0.094                           & 0.119                           & \multicolumn{1}{c|}{\textbf{0.092}} & 0.243                           & 0.446                           & 0.654                           & 0.611                           & 0.198                           & 0.149                           & \multicolumn{1}{c|}{\textbf{0.107}} & 0.588                           & 0.581                           & 0.727                           & 0.326                           & 0.503                           & 0.194                           & \textbf{0.189} \\ \bottomrule
\end{tabular}
}

\end{table*}

\begin{figure*}[t]
\vspace{-1mm}
\centering
\begin{overpic}
[width=0.99\linewidth]{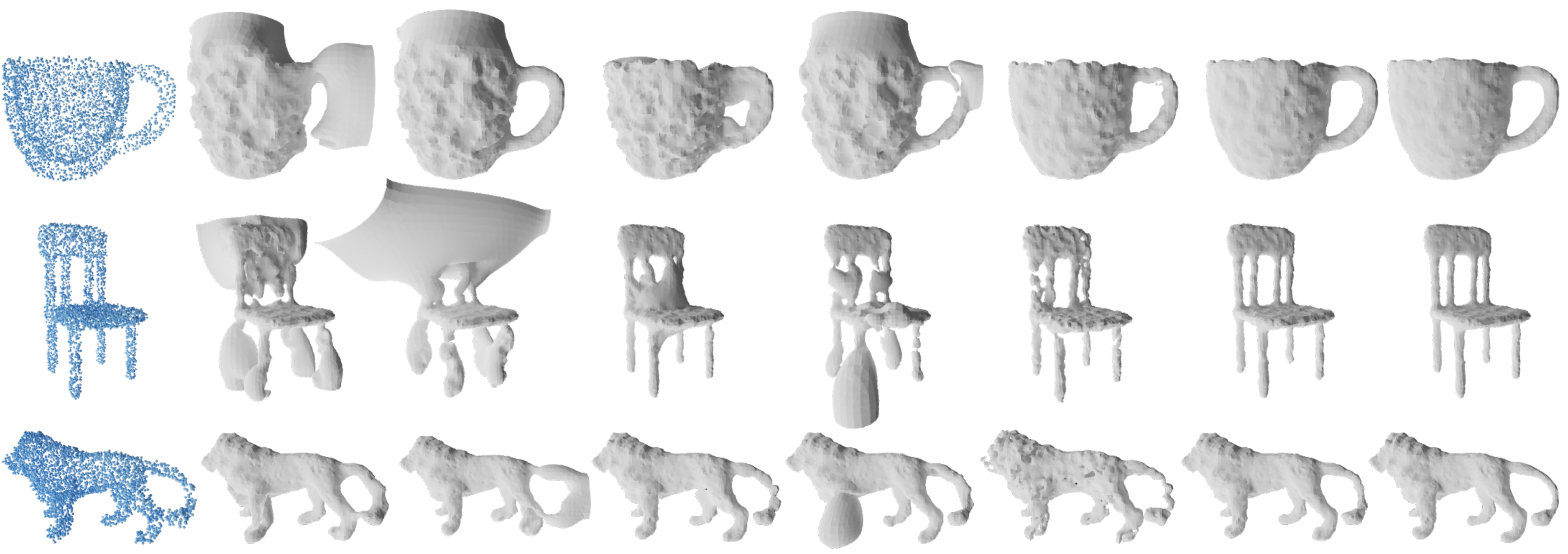}
\put(4, -1){\textbf{Input}}
\put(16, -1){\textbf{Hoppe}}
\put(28, -1){\textbf{K{\"o}nig}}
\put(40, -1){\textbf{PCPNet}}
\put(53, -1){\textbf{Dipole}}
\put(68, -1){\textbf{PGR}}
\put(80, -1){\textbf{Ours}}
\put(92, -1){\textbf{GT}}
\end{overpic}
\caption{
% Comparing the reconstruction quality 
% by feeding the oriented normals given by various approaches. 
Comparing the reconstruction quality on point clouds with $0.5\%$ Gaussian noise. 
% Note that the points are generated by white noise sampling,
% and then perturbed by adding $0.5\%$ Gaussian noise. 
}
% \vspace{-5mm}
\label{fig:comp_noise}
\end{figure*}

\begin{figure}[!h]
    \centering
    \vspace{-2mm}
    \begin{overpic}[width=0.98\linewidth]{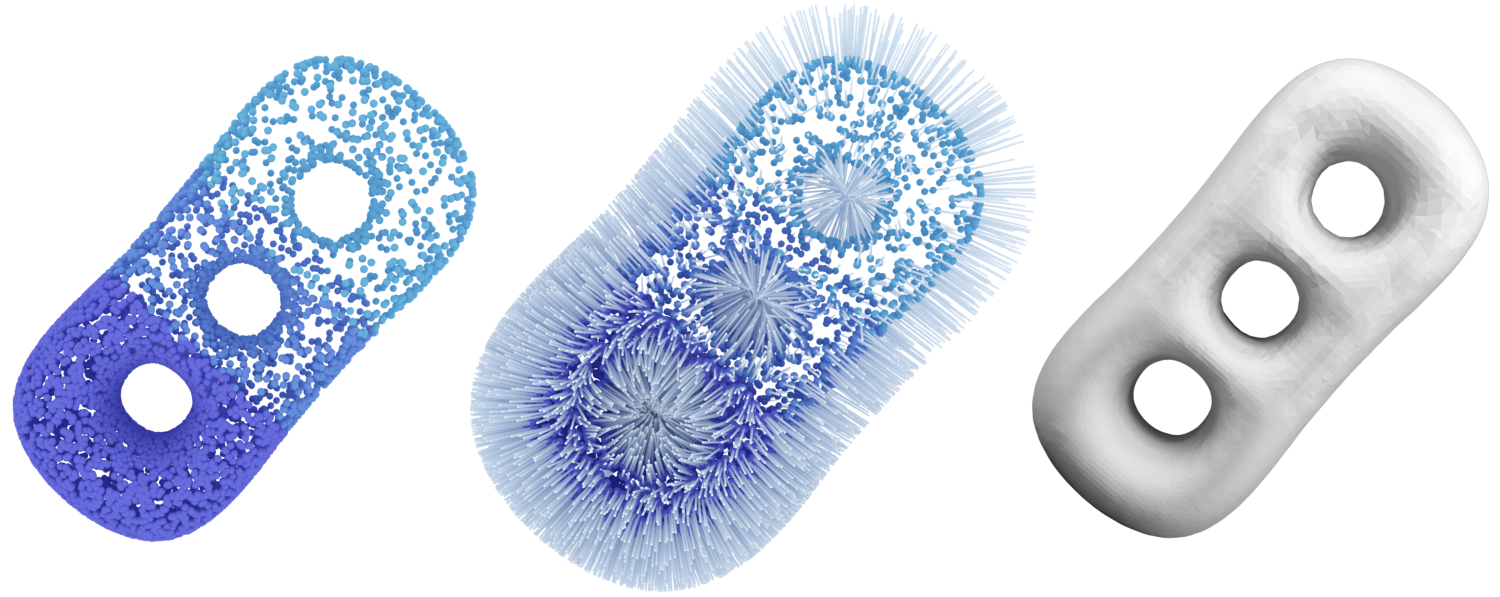}
    \put(3, -3){\textbf{(a)~Input points}}
    \put(37, -3){\textbf{(b)~Our result}}
    \put(68, -3){\textbf{(c)~Reconstruction}}
    \end{overpic}
    \caption{
    We construct a point cloud of the genus-3 torus with varying point densities (colored in varying darkness).
    Both the predicted normals and the reconstructed surface
    show that our approach is robust to the point density. 
    % Performance of our method for orientating normal vectors on a point cloud with varying point densities. From darker to lighter blue, the point density decreases from high to low. We test with the model genus-3 torus.
    }
    \vspace{-5pt}
    \label{fig:comp_density}
\end{figure}

\paragraph{Quality of reconstructed surfaces }
We further take the SPR solver as a blackbox 
to observe the reconstruction quality. 
For a fair comparison,
we capture the normals of PGR~\cite{PGR2022Siyou}
and feed the oriented point set into the SPR solver.
In fact, the original reconstruction strategy of PGR 
uses iso-surfacing to extract reconstructed surfaces
and tends to produce over-smooth results.
By comparison, SPR is better than iso-surfacing in preserving geometric details for normals of the same quality.
% \NW{In face, the original reconstruction strategy of PGR, which uses iso-surfacing to extract reconstructed surfaces, are generally over-smooth compared with the ground truth. By contrast, SPR is better than iso-surfacing in preserving geometric details for normals of the same quality.}
% (In fact, the reconstruction surfaces of PGR, if following the original reconstruction strategy, are generally over-smooth compared with the ground truth. By contrast, SPR is better than PGR in preserving geometric details for normals of the same quality.)
% \XR{It is worth noting that, for the sake of fairness, we only used the normal result of PGR~\shortcite{PGR2022Siyou} for the SPR solver, but did not use the reconstruction result of PGR itself.}
A basic fact is that better normals lead to better-reconstructed surfaces. 
We record the statistics about the reconstruction quality in 
Table~\ref{tab:recon_poisson}.
Note that the Chamfer Distance between the reconstructed surface
and the ground-truth surface is scaled by $100$ times for a better presentation.
The statistics show that for most of the 18 models, 
our predicted normals produce the best reconstruction quality. 
For example, when the point sets are added by $0.5\%$ Gaussian noise, our method has the best scores on $55\%$ of the models. 
Based on the scores, 
the three top-ranked approaches are 
ours, K\"{o}nig~\cite{konig2009consistent} and Hoppe~\cite{hoppe1992surface}, respectively. 
% We summarize the quantitative results in Table~\ref{tab:recon_poisson} using the same $18$ models, sampling conditions and noise levels as before.
% It shows that our method produces the most accurate surface reconstruction results, which also indicates that our approach yields much more accurate and reliable normal orientation results. For point sets fed with $0.5\%$ of white noise, for example, our method outperforms other methods over $60\%$ of the models. 

Furthermore, 
we use Fig.~\ref{fig:comp_recon} to
visually compare reconstruction results 
on the Vase model
at various sampling conditions and noise levels.
It can be seen that
from the reconstructed surfaces
our approach can infer the normals,
with the highest fidelity. 
Especially, even if the noise level amounts to 0.5\%, 
our method can still produce a faithful result;
See the handles of the Vase model. 
% \SQ{please add some words...}

\begin{figure*}[h]
%\vspace{5px}
\centering
\begin{overpic}
[width=0.99\linewidth]{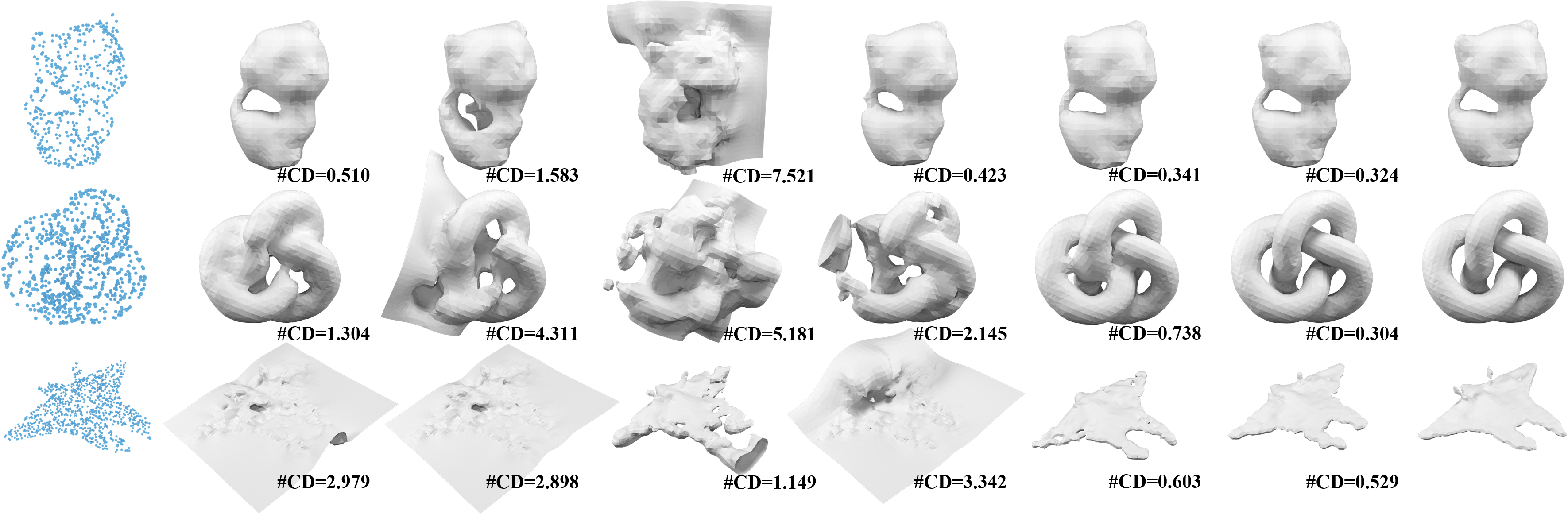}
\put(3, -1){\textbf{Input}}
\put(15, -1){\textbf{Hoppe}}
\put(29, -1){\textbf{K{\"o}nig}}
\put(42, -1){\textbf{PCPNet}}
\put(56, -1){\textbf{Dipole}}
\put(69, -1){\textbf{PGR}}
\put(81, -1){\textbf{Ours}}
\put(94, -1){\textbf{GT}}
\put(-1.5,22){\rotatebox{90}{\textbf{500 points}}}
\put(-1.5,12){\rotatebox{90}{\textbf{750 points}}}
\put(-1.5,1.5){\rotatebox{90}{\textbf{1K points}}}
\end{overpic}
\caption{Tests are made on sparse point clouds: $500$ points, $750$ points and $1$K points.
Our results are close to the ground truth for each of the three inputs.
The comparison shows that
our algorithm has a big advantage on sparse raw data.   \ZY{We also mark clearly the Chamfer Distance~(CD) scores between the reconstructed surface and the ground-truth surface for a quantitative comparison. Note each CD score is scaled by a factor of 100.}
% \SQ{need to update the last row}\XR{Done.}
% We choose models with tubular or thin-plate structures: \textit{kitten}, \textit{knot}, and \textit{plane-1}. \NW{Updated}
}
\label{fig:comp_lowsampling}
\end{figure*}

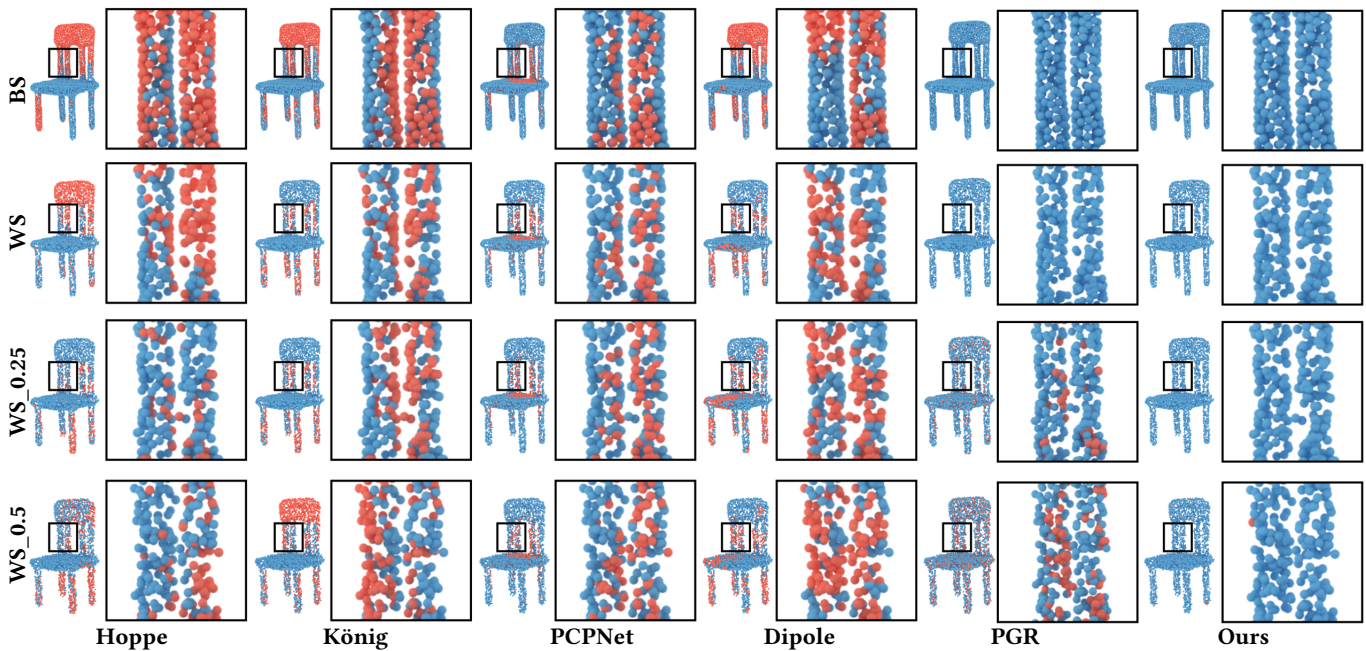
\begin{figure*}[!t]
%\vspace{5px}
\centering
\begin{overpic}
[width=0.99\linewidth]{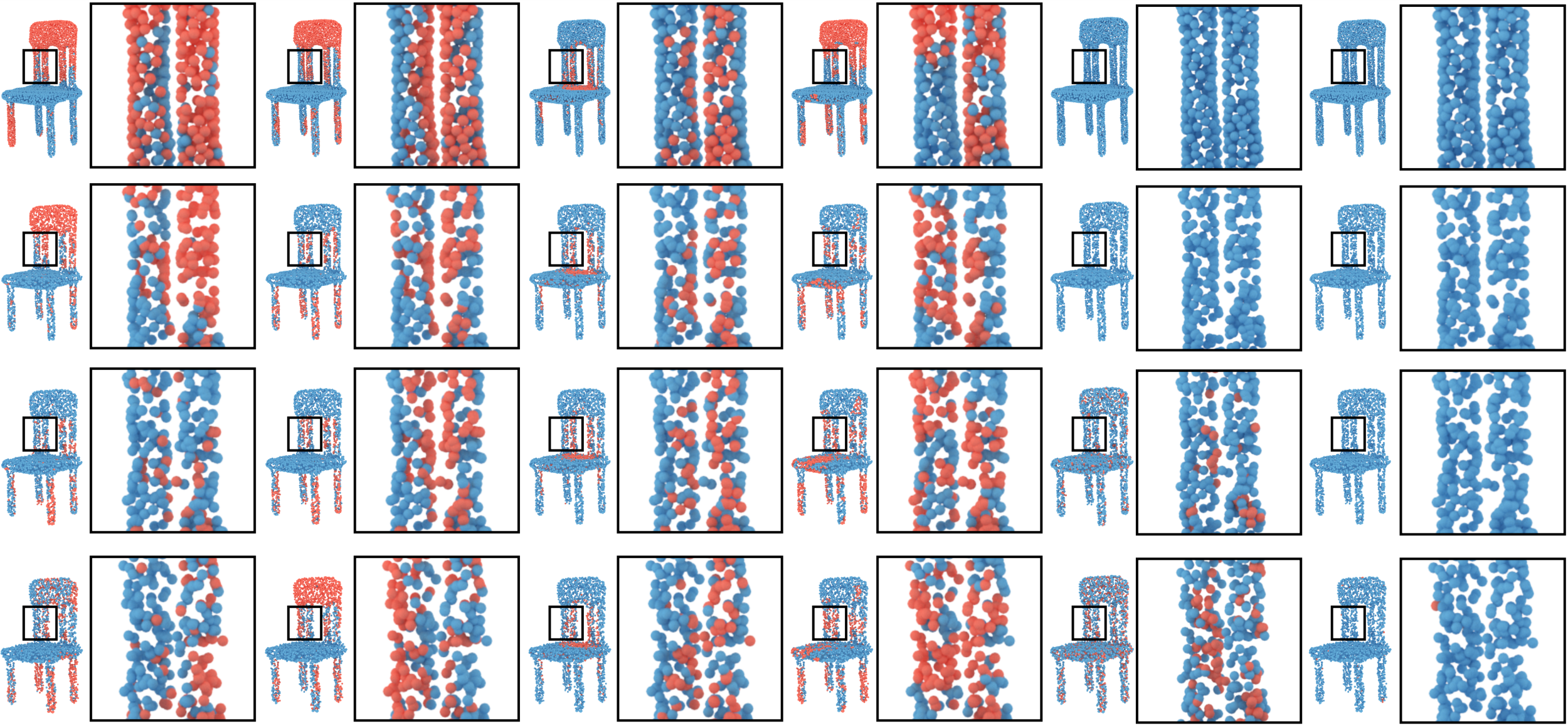}
\put(5, -1.5){\textbf{Hoppe}}
\put(22, -1.5){\textbf{K{\"o}nig}}
\put(39, -1.5){\textbf{PCPNet}}
\put(55, -1.5){\textbf{Dipole}}
\put(72, -1.5){\textbf{PGR}}
\put(89, -1.5){\textbf{Ours}}
\put(-1.5,39){\rotatebox{90}{\textbf{BS}}}
\put(-1.5,28){\rotatebox{90}{\textbf{WS}}}
\put(-1.5,14){\rotatebox{90}{\textbf{WS\_0.25}}}
\put(-1.5,3){\rotatebox{90}{\textbf{WS\_0.5}}}
\end{overpic}
\caption{
% Qualitative comparison of the normal orientations among our method and four other methods: Hoppe~\cite{hoppe1992surface}, K{\"o}nig~\cite{konig2009consistent}, PCPNet~\cite{guerrero2018pcpnet}, and Dipole~\cite{dipole_propagation}. We show results of four different sampling conditions: BS (blue noise sampling), WS (white noise sampling), WN\_0.25 (white noise sampling with 0.25\% noise) and WS\_0.5 (white noise sampling with 0.5\% noise). 
The Chair model contains thin-walled tubes and plates,
as well as nearby gaps (see the highlighted region). 
Our approach can yield the highest truth percentage among the five approaches. Note that the false predictions are colored in red.   
% Comparison with exiting methods on noise robustness abilities, where the truly predicted normals are colored in blue and the false predictions are colored in red. \NW{Updated}
}
\label{fig:comp_gap}
\end{figure*}

\subsection{Noise, Varying Point Density and Data Sparsity}
%Varying Samplings, Data Sparsity and Point Densities}
\label{sec:comparison_varying}

\paragraph{Noise}
In Fig.~\ref{fig:comp_noise},
we add $0.5\%$ Gaussian noise to the point clouds
of the Cup model, the Chair model, and the Lion model,
to test the noise-resistant ability. 
%see if our approach is resistant to noise. 
% to compare the reconstruction quality on point clouds with $0.5\%$ Gaussian noise. 
% on three point sets with $0.5\%$ Gaussian noise. 
It can be clearly seen from the visual comparison 
that our algorithm has a better noise-resistant ability.
Specially, our algorithm can provide faithful normals 
on the back and the legs of the Chair model, even in presence of serious noise. 
Two reasons account for the noise-resistant property. 
First, we examine the winding number at the Voronoi vertices
whose positions are robust to small variations of the original point cloud, especially for those Voronoi vertices distant to the surface.
Second, the whole optimization framework is built on the regularization of the winding number,
and thus can capture the normal consistency from a global perspective.

\paragraph{Varying point density}
% \NW{Updated}
In Fig.~\ref{fig:comp_density},
we construct a point cloud with varying point density (colored in varying darkness).
We intend to use this example 
to test if our algorithm can deal with irregular point distributions. 
Recall that Eq.~\eqref{eq:windingnum_ori} 
includes an area weight~$a_i$,
which has a serious influence on the estimation accuracy of the winding number. 
We give an intuitive technique for estimating~$a_i$
based on the Voronoi diagram; See Section~\ref{sec:Implementation}. 
The technique is parameter-free and computationally efficient. 
It can be seen from Fig.~\ref{fig:comp_density}
that both the predicted normals and the reconstructed surface 
have a high quality,
which shows that the estimation of~$a_i$ is independent of the point density. 

\paragraph{Data sparsity}
In Fig.~\ref{fig:comp_lowsampling},
we have three sparse point clouds,
and the numbers of points are respectively 
500, 750, and 1K. 
We intend to use this example
to test the performance on sparse inputs
, since when there are nearby gaps and thin-walled tubes/plates, data sparsity will inevitably double the difficulty of predicting normals.
It can be clearly seen from the visual comparison that
our results are close to the ground-truth for each of the three inputs.

\subsection{Nearby Gaps, Thin Plates/Tubes, and Sharp Angles}

\paragraph{Nearby gaps and thin plates/tubes}
% \XR{Talk about Fig.~\ref{fig:comp_gap} here. }
For our approach, the strength in dealing with thin tubes 
has been validated on the Vase model shown in Fig.~\ref{fig:comp_recon}.
In Fig.~\ref{fig:comp_gap},
we give four versions of the Chair point cloud
to evaluate how well it handles nearby gaps and thin plates/tubes. 
As can be observed, our approach noticeably outperforms the SOTA methods in addressing these flaws (see the highlighted region),
which is due to the global property inherited from the winding number. 

% Specially, our algorithm is strong in deal with 
% thin plates 
% of the the Chair model
% add the point cloud of the Chair model
% with different levels of noise,
% The Chair model contains thin-walled tubes and plates,
% as well as nearby gaps (see the highlighted region). 
% Our approach can yield the highest truth percentage among the five approaches. Note that the false predictions are colored in red.  
% In Fig.~\ref{fig:comp_gap}, we conduct a comparison on normal orientation with point cloud at different levels of noises: 1) uniform sampled points without noise (blue noise sampling); 2) unevenly sampled points cloud without noise (white noise sampling); 3) uneven sampled points cloud with $0.25\%$ noises; 4) uneven sampled points cloud with $0.5\%$ noises. It is clear to see that our method generates the most faithful normals with minimum number of false predictions (red points) for all sampling conditions in Fig.~\ref{fig:comp_gap}. \XR{Especially in the dense thin tubes on the back of the chair(See the zoom in window).}
% In Fig.~\ref{fig:comp_recon}, we show our ability of handling thin tubes comparing to SOTA methods using various noisy input point clouds. We use the model $vase$ with many thin tubular parts, \ie the highlighted region. \XR{It can be seen that the reconstruction results of other methods have bulges or breaks under different noise conditions.} Our method is the only method that gives remarkable results with minimum influence of noises. \XR{Especially under 0.5\% noise, only our method maintains the original shape.}

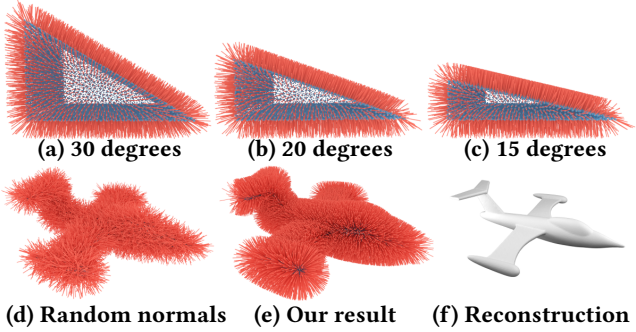
\begin{figure}[t]
    \centering
    \begin{overpic}[width=0.99\linewidth]{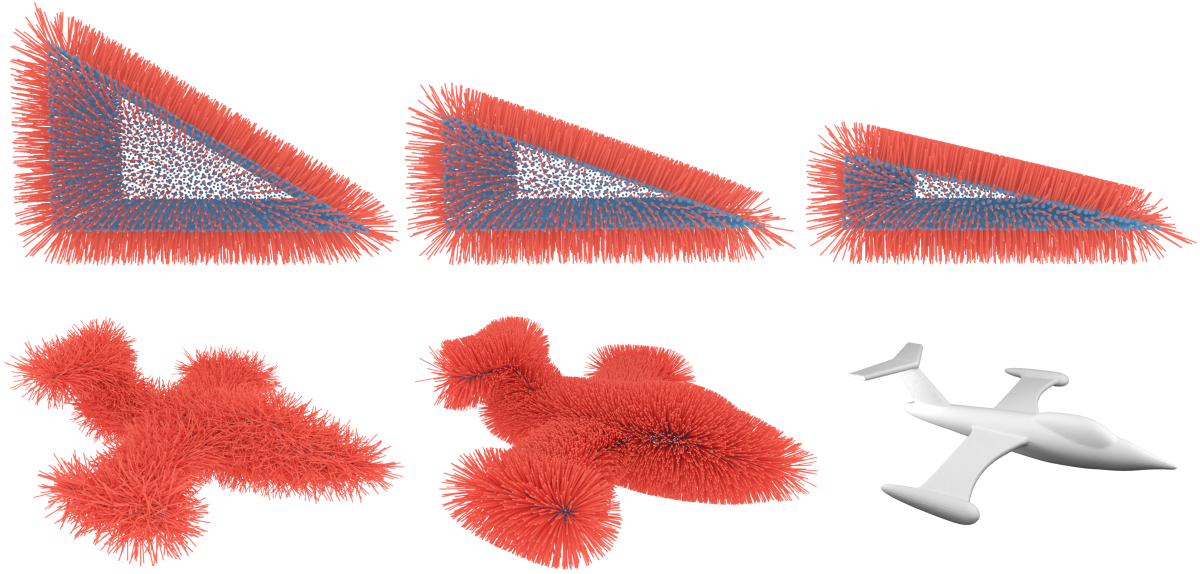}
    \put(6, 23){\textbf{(a)~30 degrees}}
    \put(39, 23){\textbf{(b)~20 degrees}}
    \put(73,23){\textbf{(c)~15 degrees}}
    \put(1, -3){\textbf{(d)~Random normals}}
    \put(40, -3){\textbf{(e)~Our result}}
    \put(68, -3){\textbf{(f)~Reconstruction}}
    \end{overpic}
    \caption{
 (a-c) 
    Our approach can estimate normals accurately even if the angles are as small as 15 degrees. 
    % The normal orientation for point clouds with different sharp angles. 
    (d-f) For the Airplane model with sharp angles, 
    we show the initial normals, the optimized normals, and the faithfully reconstructed result.}
    \label{fig:sharp}
    %\vspace{-5pt}
\end{figure}

\paragraph{Sharp angles}
%\label{sec:sharp_angles}
The existence of sharp angles 
is one of the challenges for orienting a raw point cloud. 
% poses a challenge during normal orientation. 
% Actually, the sharp feature of a shape is ill-defined at different scales. 
In the top row of Fig.~\ref{fig:sharp},
we show three toy models with different dihedral angles. 
It can be seen that our approach can estimate normals accurately even if the angles are as small as 15 degrees. 
We also use the Airplane model 
to test the ability to deal with sharp angles.
Both the optimized normals and the faithfully reconstructed result
show that our algorithm can produce a desirable result 
for point clouds with sharp angles.
The contrast in the bottom row of Fig.~\ref{fig:comp_lowsampling} also validates the effectiveness of our approach in coping with sharp angles.
It's worth pointing out that the propagation-based methods~\cite{dipole_propagation} 
rely on the assumption of spatial coherence, 
which does not hold when sharp angles exist,
and thus fail to fully capture the global context of the shape
in presence of sharp angles. 
    
% For techniques heavily relying on the local features, their methods inevitably degenerate on sharp angles. Different from these methods, our approach puts the sharpness in a global context of the shape while taking advantage of the local features. Therefore, the proposed method effectively achieves globally consistent normal orientation results for this cumbersome case. We show our results on some point cloud with sharp features in Fig.~\ref{fig:sharp}.\XR{Even if the dihedral angle of the input shape is reduced to 15 degrees, our method can still output a globally consistent normal.}

\begin{figure}[!t]
    \centering
    \begin{overpic}[width=\linewidth]{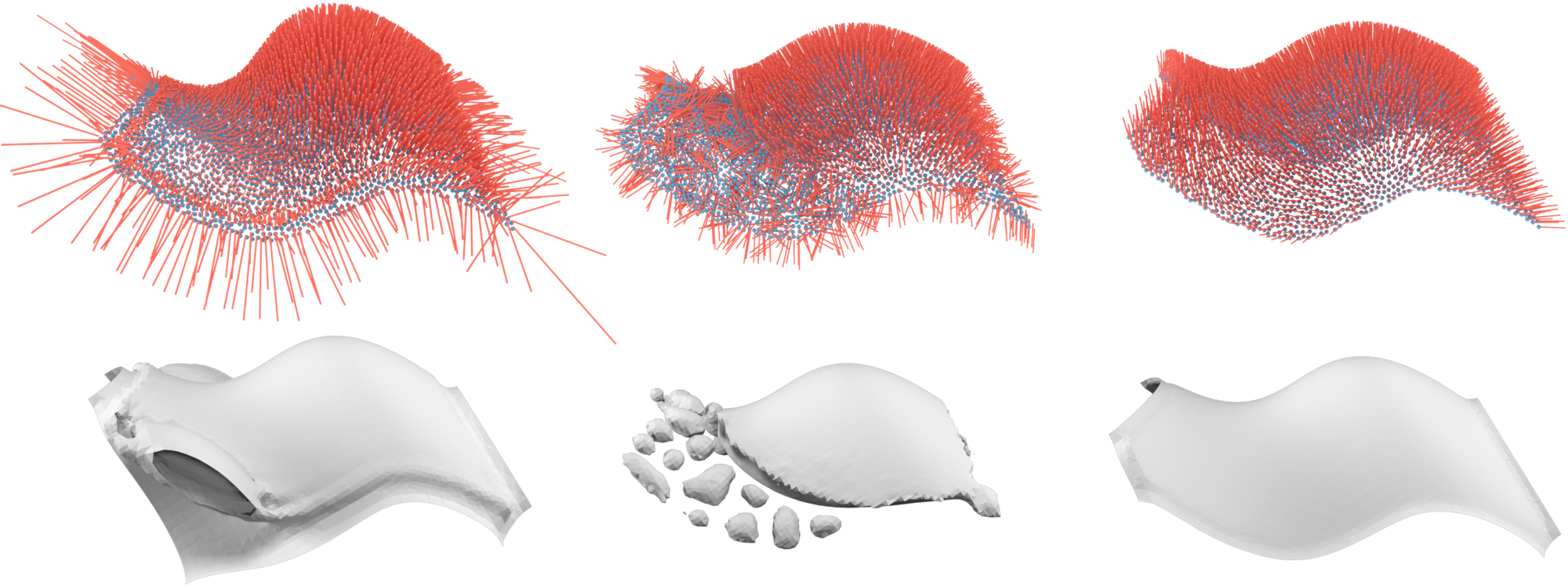}
    \put(13, -3){\textbf{(a)~PGR}}
    \put(45, -3){\textbf{(b)~iPSR}}
    \put(78, -3){\textbf{(c)~Ours}}
    \end{overpic}
    \caption{Comparing PGR, iPSR, and ours on estimating normals for an open-surface point cloud.     
    iPSR does not support open surfaces.
    PGR fails to report reliable normals for the boundary points.
    However, our estimated normals comply with the real shape
    at both the interior points and the boundary points.
    }
    \label{fig:comp_incomplete}
    %\vspace{-2pt}
\end{figure}

\begin{figure}[!t]
    \centering
    %\vspace{-4mm}
    \begin{overpic}[width=0.99\linewidth]{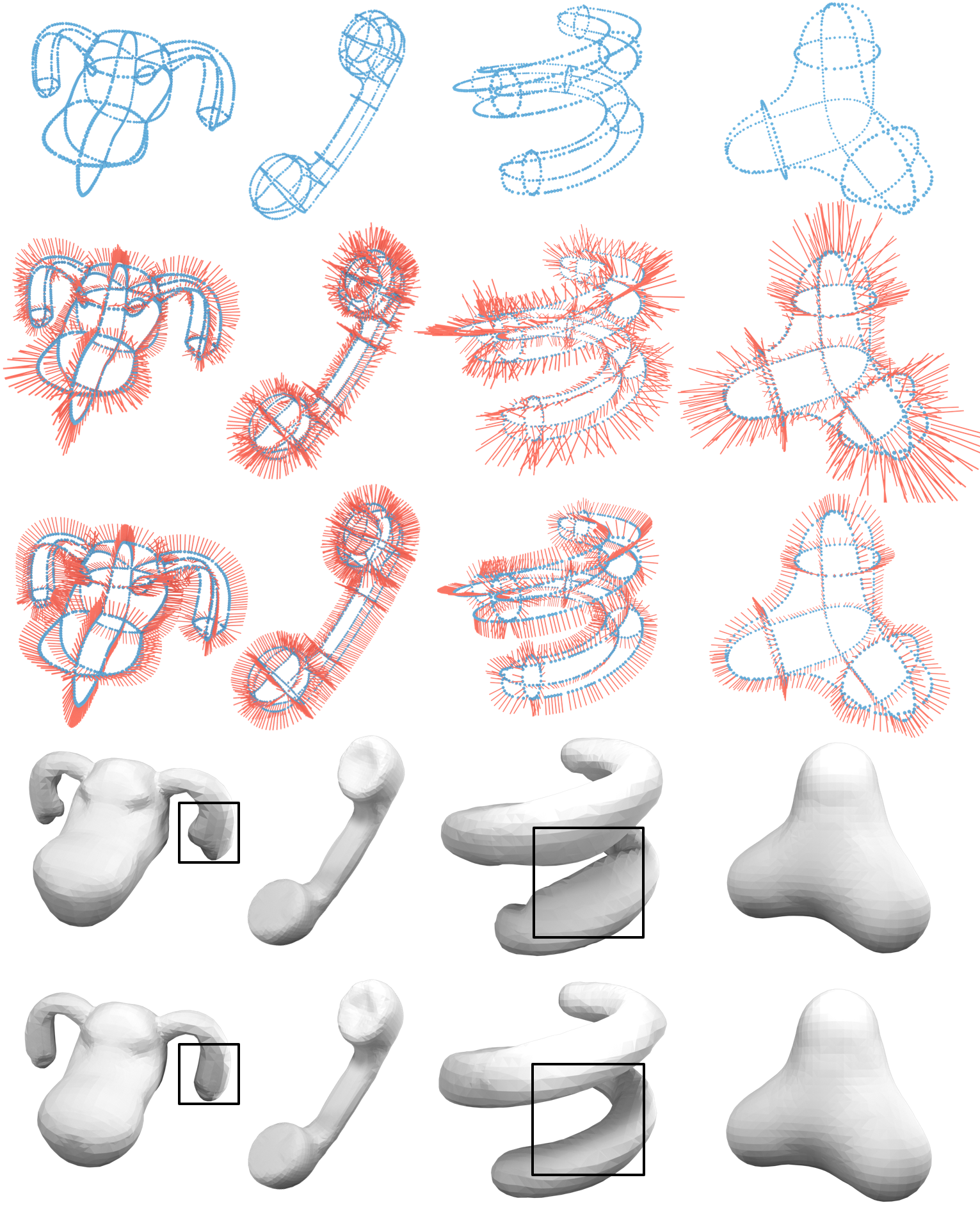}
    \put(-2, 88){\rotatebox{90}{\textbf{Input}}}
    \put(-2, 63){\rotatebox{90}{\textbf{PGR Normal}}}
    \put(-2, 41){\rotatebox{90}{\textbf{Our Normal}}}
    \put(-2, 22){\rotatebox{90}{\textbf{PGR Recon}}}
    \put(-2, 3){\rotatebox{90}{\textbf{Our Recon}}}
    \end{overpic}
    \caption{Normal estimation for wireframe-type point clouds. All the models are from VIPSS~\cite{huang2019variational}.
    PGR may produce bulges around thin tubular structures. 
    }
    \label{fig:wireframes}
%    \vspace{-10pt}
\end{figure}

\begin{figure*}[h]
%\vspace{5px}
\centering
%\vspace{-2mm}
\begin{overpic}
[width=0.99\linewidth]{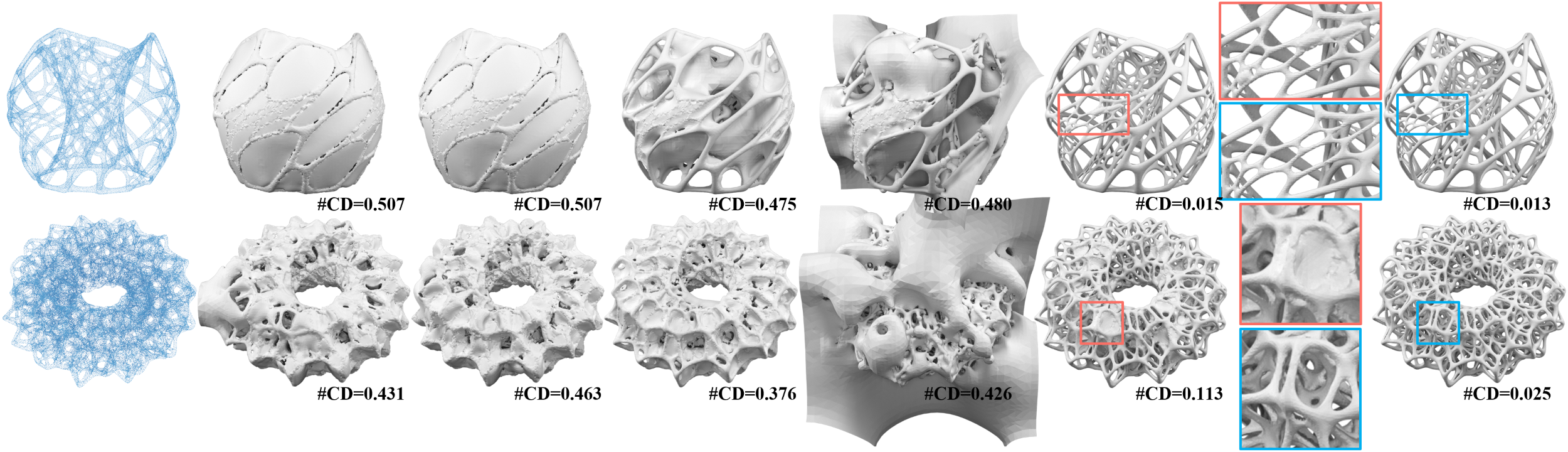}
\put(5, 0){\textbf{Input}}
\put(17, 0){\textbf{Hoppe}}
\put(30, 0){\textbf{K{\"o}nig}}
\put(42, 0){\textbf{PCPNet}}
\put(57, 0){\textbf{Dipole}}
\put(71, 0){\textbf{iPSR}}
\put(92, 0){\textbf{Ours}}
\end{overpic}
\vspace{-2mm}
\caption{
Tests on point clouds with highly complex topology/geometry.
The model in the top row has 80K points while the model in the bottom row has 100K points.
Note that PGR runs out of GPU memory on an NVIDIA GeForce RTX 3090 graphics card. \ZY{Chamfer Distance~(CD) between the reconstructed surface and the ground-truth surface is marked for a quantitative comparison. Note each CD value is scaled by a factor of 100.}%due to the Graphics card memory limit.
}
\vspace{-2mm}
\label{fig:comp_diffcult}
\end{figure*}

\begin{figure}[!h]
    \centering
    \begin{overpic}[width=0.9\linewidth]{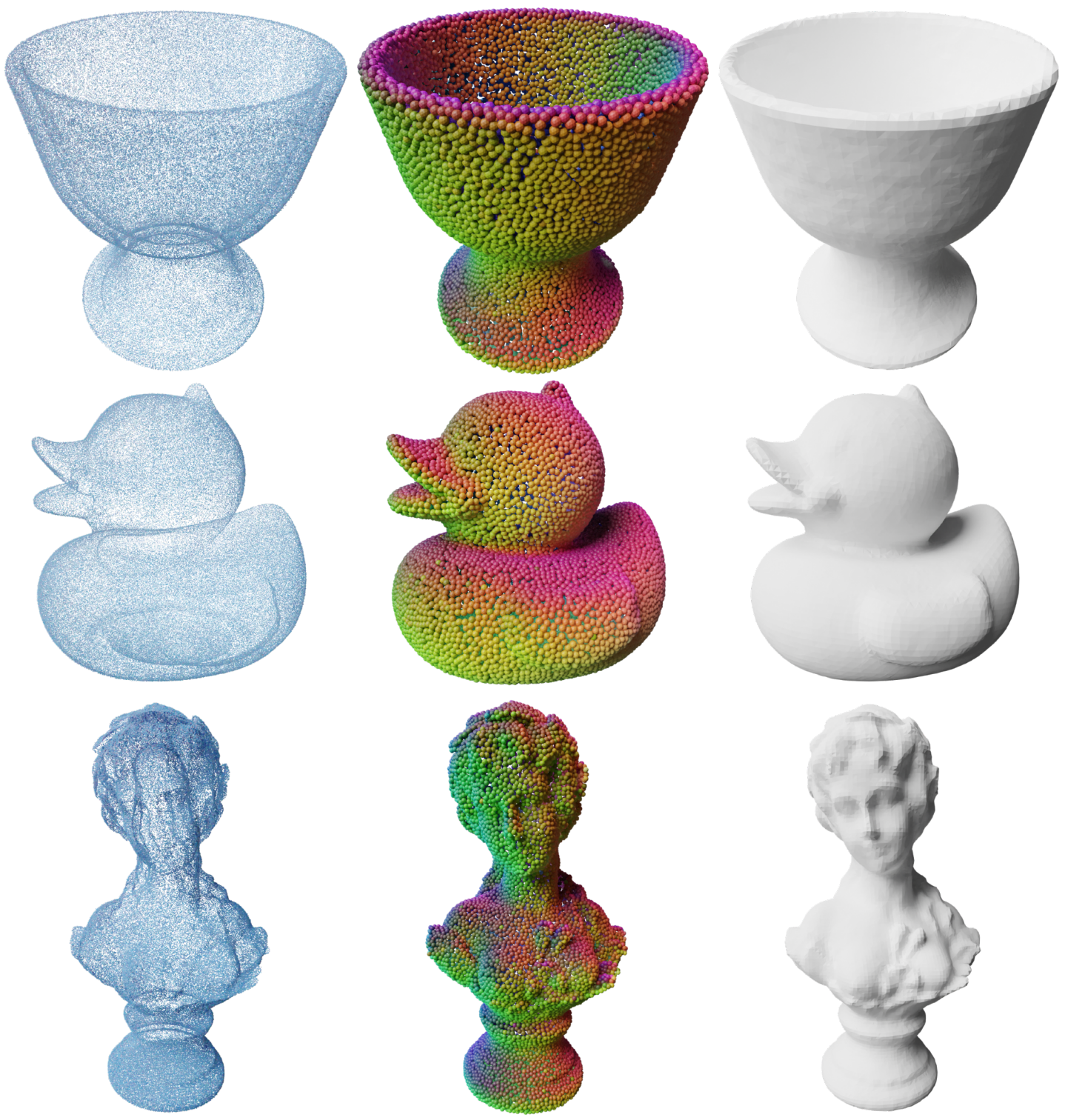}
    \put(10, -4){\textbf{Input}}
    \put(33, -4){\textbf{Normal Encoding}}
    \put(67, -4){\textbf{Reconstruction}}
    \end{overpic}
    \vspace{5pt}
    \caption{
    The three raw point clouds are real scans downloaded from~\cite{huang2022surface} dataset. 
    Each of them is downsampled to 10K points. 
    From the color-coded visualization of normals, 
    as well as the reconstructed surfaces, 
    we can see that our method can deal with real scans. 
    }
    \label{fig:real_scan}
    %\vspace{-8pt}
\end{figure}

% \begin{figure}[!h]
%     \centering
%     \begin{overpic}[width=1.0\linewidth]{figures/4_result/pgripsr.png}
%     \put(10, -4){\textbf{Input}}
%     \put(35, -4){\textbf{iPSR}}
%     \put(59, -4){\textbf{PGR}}
%     \put(85, -4){\textbf{Ours}}
%     \end{overpic}
%     %\vspace{5pt}
%     \caption{
%     Comparison with two globally methods iPSR~\cite{hou2022iterative} and PGR~\cite{PGR2022Siyou}.
%     }
%     \label{fig:pgripsr}
%     %\vspace{-15pt}
% \end{figure}

\subsection{Open Surfaces, Wireframes, Complex Topology and Real Scans}
\label{sec:comparison_more}

\paragraph{Open-surface Point Cloud}
In Fig.~\ref{fig:comp_incomplete},
we sample a point set from an open surface. 
We compare PGR, iPSR, and ours on estimating normals 
on the open-surface point cloud.
iPSR does not support open surfaces. PGR fails to report
reliable normals for the boundary points since it assumes the closed surface.
In contrast, our estimated normals
comply with the real shape at both the interior points and the boundary
points.

On one hand, the winding number is still indicative for open surfaces~\cite{jacobson2013robust, Chi_2021_ICCV}.
On the other hand, the three terms in our objective function 
do not assume the closedness of the surface.
Recall that we include the intersections 
between the 1.3x bounding box and the Voronoi diagram 
as examination points. 
If the input point set encodes a closed surface,
it is proper to deem the intersections as outside points
and enforce the winding number at the intersections to be 0. 
But for open surfaces, the constraint cannot be specified. 
Therefore, in our implementation, 
we do not specify the requirements in all our experiments. 
    % \caption{Computing consistent normals for the point set of
    % an open surface.}
    % \label{fig:comp_incomplete}
    
% Our method is also capable of processing incomplete point cloud that encodes an open surface. Fig.~\ref{fig:comp_incomplete} demonstrates the normal orientating process of our algorithm given a raw point cloud of a shape with an open boundary, in which the surface normals are gradually oriented during the optimization. This is because the \textit{winding-number field} is well-defined for both closed and open-boundary shapes~\cite{jacobson2013robust,Chi_2021_ICCV}, enabling a consistent normal orientation on the surface.

\paragraph{Wireframes}
Wireframes serve as a kind of compact skeletal representation of a real-world object.
Due to the extreme data sparsity, 
the SOTA methods fail to correctly predict the normals. 
Unlike patch-based normal fitting~\cite{dipole_propagation},
our approach aims at 
evaluating the global normal consistency by
computing the overall contribution of each point.
The experimental results in Fig.~\ref{fig:wireframes}
show that both our method and PGR are capable of handling the wireframe-type inputs,
but PGR may produce bulges around thin
tubular structures (see the highlighted window). 
% \XR{and outperforms PGR~\cite{PGR2022Siyou} }.
% \XR{Although PGR~\cite{PGR2022Siyou} can also handle wireframe input, it is difficult to perfectly distinguish complex tubular objects, which leads to the generation of bulges, such as the dog head and spring models in Fig.~\ref{fig:wireframes}.
% }
% \XR{Because our approach is to optimize the entire winding-number field globally, even wireframe-type inputs will work perfectly.} In Fig.~\ref{fig:wireframes}, we demonstrate a set of examples of the normal orientation and surface reconstruction results.

\paragraph{Highly complex structures}
Fig.~\ref{fig:comp_diffcult}
shows two nest-like models with complex topology/geometry.
The point cloud in the top row has 80K points
while the point cloud in the bottom row has 100K points. 
It can be seen that all five SOTA methods fail on the two 
highly complex models.
%because they are too complicated in topology/geometry. 
iPSR~\cite{hou2022iterative} depends on the initialization of normals. 
For a shape with complicated topology/geometry,
iPSR cannot reverse the false normals to the correct configuration, 
and thus easily cause disconnection or adhesion,
especially around thin structures.
PGR is not GPU-memory friendly (superlinear growth w.r.t. the number of points) and runs out of memory when the input point cloud
reaches 80K points (note that we test PGR on an NVIDIA GeForce RTX 3090 graphics card with 24GB of GPU memory). 
In contrast, our method can deal with complicated geometry/topology
and faithfully recover the normal vectors.

\paragraph{Real scans}
% \label{sec:real_scan}
% \ZY{Done.}
Fig.~\ref{fig:real_scan}
shows three raw point clouds,
each of which is down-sampled to 10K points. 
From the color-coded visualization of normals, as well as the reconstructed surfaces, it can be seen that our method can effectively orient the normals for real-life objects,  which validates the usefulness of our algorithm in practical scenarios.

% various shapes and is robust to input noise. With the re-oriented point normal vectors, the surfaces are faithfully reconstructed from real-scan, which validates the usefulness of our algorithm in practice scenarios.

% We also demonstrate the capability of our method in handling real scanned point clouds. We evaluate our algorithm on a set of scans of real-life objects from \cite{huang2022surface} dataset. All point clouds tested are randomly down-sampled to $10$k points. As shown in Fig.~\ref{fig:real_scan}, our method effectively orients the normals for various shapes and is robust to input noise. With the re-oriented point normal vectors, the surfaces are faithfully reconstructed from real-scan, which validates the usefulness of our algorithm in practice scenarios.
\begin{figure}[!t]
    \centering
    \begin{overpic}[width=1.0\linewidth]{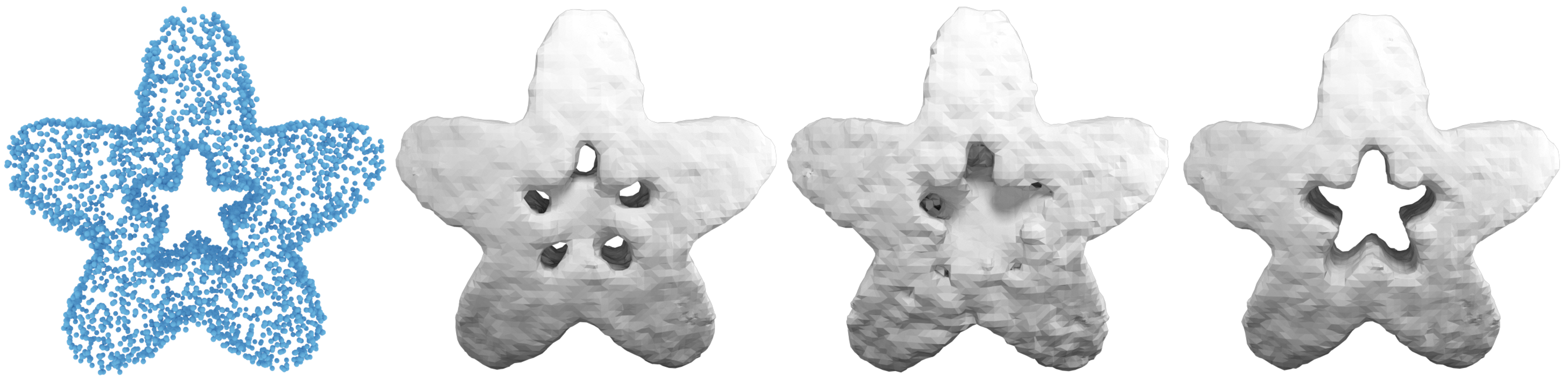}
    \put(8, -4){\textbf{Input}}
    \put(33, -4){\textbf{PGR\_1}}
    \put(58, -4){\textbf{PGR\_2}}
    \put(84, -4){\textbf{Ours}}
    \end{overpic}
    %\vspace{5pt}
    \caption{
    Two parameter settings of PGR~\cite{PGR2022Siyou}.
    PGR\_1: wmin=0.04, alpha=2.0. PGR\_2: wmin=0.0015, alpha=1.05. 
    }
    \label{fig:pgr}
    %\vspace{-15pt}
\end{figure}
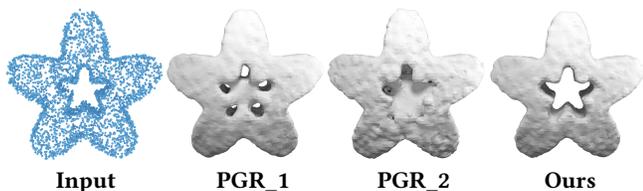

\subsection{Discussion on Global Methods}
In the following, 
we make a discussion on the global methods including
iPSR~\cite{hou2022iterative}, PGR~\cite{PGR2022Siyou} and ours. 
\paragraph{Ours v.s. iPSR}
iPSR, as a global method, 
is excellent in estimating oriented normals. 
Benefiting from Poisson surface reconstruction,
it has many nice features.
In its nature, iPSR gets more and more prior during the iterations of the reconstruction surface.
If the given raw data does not have serious imperfections or challenging structures,
iPSR can produce desirable normals, as well as a high-quality reconstruction surface.
On the flip side, iPSR inherits some disadvantages of Poisson surface reconstruction. 
For example, %it is easy to imagine that 
iPSR cannot deal with the point clouds of an open surface,
as shown in Fig.~\ref{fig:comp_incomplete}.
Additionally, 
when the point clouds are as complex as Fig.~\ref{fig:comp_diffcult},
iPSR cannot reverse the false normals to the correct configuration 
and easily cause disconnection or adhesion,
especially around thin structures.
To summarize, the biggest weakness of iPSR lies in that 
if the initial surface is much different from the target surface, 
the structural/topological issues are hard to be fixed.
% which shows that iPSR depends on the initialization of normals. 
% For a shape with complicated topology/geometry,

\paragraph{Ours v.s. PGR} 
First, in the original paper of PGR, 
the authors recommend several groups of parameters,
depending on the number of points in the raw data. 
In contrast, our parameters remain the same for all the experiments,
independent of the size of the raw data. 
Second, the statistics (available in the supplementary material) show that
our method has better accuracy in predicting normals due to the alignment term that enforces the normals to point toward outside Voronoi poles, whereas, the inaccurate normals produced by PGR 
weaken the ability of fidelity preserving.
Fig.~\ref{fig:pgr} shows that 
the inaccurate normals of PGR cause a failure in recovering the center hole of the star shape, and any recommended parameters. % due to the inaccuracy of predicted normals. 
Finally, PGR incurs a quadratic complexity of computational time and memory footprint, which limits its practical usage
, especially on large models.
For example, PGR fails to deal with the complex shapes shown in Fig.~\ref{fig:comp_diffcult}.

\subsection{Run-time Performance}
\label{sec:runtime}
We provide the run-time performance statistics in Table~\ref{tab:time}. 
% We report the detailed time statistics of our method \XR{and other methods} in Table~\ref{tab:time}.
The tests are made on the torus model
with different resolutions ranging from $0.5K$ points to $10K$ points. 
The total running time mainly consists of the construction of the Voronoi diagram and the optimization. 
It can be seen that
optimization is the most time-consuming stage
due to (1)~the number of variables is twice as large as the number of the points, and (2)~the objective function has to be evaluated by a double loop, i.e., over each~$\sample_i$ and each~$\query_j$, leading to a non-linear climbing 
in the computational overhead. 
But we must point out that 
even for the Torus model with 10K points, generally, 50 iterations, computed in 10 minutes, suffice to arrive at the termination. 
The overhead is acceptable for many non-real-time geometry processing tasks.

\begin{table}[h]
\caption{Running time (in seconds) of different methods w.r.t. the number of points \#V. We test with the Torus model. %The total timing cost mainly consists of the construction of the Voronoi diagram and the optimization. 
% \XR{add others...}
% \SQ{be risky to say "the LBFGS's max iteration time to be $50$ steps"}
%and set the LBFGS's max iteration time to be $50$ steps. 
}
\vspace{-3mm}
\label{tab:time}
\resizebox{\linewidth}{!}{
\begin{tabular}{l|cccccc}
\toprule
\#V     & 0.5K   & 1K   & 3K   & 5K    & 7K    & 10K   \\ \midrule
Hoppe     & 0.324 & 0.477  & 0.625  & 0.967   & 1.112   & 1.569  \\
K\"{o}nig     & 0.306 & 0.353  & 0.625  & 0.815   & 1.017   & 1.185  \\
PCPNet     & 4.226 & 5.446  & 6.388  &  8.753  & 11.015   & 12.581  \\
Dipole     & 3.277 & 3.565  & 5.489  & 8.411   & 11.602   & 14.517  \\
PGR     & 0.228 & 0.260  & 0.489  & 0.612   & 0.823   & 1.020  \\
iPSR     & 2.801 & 3.535  & 4.321  & 4.543   & 4.476   & 5.173  \\
% Ours_{Voronoi}     & 0.584 & 1.181  & 2.743  & 5.267   & 6.034   & 10.132  \\
% Ours_{Opt}       & 4.656 & 14.318 & 69.691 & 169.274 & 276.260  & 549.722 \\
Ours       & 5.240  & 15.499 & 72.434 & 174.541 & 282.294 & 559.854 \\ \bottomrule
\end{tabular}
}
\end{table}
% The tests are made on the 
% Running time (in seconds) w.r.t the number of points \#V. We use the torus model for testing and set the LBFGS's max iteration time to be $50$ steps.
% different numbers of sample points. Our whole pipeline is composed of two main steps: (1) Voronoi diagram computation and (2) optimization. 
% % We show the runtime performance w.r.t. different numbers of sample points in Table~\ref{tab:time}. 
% % As summarized in Table~\ref{tab:time}, 
% The main computation overhead of our method is the optimization stage (noted by LBFGS) due to the complexity of global optimization. The computational time increases polynomially when the size of the input point cloud gets larger. However, our method can\XR{have the ability to?} be extended to GPU-based computation, which is discussed in the limitation and future work in Sec.~\ref{sec:limit_future}.

\begin{figure}[!t]
    \centering
    \begin{overpic}[width=\linewidth]{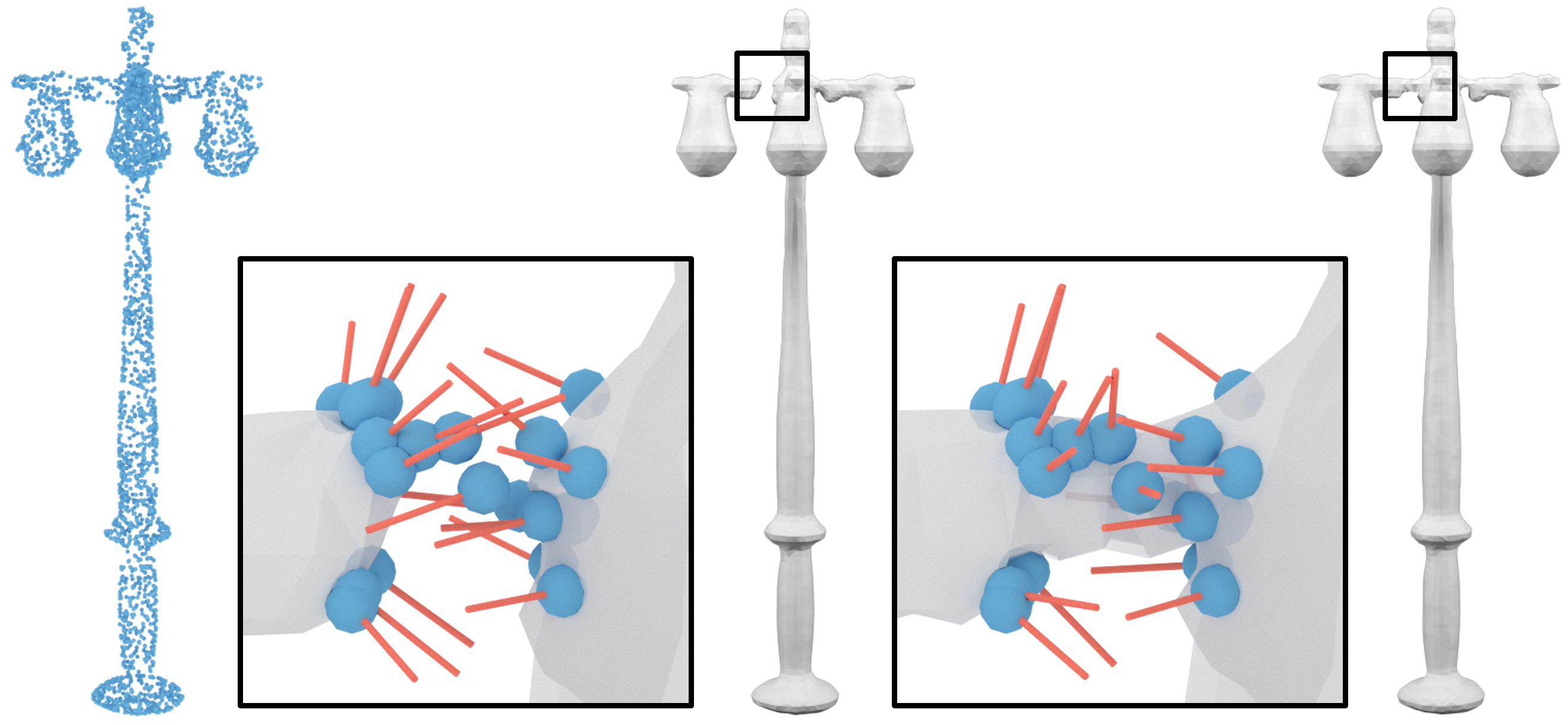}
    \put(2, -3.5){\textbf{(a)~Input}}
    \put(43, -3.5){\textbf{(b)~iPSR}}
    \put(84, -3.5){\textbf{(c)~Ours}}
    \end{overpic}
    \caption{
    We sample 3K points from the Lamp model ShapeNet~\cite{chang2015shapenet}. 
    Compared with iPSR~\cite{hou2022iterative},
    our method produces better reconstruction quality.
    However, the highlighted part shows that 
    our predicted normals are not very accurate, 
    leading to a conspicuous artifact in the reconstructed surface. 
    % , which shows that    
    % Comparison with iPSR~\cite{hou2022iterative} on surface reconstruction from unoriented point cloud. We adopt the lamp model from ShapeNet~\cite{chang2015shapenet} with the number of points being $3$k.
    }
    \label{fig:Comp_ispr}
    %\vspace{-10pt}
\end{figure}
\section{Limitations and Future Work}
\label{sec:limit_future}

The first limitation lies in the run-time performance. 
As we have to repeatedly evaluate the winding number
for each data point and each query point, 
the timing cost spent in a single computation of the objective function amounts to~$O(NM)$, where $N$ and $M$ are respectively the number of data points and the number of query points. 
To alleviate this, one could downsample the input point set.
After the normals of the subset are estimated, 
the un-oriented points can get normals by a simple propagation. 
Another direction of boosting the run-time performance 
is to develop a GPU version to further improve the parallelism.

The second limitation is that 
there is room for further improvement in the accuracy of predicted normals. 
In Fig.~\ref{fig:Comp_ispr}, 
we sample 3K points on the Lamp model from ShapeNet~\cite{chang2015shapenet}. 
Our predicted normals are much different from the ground-truth normals, 
in spite of being better than iPSR~\cite{hou2022iterative}.
The inaccurate normals lead
to a conspicuous artifact in the reconstructed surface.
\XR{In the future, we shall further improve the prediction accuracy based on prior knowledge about the geometry/topology.}

% \XR{When the input point cloud (dense or sparse) has small noise, the points still have the balance property. Our method can deal with dense point clouds (e.g. 50K points) with 1\% Gaussian noise. But when the noise increased to a higher level, our method may fail, the balance property does not hold anymore under harsh noise. }

\ZY{Finally, our method may fail when
there are many points scattered inside the volume,
or a high-density point cloud is coupled with high-level noise. 
Both situations may violate the $0-1$ balance requirement, potentially resulting in a failure case. To address these challenges, 
it is necessary to develop some pre-processing techniques to filter out those points that do not contribute to the underlying surface at all.}

% \ZY{@Rui? only when the points are inside the volume? what if outside the volume?}
% \XR{That will be outliers, it doesn't matter.}

% Another limitation lies in the accuracy of the estimated normal vectors. This is because our method is mainly designed for achieving globally consistent normal orientation where normals are over-smoothed during the regularizing the winding-number field in sharp areas causing errors. The further direction can be introducing more location constraints according to the geometric features to improve the estimation accuracy. 

% \ZY{We also compare with the recently proposed method iterative Poisson Surface Reconstruction~(iPSR)~\cite{hou2022iterative} that reconstructs the surface from the unoriented point cloud iteratively and updates normals using the generated surface of the last iteration. The result can be found in Fig.~\ref{fig:Comp_ispr}. Compared with iPSR~\cite{hou2022iterative}, our algorithm produces a more faithful reconstruction result, whereas iPSR yields a result with a destroyed topology.}\XR{Move to limitation.}

\section{Conclusion}
This paper presents a globally consistent normal orientation method by regularizing the winding-number field.
We formulate the normal orientation problem
into an optimization-driven framework that considers three requirements in the objective function,
two of which specify requirements on the winding-number field and the other term constraining the alignment with Voronoi poles.
We conduct extensive experiments on point clouds with various imperfections and challenges, such as noise, data sparsity, nearby gaps, thin-walled plates, and highly complex geometry/topology.
Experimental results exhibit the advantage of the proposed approach.

% Compared with SOTA methods, our method produces supreme normal orientation results while exhibiting impressive results in handling sparse, noisy point clouds and shapes with highly complex geometry/topology. We also test with various inputs, which validates the effectiveness and usefulness
% of the proposed method.

\begin{acks}
The authors would like to thank the anonymous reviewers for their valuable comments and suggestions. This work is supported by the National Key R\&D Program of China (2021YFB1715900), the National Natural Science Foundation of China (62002190, 62272277, 62072284), and the Natural Science Foundation of Shandong Province (ZR2020MF036, ZR2020MF153). Ningna Wang and Xiaohu Guo were partially supported by National Science Foundation (OAC-2007661).
\end{acks}

\appendix
\section*{Appendix}

\section{$f$ Has a Lower Bound}

In order to show that the minimization of~$f(\normal)$ (see Eq.~(2) in the paper) can arrive at the termination, we need to prove why the objective function has a lower bound. 

Recall that $f(\normal)$ has three terms $f_{01}$, $f_A$ and $f_B$.
By assuming that the maximum of $\|\query_{k}^i- \sample_i\|$
is $L$, the diagonal length of the enclosing box, we have
\begin{equation}
\begin{split}
|f_{A}(\normal)| 
&\leq \sum_i^N\left |\frac{1}{\numV_i}\sum_k^{\numV_i}{w_{k}^{i}}\normal_i\cdot (\query_{k}^i- \sample_i)\right | \\
&\leq \sum_i^N\left(\frac{L}{\numV_i}\sum_k^{\numV_i}{|w_{k}^{i}|}  \right).
\end{split}
\end{equation}
Therefore, it is easy to show that $f_{01}$ is quartic about $w_j$ 
while $f_A$ and $f_B$ can be bounded by a lower-degree polynomial function about $w_j$.
Suppose that $w_j$ goes to $+\infty$ or $-\infty$.
$f_{01}$ must approach $+\infty$ in either case. 
Considering that $f_{01}$ has a higher rate of change
than $f_A$ and $f_B$,
we can conclude that 
when $w_j$ goes to $+\infty$ or $-\infty$, 
the overall value of $f(\normal)$ must approach $+\infty$.
As our goal is to minimize $f(\normal)$, 
$w_j$ must be naturally constrained to a limited range of $[W_1, W_2]$.
The boundedness of $f(\normal)$
can be immediately verified based on the fact that $f(\normal)$ is a continuous function in the closed interval $[W_1, W_2]$.

% all other functions have lower degrees. 
% Specially, 
% Since $f_{01}$ is a \textit{double well function}, whose graph is shown in Fig.~6, it naturally has a lower bound. By minimising $f_{01}$, we constraint the value of $w_j$ in a range of $[W_1, W_2]$. Suppose $\vecdiagonal$ is a vector whose size $|\vecdiagonal|$ is the same as the length of the bounding box diagonal, then $|\query_{k}^i- \sample_i| \leq |\vecdiagonal|$ in $f_A$ (see Eq.~6). And the minimum value of $\normal_i \cdot \vecdiagonal$ is $-|\vecdiagonal|$, so the $f_A$ can be approximated as:
% \begin{equation}
% \begin{split}
% f_{A}(\normal) 
% &= \sum_i^N\left(\frac{1}{\numV_i}\sum_k^{\numV_i}{w_{k}^{i}}\normal_i\cdot (\query_{k}^i- \sample_i)\right) \\
% &\geq - \sum_i^N\left(\frac{|\vecdiagonal|}{\numV_i}\sum_k^{\numV_i}{w_{k}^{i}}  \right)
% \end{split}
% \end{equation}
% Moreover, $f_B$ is a quadratic function of $w_j$ and decrease much slower than $f_{01}$. Since $w_j$ is constrained by $f_{01}$, functions $f_A$ and $f_B$ must have lower bound when $w_j=W_2$.

\section{More Comparison}
% \subsection{Experimental Setting}

% \paragraph{Platform}
% Our experiments are conducted  on a computer with an AMD Ryzen 9 5950X CPU and 32 GB memory.  We run the learning-based approaches 
% \cite{guerrero2018pcpnet,zhu2021adafit,li2022neaf,dipole_propagation} on an NVIDIA GeForce RTX 3090 card.

% \paragraph{Normalization} For the comparison with  All the point cloud models are normalized to a range of $[-0.5,0.5]^3$. We test our method with various sampling conditions by taking both point density and noise into consideration; See Sec~\ref{sec:normal_ori_recon} for more details. 

% \paragraph{Parameters}
% In all the experiments, we adopt the same parameter setting: $\lambda_N = 10.0$, $\lambda_V = 50.0$ and $D = 4.0$.
% We use the L-BFGS algorithm implemented by C++ for solving the optimization defined in Eq.~(\ref{eq:all}). 
% The termination condition is set by requiring the difference of the objective function value at two consecutive steps not to exceed a threshold $1.0$.
% %In this paper, we set $\tau=1.0$ for all experiments. %with a maximum step set to be $50$ steps.

% \subsection{Comparisons}
\label{sec:more_comp}

\begin{table*}[!htp]
\centering
\caption{Comparison of the angle RMSE with the SOTA methods at different sampling conditions.}
\vspace{-3mm}
\label{tab:normal_rmse}
\resizebox{1.0\linewidth}{!}{
\begin{tabular}{l|cccccc|cccccc|cccccc|cccccc}
\toprule
Sampling    & \multicolumn{6}{c|}{Blue Noise Sampling}                                                                                                                                             & \multicolumn{6}{c|}{White Noise Sampling}                                                                                                                                             & \multicolumn{6}{c|}{White Noise Sampling With 0.25\% noise}                                                                                                 & \multicolumn{6}{c}{White Noise Sampling With 0.5\% noise}                                                                                                    \\ \cmidrule{1-25}
Models      & Hoppe                           & K\"{o}nig                           & PCPNet                           & Dipole  & PGR                              & Ours                             & Hoppe                           & K\"{o}nig                            & PCPNet                           & Dipole  & PGR                              & Ours                             & Hoppe                           & K\"{o}nig                            & PCPNet                           & Dipole  & PGR    & Ours                             & Hoppe                            & K\"{o}nig                            & PCPNet                           & Dipole  & PGR    & Ours                             \\ \midrule
82-block    & 117.870                         & 20.191                          & 52.218                           & 24.837  & 30.611                           & \textbf{18.631 } & 22.972                          & 21.895                           & 57.822                           & 36.191  & 30.903                           & \textbf{19.890 } & 67.967                          & 22.214                           & 58.451                           & 32.666  & 55.719 & \textbf{22.205 } & 27.425                           & 23.005                           & 60.602                           & 32.170  & 77.330 & \textbf{27.192 } \\
bunny       & 25.051                          & 33.147                          & 45.505                           & 35.795  & \textbf{19.388 } & 23.581                           & 17.652                          & 30.446                           & 48.111                           & 41.014  & 21.082                           & \textbf{18.473 } & 24.560                          & 33.488                           & 49.925                           & 38.942  & 51.351 & \textbf{17.334 } & 28.607                           & 32.378                           & 50.139                           & 34.774  & 73.198 & \textbf{24.735 } \\
chair       & 98.024                          & 96.834                          & 60.557                           & 105.177 & 18.597                           & \textbf{9.324 }  & 92.216                          & 60.172                           & 62.560                           & 74.170  & 20.653                           & \textbf{12.178 } & 57.527                          & 60.761                           & 62.651                           & 86.063  & 65.082 & \textbf{16.434 } & 84.449                           & 96.390                           & 63.778                           & 77.731  & 86.572 & \textbf{32.448 } \\
cup-22      & 18.583                          & 107.427                         & 88.692                           & 109.907 & 17.573                           & \textbf{9.227 }  & 103.229                         & 103.407                          & 89.561                           & 107.211 & 24.954                           & \textbf{13.710 } & 44.816                          & 104.835                          & 90.176                           & 106.561 & 48.635 & \textbf{14.220 } & 109.334                          & 103.585                          & 90.580                           & 104.788 & 69.075 & \textbf{20.870 } \\
cup-35      & 10.453                          & 89.978                          & 67.295                           & 120.922 & 19.711                           & \textbf{9.610 }  & 15.208                          & 108.913                          & 66.949                           & 112.636 & 21.614                           & \textbf{11.080 } & 17.309                          & 107.568                          & 67.185                           & 114.834 & 49.497 & \textbf{12.680 } & 21.296                           & 106.810                          & 69.896                           & 113.894 & 71.405 & \textbf{19.041 } \\
fandisk     & 111.246                         & 19.243                          & 32.898                           & 31.515  & 24.190                           & \textbf{15.095 } & 25.464                          & 21.070                           & 37.061                           & 63.276  & 24.943                           & \textbf{18.852 } & 23.148                          & 21.158                           & 37.920                           & 63.270  & 53.196 & \textbf{20.392 } & 31.770                           & 27.474                           & 40.835                           & 39.086  & 76.762 & \textbf{22.594 } \\
holes       & \textbf{5.428 } & \textbf{5.428 } & 39.000                           & 42.304  & 21.160                           & 20.879                           & \textbf{6.225 } & \textbf{6.225 }  & 42.841                           & 55.063  & 21.032                           & 9.730                            & \textbf{6.667 } & \textbf{6.667 }  & 44.039                           & 52.910  & 59.078 & 12.778                           & 10.283                           & \textbf{7.836 }  & 47.739                           & 52.236  & 83.886 & 24.310                           \\
horse       & 33.319                          & 47.994                          & 31.786                           & 46.661  & 24.270                           & \textbf{23.428 } & 36.567                          & 51.363                           & 37.052                           & 49.740  & 22.636                           & \textbf{17.222 } & 45.777                          & 50.092                           & 39.566                           & 44.637  & 66.504 & \textbf{21.637 } & 42.982                           & 51.202                           & 41.717                           & 53.306  & 88.838 & \textbf{39.494 } \\
kitten      & 16.164                          & \textbf{9.882 } & 36.495                           & 28.360  & 18.149                           & 12.834                           & 13.994                          & \textbf{11.001 } & 43.810                           & 25.569  & 19.388                           & 12.335                           & 13.970                          & \textbf{11.249 } & 44.775                           & 24.671  & 59.984 & 14.473                           & 14.873                           & \textbf{11.980 } & 45.921                           & 27.288  & 80.554 & 23.598                           \\
knot        & \textbf{4.634 } & 6.012                           & 71.819                           & 87.369  & 31.773                           & 5.291                            & 8.174                           & 6.838                            & 70.257                           & 116.168 & 29.990                           & \textbf{6.121 }  & 7.929                           & \textbf{7.773 }  & 69.993                           & 95.297  & 57.246 & 10.086                           & \textbf{10.336 } & 44.105                           & 71.110                           & 93.984  & 77.715 & 17.086                           \\
lion        & 39.775                          & 48.542                          & 40.831                           & 55.493  & 47.223                           & \textbf{32.986 } & 42.277                          & 52.103                           & 44.106                           & 55.818  & 44.638                           & \textbf{23.621 } & 46.637                          & 49.633                           & 45.442                           & 48.789  & 75.879 & \textbf{30.074 } & 59.319                           & 50.398                           & \textbf{49.112 } & 59.108  & 91.063 & 52.139                           \\
mobius      & 26.234                          & 122.025                         & 53.639                           & 120.832 & \textbf{21.448 } & 29.550                           & 27.995                          & 121.133                          & 55.871                           & 118.247 & \textbf{26.797 } & 28.346                           & 102.424                         & 120.963                          & 61.251                           & 117.930 & 60.265 & \textbf{49.081 } & 119.506                          & 120.518                          & 71.101                           & 117.068 & \textbf{73.098} & 88.530 \\
mug         & 12.380                          & 99.870                          & 78.598                           & 94.666  & 20.618                           & \textbf{9.458 }  & 22.274                          & 99.145                           & 77.615                           & 92.522  & 26.271                           & \textbf{11.399 } & 22.743                          & 97.616                           & 77.761                           & 93.732  & 47.326 & \textbf{12.412 } & 95.910                           & 96.303                           & 77.940                           & 95.133  & 65.508 & \textbf{17.144 } \\
octa-flower & 115.579                         & 63.804                          & \textbf{21.108 } & 31.533  & 27.405                           & 30.227                           & 115.992                         & 60.782                           & \textbf{27.439 } & 41.275  & 28.288                           & 42.195                           & 112.586                         & 105.038                          & \textbf{30.186 } & 41.486  & 59.050 & 88.270                           & 56.832                           & 62.286                           & \textbf{35.594 } & 40.633  & 80.510 & 39.684                           \\
sheet       & 121.675                         & 121.682                         & 69.920                           & 117.485 & 29.288                           & \textbf{26.978 } & 119.934                         & 119.905                          & 68.632                           & 110.042 & 19.913                           & \textbf{16.290 } & 23.038                          & 119.754                          & 69.260                           & 109.920 & 61.542 & \textbf{20.229 } & 119.582                          & 119.680                          & 73.599                           & 102.809 & 87.584 & \textbf{32.310 } \\
torus       & \textbf{3.239 } & \textbf{3.239 } & 30.858                           & 19.187  & 13.481                           & 12.493                           & \textbf{3.723 } & \textbf{3.723 }  & 35.484                           & 5.376   & 14.901                           & 7.455                            & \textbf{4.167 } & \textbf{4.167 }  & 36.483                           & 4.775   & 48.894 & 11.288                           & \textbf{5.510 }  & \textbf{5.510 }  & 38.420                           & 9.096   & 72.588 & 19.166                           \\
trim-star   & 24.069                          & 18.808                          & 48.281                           & 43.216  & 29.519                           & \textbf{17.774 } & 34.325                          & 20.377                           & 51.527                           & 35.829  & 31.829                           & \textbf{17.275 } & 25.209                          & 20.848                           & 52.233                           & 43.734  & 61.031 & \textbf{19.974 } & 26.540                           & \textbf{21.339 } & 53.525                           & 42.390  & 81.154 & 27.245                           \\
vase        & 39.500                          & 48.036                          & 66.198                           & 59.083  & 21.234                           & \textbf{13.957 } & 38.677                          & 52.584                           & 66.170                           & 80.204  & 23.450                           & \textbf{14.656 } & 41.840                          & 53.557                           & 66.856                           & 69.580  & 62.990 & \textbf{18.694 } & 61.422                           & 57.240                           & 68.643                           & 52.113  & 86.423 & \textbf{29.463} \\
\bottomrule
\end{tabular}
}
\end{table*}

\begin{table*}[!htp]
\centering
\caption{Comparison of normal orientation with three normal estimation methods.}
\vspace{-3mm}
\label{tab:normal_est}
\resizebox{0.80\linewidth}{!}{
\begin{tabular}{l|cccc|cccc|cccc|cccc}
\toprule
Sampling                 & \multicolumn{4}{c|}{Blue Noise Sampling}                  & \multicolumn{4}{c|}{White Noise Sampling}                      & \multicolumn{4}{c|}{0.25\% noise White Noise Sampling}                  & \multicolumn{4}{c}{0.5\% noise White Noise Sampling }                  \\ \cmidrule{1-17}
 Models                    & PCA     & AdaFit  & NeAF   & Ours  & PCA     & AdaFit  & NeAF   & Ours & PCA     & AdaFit  & NeAF   & Ours & PCA     & AdaFit  & NeAF   & Ours  \\ \midrule
 82-block        & 75.950 & 50.150 & 51.800 & \textbf{100.000} & 75.675 & 50.675 & 51.225 & \textbf{99.980 } & 75.625 & 51.225 & 52.250 & \textbf{99.930 } & 75.825 & 52.050 & 53.650 & \textbf{99.880 } \\
 bunny           & 89.750 & 51.375 & 50.375 & \textbf{100.000} & 89.800 & 51.375 & 50.125 & \textbf{99.750 } & 89.725 & 50.475 & 50.300 & \textbf{99.980 } & 89.775 & 50.850 & 50.800 & \textbf{99.580 } \\
 chair           & 62.775 & 53.125 & 50.625 & \textbf{100.000} & 64.250 & 52.750 & 50.100 & \textbf{100.000} & 63.575 & 52.825 & 54.000 & \textbf{100.000} & 64.825 & 52.125 & 51.050 & \textbf{99.400 } \\
 cup-22          & 58.425 & 52.225 & 52.000 & \textbf{100.000} & 58.175 & 50.075 & 51.550 & \textbf{99.950 } & 58.550 & 50.150 & 51.325 & \textbf{99.950 } & 58.450 & 50.550 & 50.175 & \textbf{99.850 } \\
 cup-35          & 66.775 & 50.775 & 51.550 & \textbf{100.000} & 67.725 & 50.525 & 50.350 & \textbf{100.000} & 68.025 & 50.200 & 52.250 & \textbf{100.000} & 67.600 & 51.450 & 53.650 & \textbf{100.000} \\
 fandisk         & 90.200 & 54.950 & 50.700 & \textbf{100.000} & 90.000 & 55.075 & 50.975 & \textbf{100.000} & 90.050 & 54.675 & 50.675 & \textbf{99.950 } & 89.975 & 54.775 & 51.650 & \textbf{99.750 } \\
 holes           & 79.275 & 50.575 & 51.325 & \textbf{100.000} & 79.350 & 51.200 & 51.175 & \textbf{100.000} & 79.650 & 51.575 & 50.675 & \textbf{100.000} & 79.075 & 50.900 & 51.650 & \textbf{100.000} \\
 horse           & 81.875 & 50.200 & 50.600 & \textbf{99.500 } & 80.275 & 51.050 & 50.500 & \textbf{99.800 } & 80.800 & 50.350 & 51.800 & \textbf{99.750 } & 80.625 & 51.350 & 51.025 & \textbf{97.500 } \\
 kitten          & 90.977 & 55.011 & 51.937 & \textbf{100.000} & 91.075 & 57.325 & 51.750 & \textbf{99.980 } & 90.975 & 57.625 & 53.875 & \textbf{100.000} & 90.950 & 57.100 & 51.250 & \textbf{99.980 } \\
 knot            & 72.975 & 50.100 & 51.050 & \textbf{100.000} & 72.750 & 50.825 & 50.250 & \textbf{100.000} & 72.875 & 50.925 & 52.100 & \textbf{100.000} & 72.275 & 50.775 & 51.850 & \textbf{99.980 } \\
 lion            & 85.275 & 53.025 & 52.125 & \textbf{99.380 } & 84.900 & 54.575 & 50.275 & \textbf{99.700 } & 85.200 & 54.750 & 52.625 & \textbf{99.550 } & 85.475 & 55.775 & 51.850 & \textbf{93.830 } \\
 mobius          & 55.225 & 53.700 & 53.050 & \textbf{100.000} & 55.425 & 54.575 & 54.625 & \textbf{100.000} & 55.500 & 53.800 & 52.500 & \textbf{97.380 } & 54.875 & 53.825 & 52.650 & \textbf{85.780 } \\
 mug             & 64.775 & 50.575 & 53.525 & \textbf{100.000} & 67.250 & 50.650 & 50.875 & \textbf{100.000} & 67.225 & 50.825 & 52.125 & \textbf{100.000} & 66.825 & 51.000 & 52.950 & \textbf{100.000} \\
 octa-flower     & 98.800 & 50.275 & 51.750 & \textbf{100.000} & 94.750 & 51.550 & 53.050 & \textbf{99.330 } & 94.650 & 51.525 & 50.900 & \textbf{98.800 } & 94.775 & 51.700 & 51.800 & \textbf{98.550 } \\
 sheet           & 83.600 & 54.425 & 56.400 & \textbf{100.000} & 71.575 & 51.925 & 51.625 & \textbf{100.000} & 71.750 & 51.550 & 51.775 & \textbf{99.950 } & 71.325 & 51.775 & 52.000 & \textbf{98.980 } \\
 torus           & 89.225 & 50.125 & 53.800 & \textbf{100.000} & 89.950 & 50.500 & 52.100 & \textbf{100.000} & 89.950 & 50.350 & 51.000 & \textbf{100.000} & 90.125 & 50.225 & 50.400 & \textbf{100.000 } \\
 trimstar        & 80.975 & 50.300 & 52.625 & \textbf{100.000} & 81.150 & 50.325 & 53.475 & \textbf{100.000} & 81.125 & 51.250 & 50.750 & \textbf{100.000} & 81.250 & 51.375 & 51.175 & \textbf{100.000 } \\
 vase            & 84.575 & 53.400 & 50.375 & \textbf{100.000} & 86.350 & 51.825 & 50.250 & \textbf{100.000} & 86.725 & 52.050 & 51.475 & \textbf{100.000} & 86.275 & 52.500 & 50.450 & \textbf{99.650 } \\ \bottomrule
 
\end{tabular}
}
\end{table*}

\paragraph{Angle RMSE}
In the main paper, 
we give the statistics about the ratio of true normals. 
Here we further give the statistics about the Root Mean Square Error (RMSE) of the angles between the estimated normals and the ground truth normals.% as a new indicator for further evaluations.
% To further evaluating the re-oriented normals generated using our method, we use another metric, the Root Mean Square Error (RMSE) of angles between the estimated normal and the ground truth normal, as an additional indicator.

%\paragraph{Noise with angle RMSE}
\paragraph{Normal Evaluation Using Angle RMSE}
In Sec.~5.4, we give a visual comparison to exhibit the noise-resistant ability of our approach compared with the SOTA methods~\cite{dipole_propagation,hoppe1992surface,konig2009consistent,guerrero2018pcpnet,PGR2022Siyou}. 
We report the angle RMSE statistics under four different sampling conditions in Table~\ref{tab:normal_rmse}. 
It can be clearly seen that our method surpasses the other methods 
in terms of normal orientations.
% regarding normal orientations.
% Here we give a statistics in Table~\ref{tab:normal_rmse} using angle RMSE for four types of noise sampling conditions. It it clearly to see that our method surpass other methods regarding normal orientations.
Even if the noise level amounts to $0.5\%$,
our method can still get the best score for $55\%$ of the models,
and a competitive score for another $40\%$. 
For example, 
the best score ($10.336$) is given by Hoppe 
on the Knot model, and ours is the second best ($17.086$),
which is much better than the remaining scores $44.105$, $71.110$ and $93.984$. 

There are some methods such as PCA~\cite{Rusu_ICRA2011_PCL}, AdaFit~\cite{zhu2021adafit} and NeAF~\cite{li2022neaf} 
that focus on normal estimation.
We also include them for comparison; 
See the statistics in Table~\ref{tab:normal_est}. 
% Even though our method is designed for normal orientation, there are some other works focusing on the normal estimation, which could be considered for orientating normals. We compare with representative methods PCA~\cite{Rusu_ICRA2011_PCL}, AdaFit~\cite{zhu2021adafit} and NeAF~\cite{li2022neaf} and report the statistics in \ref{tab:normal_est}. 
The statistics show that our algorithm has a big advantage of prediction accuracy over the SOTA methods, on all the $18$ models and under all the $4$ noise sampling conditions.

\begin{figure}[!h]
%\vspace{5px}
\centering
\begin{overpic}
[width=.99\linewidth]{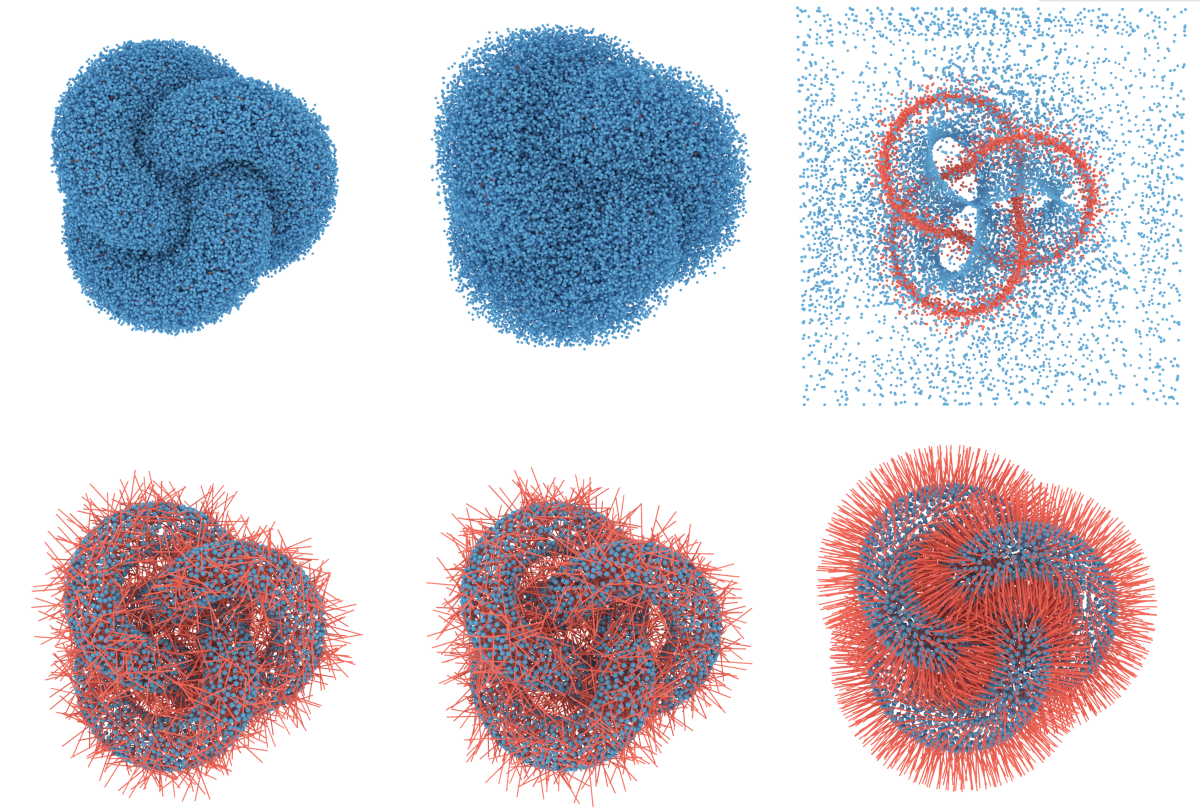}
\put(3, -3){Random $r=0.1$}
\put(37, -3){Random $r=0.2$}
\put(70, -3){Voronoi vertices}
\end{overpic}
\caption{
In order to test different sampling strategies,
we generate Gaussian noise near the point cloud (radius $r=0.1$ or $0.2$)
to produce examination points.
First column: $r=0.1$ (top: examination points; bottom: normals).
Second column: $r=0.2$ (top: examination points; bottom: normals).
Last column: Voronoi vertices as examination points. 
It can be seen that Voronoi vertices are more suitable for serving as the examination points.
Note that we iterate $200$ steps for $r=0.1,0.2$ and $40$ steps for 
the situation of Voronoi vertices as examination points. 
% Voronoi strategy.
% the strategy of taking Voronoi vertices as examination points, 
% We compare results of two different sampling strategies using either random vertices (radius $r=0.1$ or $0.2$) and Voronoi vertices. We iterate $200$ steps for random strategy and $40$ steps for Voronoi strategy. Top: Query points with winding number $~0$ in blue and $~1$ in red. Bottom: Optimization result with normals indicated in red.
% \XR{TODO.} 
% \NW{Updated}
% We experiment with the cup model.
}
\label{fig:ablation_voronoi}
\end{figure}

\section{Strategies for Generating Examination Points}
We conduct the ablation study about different strategies for generating examination points. Specially, we compare our Voronoi-based sampling strategy with the off-surface random sampling strategy.
As shown in Fig.~\ref{fig:ablation_voronoi}, 
% we use two sampling radius $r=0.1$ and $0.2$ for near surface sampling strategy, 
we randomly sample points around the surface of the shape with two sampling radii $r=0.1$ and $r=0.2$. 
%\XG{How do you define $f_A$ and $f_B$ in this case? Since we are not using Voronoi cells here, how is the balance term defined and are you still using alignment term?}
Note that we iterate $200$ steps for $r=0.1,0.2$ and $40$ steps for 
the situation of Voronoi vertices as examination points. 
It can be seen that Voronoi vertices are more suitable for serving as the examination points. 
The superiority of Voronoi-based sampling is due to the fact that the majority of Voronoi vertices are located either deepest inside the surface, or furthest outside the surface (approximating the inner and outer medial axis~\cite{amenta2001power}). Thus their distribution of winding numbers is more likely to be pushed towards 0 and 1, compared with the random sampling strategy.

\begin{figure}[h]
    \centering
    \begin{overpic}[width=0.99\linewidth]{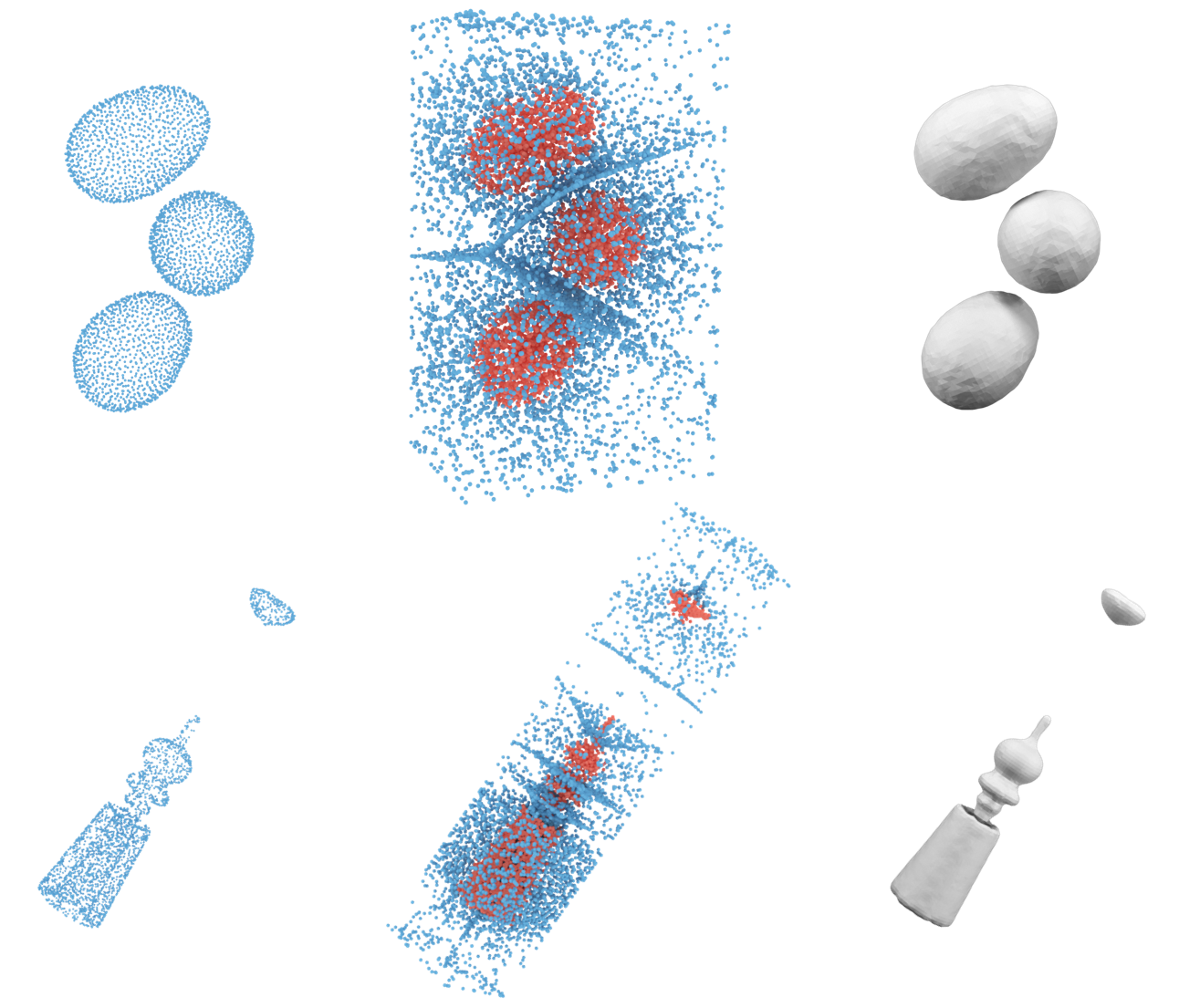}
    \put(0, -3){\textbf{(a)~Input points}}
    \put(30, -3){\textbf{(b)~Voronoi vertices}}
    \put(68, -3){\textbf{(c)~Reconstruction}}
    \end{overpic}
    \caption{Our method can handle multiple disconnected components and outliers. 
    In the middle column, the interior Voronoi vertices (whose winding number is close to 1) are colored in red
    while 
 the exterior Voronoi vertices (whose winding number is close to 0) are colored in blue. 
    %\XR{Outlier...}
    }
    \label{fig:outlier}
\end{figure}

\section{Disconnected Components and Outliers}
% \paragraph{Disconnected components}
% \paragraph{Outliers}
In this section, we show that our method can also handle multiple disconnected components and outliers. 
In the top row of Fig.~\ref{fig:outlier}, 
our Voronoi-based sampling method 
can still distinguish the interior Voronoi vertices (whose winding number is close to 1)
and the exterior Voronoi vertices (whose winding number is close to 0).
At the same time, in the bottom row of Fig.~\ref{fig:outlier}, 
we show an outlier example where the cap is completely away from the main body. 
It can be clearly seen that our approach can deal with outliers as well. 
On one hand, the Voronoi diagram can capture the proximity between data points,
thus encouraging the outlier points to be oriented independently of the main body. 
Additionally, 
the winding number field helps infer normal consistency from a global perspective,
thus unlikely to suffer from small imperfections.

% Even in this case, we show the strong capability of regularizing winding-number field, which provide us promising results for these outliers models.

% where three separate eggs form a single model, our Voronoi-based sampling method is able to sample both interior points and exterior points according to the global shape. 

% This sampling strategy robustly enhances our ability to orient normals and thus reconstruct the shape in a good manner. 

% The bottom of Fig~\ref{fig:outlier} shows an example of a outlier model where the cap is completely disconnected from the the whole shape. Even in this case, we show the strong capability of regularizing winding-number field, which provide us promising results for these outliers models.

\vfill

\section{Wind-Number Field under Different Conditions}

\begin{figure}[h]
    \centering
    \begin{overpic}[width=0.99\linewidth]{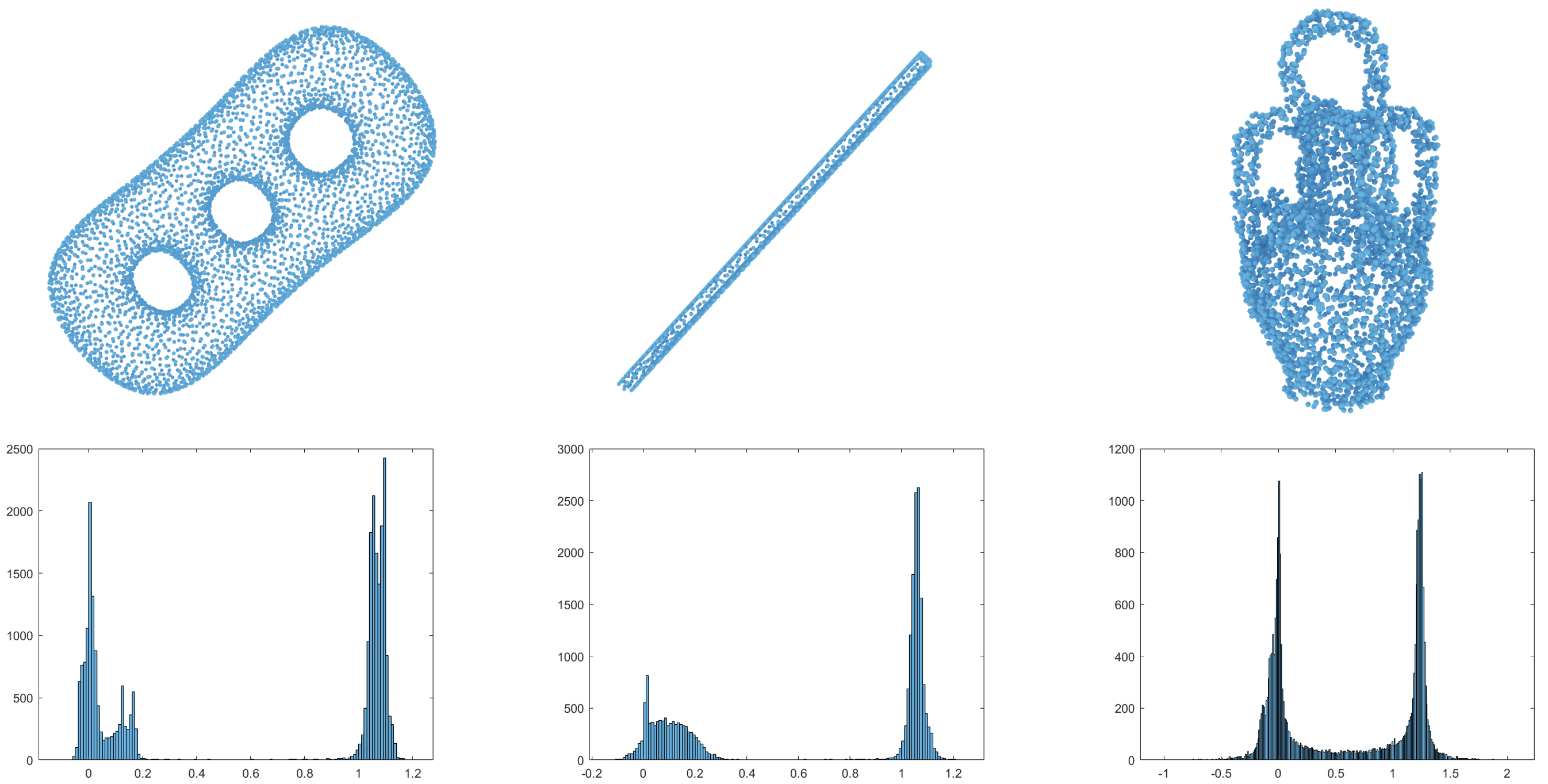}
    \put(8, -3){\textbf{(a)~Holes}}
    \put(38, -3){\textbf{(b)~Thin sheet}}
    \put(74, -3){\textbf{(c)~1.0\% Noise}}
    \end{overpic}
    \caption{\XR{The winding-number field distributions on three totally different shapes.} 
    }
    \label{fig:Distribution1}
\end{figure}

\begin{figure}[h]
    \centering
    \begin{overpic}[width=0.99\linewidth]{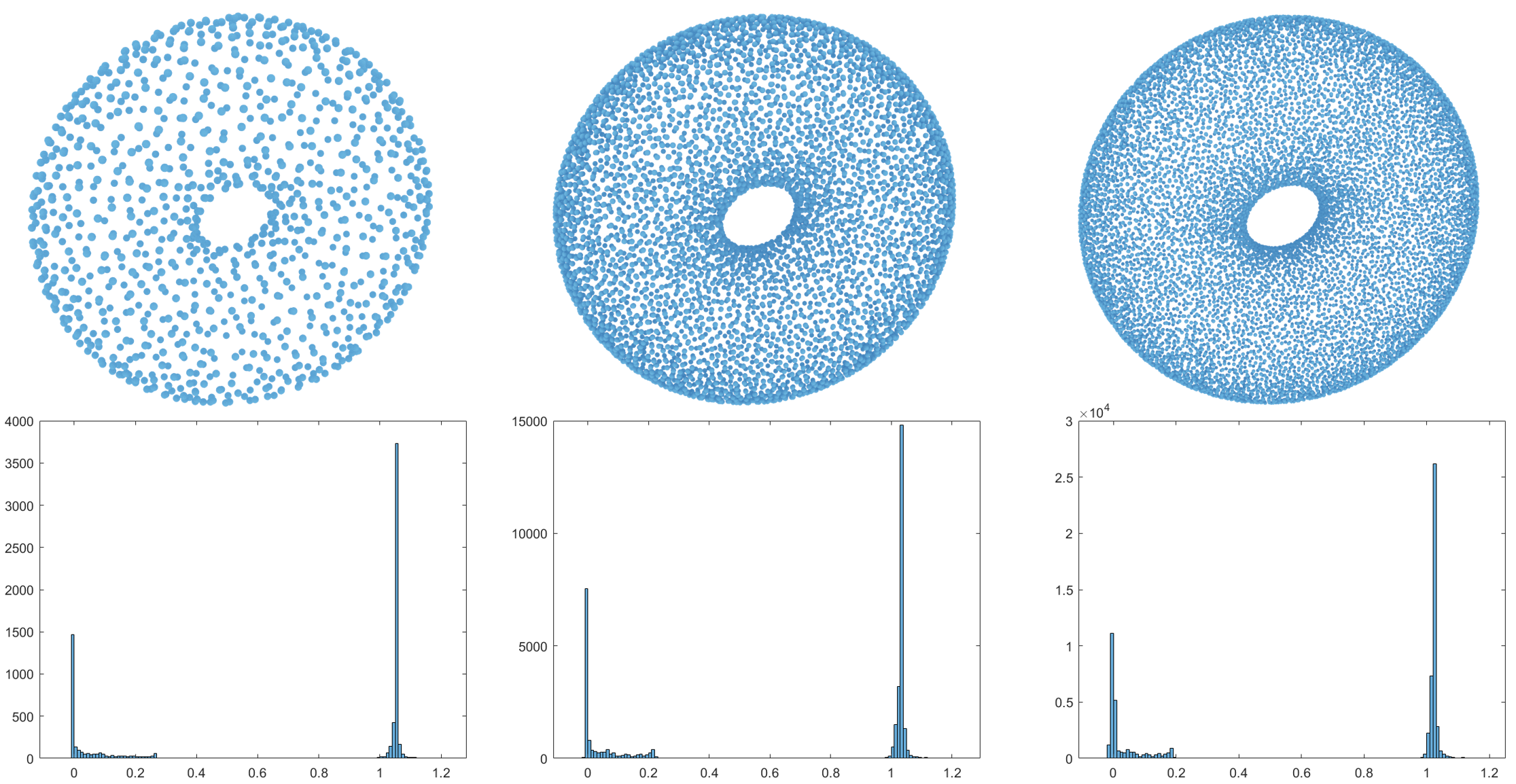}
    \put(6, -3){\textbf{(a)~1K points}}
    \put(39, -3){\textbf{(b)~5K points}}
    \put(74, -3){\textbf{(c)~10K points}}
    \end{overpic}
    \caption{\XR{The winding-number field distributions under different sampling densities.} 
    }
    \label{fig:Distribution2}
\end{figure}

\ZY{We show more winding-number fields in Fig.~\ref{fig:Distribution1}. Despite the varying topologies, all the winding-number fields are approximately binary-valued at $1$ and $0$. Moreover, we visualize how the winding-number field distribution changes with respect to the sampling density in Fig.~\ref{fig:Distribution2}.
It can be seen that the winding-number field remains binary-valued with approximate values of $1$ and $0$ as the number of points increases from $1$K to $10$K.
 }

% \XR{Here we show more distributions of the Winding number field. In Fig.~\ref{fig:Distribution1}, we show the Winding-Number Field distributions with different properties, i.e. holes, thin sheets, and noise. Although their distributions are different, they are all distributed around 0 and 1.
% Also in Fig.~\ref{fig:Distribution2}, we show how the Winding-Number Field distribution changes as the sampling density increases from 1K points to 10K points. The distribution of the Winding-Number Field is almost the same.
% }


\section{Ablation Study}
\label{sec:ablation}

% Our optimization framework is insensitive to the parameter settings in Eq~\ref{eq:all}. We conduct experiments to investigate the performance of our method with different optimization parameter settings to reveal our robustness. In this section, we report the normal orientation as well as reconstruction results with different choices of the following parameters: $\lambda_A$ and $\lambda_B$. Moreover, we provide a detailed analysis of the legitimacy of each term in Eq.~\ref{eq:all} and the choice of query points. 
%\NW{missing experiment?}. 
% Then we provide detailed analyses on each term and the $D$.

\begin{figure}[!h]
% \vspace{-5px}
\centering
\begin{overpic}
[width=\linewidth]{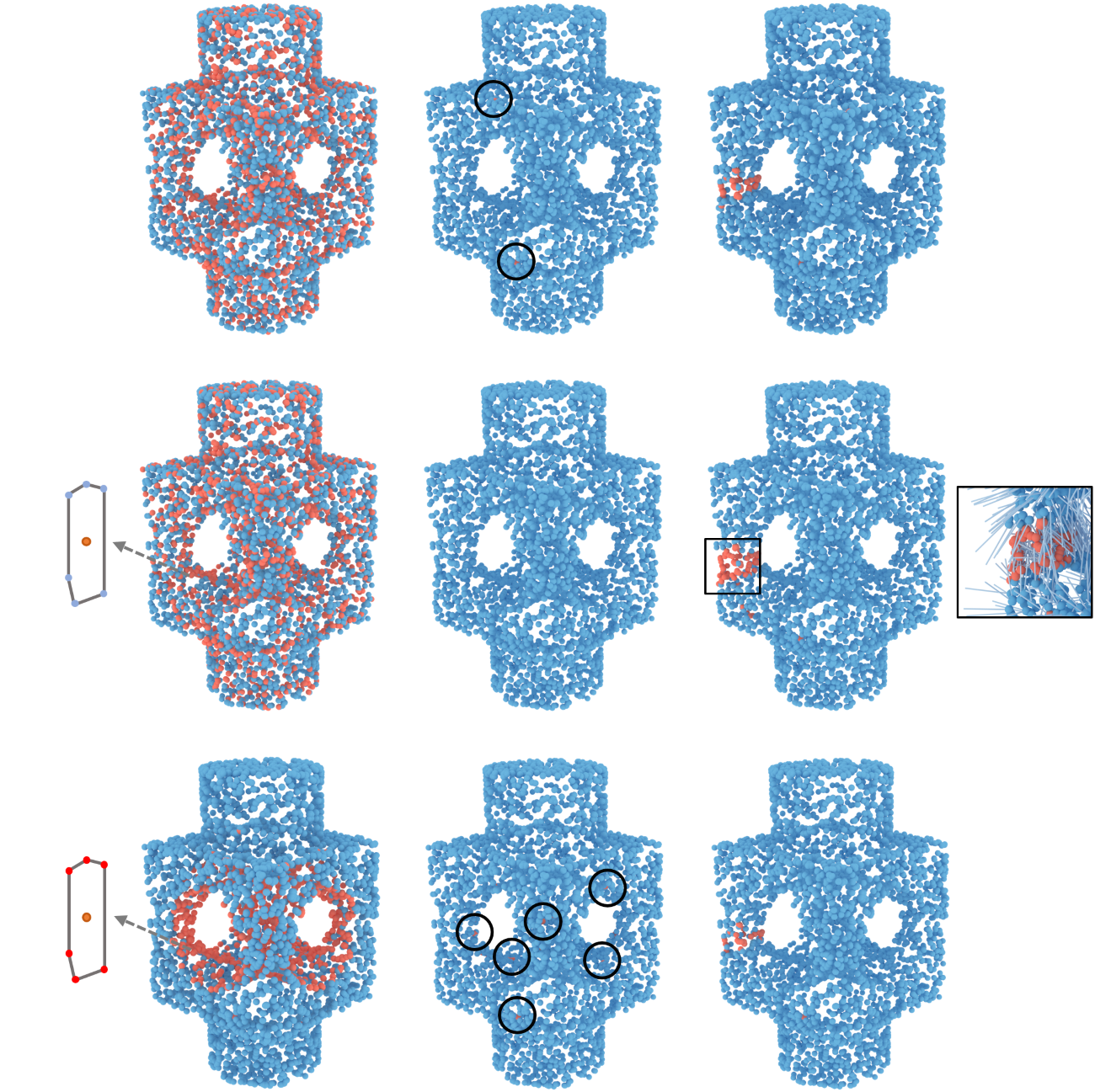}
% \put(-5,60){\rotatebox{90}{$\lambda_V=50$}}
% \put(-5,12){\rotatebox{90}{$\lambda_N=10$}}
\put(13, 66){$\lambda_A=1,\lambda_B=5$}
\put(39,66){$\lambda_A=1,\lambda_B=50$}
\put(64, 66){$\lambda_A=1,\lambda_B=500$}
\put(13, 31.5){$\lambda_A=10,\lambda_B=5$}
\put(38,31.5){$\lambda_A=10,\lambda_B=50$}
\put(64, 31.5){$\lambda_A=10,\lambda_B=500$}
\put(11, -3){$\lambda_A=100,\lambda_B=5$}
\put(37, -3){$\lambda_A=100,\lambda_B=50$}
\put(64, -3){$\lambda_A=100,\lambda_B=500$}
\put(2, 7){$w	\approx 1.0$}
\put(2, 42){$w	\approx 0.0$}
\end{overpic}
\caption{
Ablation study about the weighting coefficients $\lambda_B$ and $\lambda_A$.
We select
$\lambda_A=10,\lambda_B=50$ as the favorite combinations,
which are used in all the experiments in this paper. 
% . We test with model 82-block. The red point indicates the angle error between the oriented normal and the ground truth normal is larger than $90$ degree.
If $\lambda_B$ is too small (see the left column),
the winding numbers at the vertices of
a Voronoi cell may all stay at 0 or 1.
The predicted orientation is true if the angle between the computed normal and the ground-truth normal is less than 90 degrees. 
We colored the true predictions and false predictions in blue and red, respectively.
}
\label{fig:Parameters}
\vspace{-5pt}
\end{figure}

\paragraph{$\lambda_A$ and $\lambda_B$}
We first investigate the influence of the coefficients $\lambda_A$ and $\lambda_B$ in Eq. (2).
We test different combinations of $\lambda_A$ and $\lambda_B$ 
in Fig.~\ref{fig:Parameters}. 
The quantitative statistics  are summarized in Table~\ref{tab:parameter}. 
We have two observations:
\begin{enumerate}
    \item If $\lambda_B$ is too small,     
    the winding numbers at 
    the vertices of a Voronoi cell may not be balanced,
    all staying at~0 or 1. 
    % there are many false normal predictions
    % due to the occurrence that all the corners of a Voronoi cell 
    % are deemed inside or outside.     
    But if $\lambda_B$ is too large, 
    our algorithm may report a reverse orientation for a small point patch.
    \item If $\lambda_A$ is too small, 
    the predicted normal orientations are not accurate (see Table~\ref{tab:parameter}). 
    Instead, if $\lambda_A$ is too large,
    it may prevent the orientations from evolving to a favorite state.  %\SQ{evidence?}
\end{enumerate}
As $\lambda_A=10,\lambda_B=50$ 
gets the best scores in Table~\ref{tab:parameter}, we select
$\lambda_A=10,\lambda_B=50$ as the favorite combination,
which is used in all the experiments in this paper.

% where we make two observations: 1) $\lambda_B$ for the variance term influences more in the normal orientation result; 2) $\lambda_A$ for the normal constraint term dominates the accuracy of the normal estimation, where the estimation results rely on the output of orientation process.
% In this paper, we conduct all experiments with consistent parameters by setting $\lambda_A = 10, \lambda_B = 50$. The corresponding visual results are provided in Fig.~\ref{fig:Parameters}. 
% From Fig.~\ref{fig:Parameters}, 

\begin{figure}[!h]
% \vspace{-5px}
\centering
\begin{overpic}
[width=\linewidth]{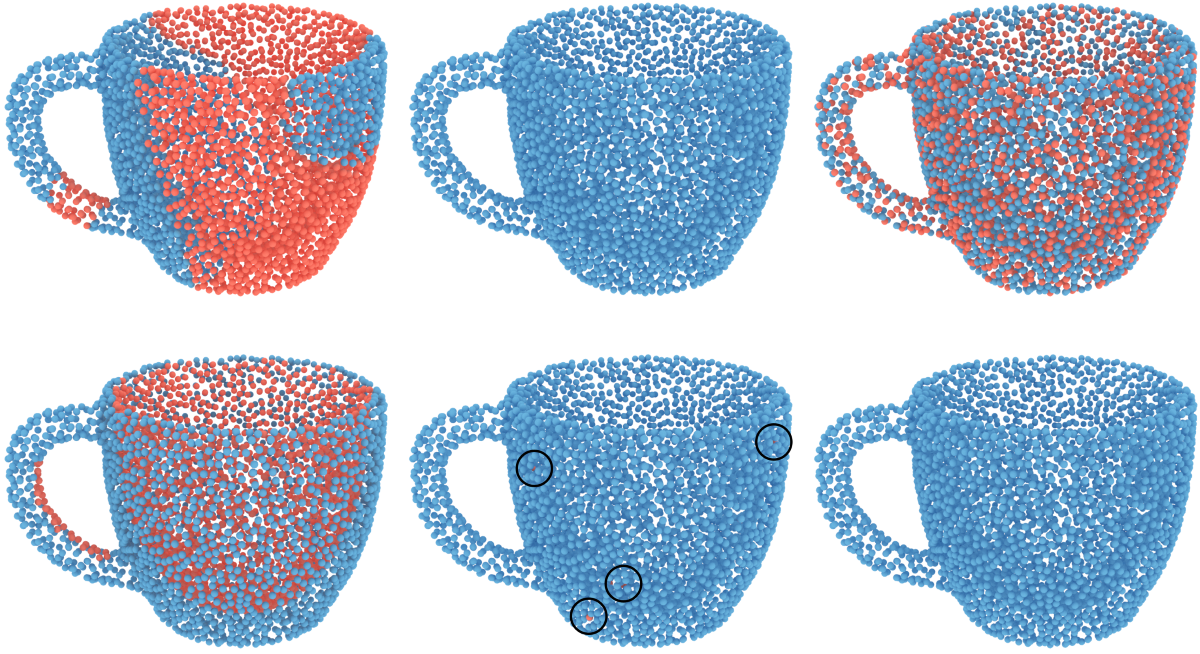}
\put(13, 26){w/o $f_{01}$}
\put(47, 26){w/o $f_A$}
\put(82, 26){w/o $f_B$}
\put(3, -3){$f_{01}=\sum_j^M (w_j-1)^2$}
\put(45, -3){w/o $w_j/D$}
\put(85, -3){All}
\end{overpic}
\caption{
Ablation study about the constituent terms in Eq. (2). 
Note that the bottom left figure 
shows the result when we replace the double well function with a single well function~$y=(x-1)^2$.
% The red point indicates the angle error between the oriented normal and the ground truth normal is larger than $90$ degree.
% We experiment with the cup model.
}
\label{fig:ablation}
\end{figure}
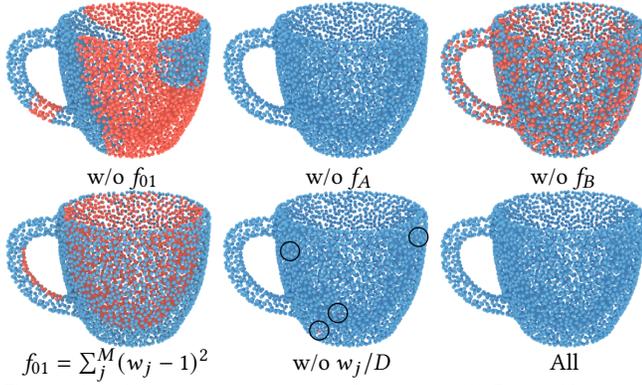

\paragraph{Optimization Terms}
We conduct the ablation study about the constituent terms in Eq. (2). 
From the top row of Fig.~\ref{fig:ablation}, we have the following observations: 
\begin{enumerate}
    \item The term~$f_{01}$ enforces the winding number to be valued at 0 or 1. Without $f_{01}$,
    the global normal consistency cannot be guaranteed. 
    Two points on the opposite sides of the thin wall may be different from the ground-truth orientations.    
    % it can be seen that the red points on the opposite sides of the thin wall are oriented toward almost the same direction (the winding number is equal to or larger than 2). 
    \item The term $f_A$ is to enforce the normals to align with Voronoi poles. Without $f_A$, the orientations remain nearly unchanged but the accuracy is decreased; See the statistics in Table~\ref{tab:Ablation}.
    \item The term $f_B$ is to eliminate the occurrence that all the vertices of a Voronoi cell 
    are inside or outside. 
    Without $f_B$, the normals 
    tend to stay at the initial random state; the winding number is 0 almost everywhere.
    % there are many false normal predictions
    % (the winding-number values are not balanced for $\sample_i$'s Voronoi vertices). 
\end{enumerate}

\begin{wrapfigure}{r}{2cm}
\vspace{-3.5mm}
  \hspace*{-5mm}
  \centerline{
  \includegraphics[width=23mm]{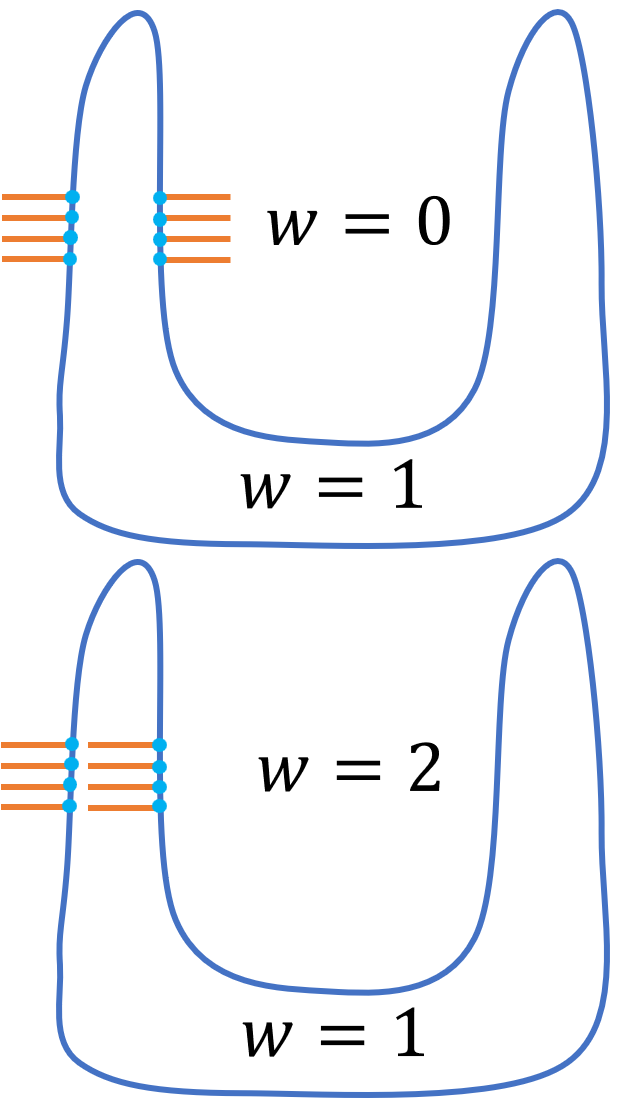}
  }
  \vspace*{-4mm}
\end{wrapfigure}
Furthermore, the double well function
is very helpful for regularizing the winding number. 
If we replace the double well function with a single well function~$y=(x-1)^2$ (see the left bottom result of Fig.~\ref{fig:ablation}), 
it will confuse 0 and 2 in inferring the winding-number values, as the effects of 0 and 2 are exactly the same; See the inset figure.
% the winding number may amount to 2.0 or a larger value 
% particularly when the thin-walled structures exist. 
% \XR{Due to the same inflence of 0.0 and 2.0. Also if we use $y=(x-0.5)^2$, which will give the same inflence to 0.0 and 1.0, but it will lead the winding numbers to 0.5. }
Besides, the shear correction term $\frac{w_j}{D}$ 
is also helpful for 
preventing the normal setting from staying in the initial random state
and pushing the winding number of some examination points 
to approach~1.

\begin{table}[!h]
\vspace{-5pt}
\caption{Quantitative results of normal orientation with different weighting schemes. Indicators are described in Sec. 5.2.}
\vspace{-3.5mm}
\label{tab:parameter}
\resizebox{0.8\linewidth}{!}{
\begin{tabular}{l|ccc}
\toprule
Parameters                      & RMSE $\downarrow$                              & $Ratio_\text{truth}$ $\uparrow$                           & $CD_{recon}$ $\downarrow$                    \\ \midrule
$\lambda_A=1,\lambda_B=5$   & 95.797                           & 63.600                           & 1.899                          \\
$\lambda_A=1,\lambda_B=50$   & 21.117                           & 99.875                           & 0.141                           \\ 
$\lambda_A=1,\lambda_B=500$   & 26.039                            & 99.050                           & 0.177                           \\ 
$\lambda_A=10,\lambda_B=5$   & 95.501                           & 63.925                           & 0.187                           \\
$\lambda_A=10,\lambda_B=50$  & \textbf{19.901}                  & \textbf{99.975}                  & \textbf{0.139}                  \\  %\cdashline{1-4}[0.5pt/3pt]
$\lambda_A=10,\lambda_B=500$ & 29.979                           & 98.225                           & 0.198                           \\ %\cdashline{1-4}[0.5pt/3pt]
$\lambda_A=100,\lambda_B=5$ & 73.566                           & 80.600                            & 0.749                           \\ %\cdashline{1-4}[0.5pt/3pt]
$\lambda_A=100,\lambda_B=50$ & 23.484                           & 99.450                            & 0.142                           \\ %\cdashline{1-4}[0.5pt/3pt]
$\lambda_A=100,\lambda_B=500$ & 24.897                           & 99.275                           & 0.172                          \\ \bottomrule
\end{tabular}
}
\vspace{-10pt}
\end{table}

Besides, 
we conduct an ablation study
about different strategies for generating examination points. 
By comparing our Voronoi-based sampling strategy 
with the off-surface random sampling strategy,
we validate the superiority of Voronoi-based sampling.
The majority of Voronoi vertices are located either deepest inside the surface or furthest outside the surface (the same reason that \textit{power crust}~\cite{amenta2001power} used them as candidates for the medial axis). Thus their distribution of winding numbers is more likely to be pushed towards 0 and 1, compared with the random sampling strategy.
%which is due to the fact that Voronoi vertices are robust to small variations of the original point cloud, especially for those Voronoi vertices distant to the surface.
% Details of the comparison results are available in the supplemental material. 
% Here we exhibit the superior of our Voronoi-based sampling strategy over random strategy. As shown in Fig.~\ref{fig:ablation_voronoi}, we sample random vertices around the surface of the shape 

\begin{table}[h]
\caption{Quantitative results of normal orientation using different terms in Eq. (2). Indicators are described in Sec. 5.2. }
\vspace{-10pt}
\label{tab:Ablation}
\resizebox{0.8\linewidth}{!}{
\begin{tabular}{l|ccc}
\toprule
Terms                     & RMSE $\downarrow$                              & $Ratio_\text{truth}$ $\uparrow$                           & $CD_{recon}$ $\downarrow$                    \\ \midrule
w/o $f_{01}$   & 106.162 & 54.700  & 1.753     \\
w/o $f_A$ & 13.810 & \textbf{100.000}   & 0.069    \\
w/o $f_B$   & 98.908 & 57.600  & 2.160     \\
$f_{01}=\sum_j^M (w_j-1)^2$ & 110.814 & 59.150 & 1.920     \\ 
w/o $w_j/D$  & 17.556 & 99.850 & 0.094    \\
All  & \textbf{10.632} & \textbf{100.000}   & \textbf{0.067}   \\ \bottomrule
\end{tabular}
}
\vspace{-10pt}
\end{table}

% \section{Different Initialization}
\section{The influence of Initialization Strategies}

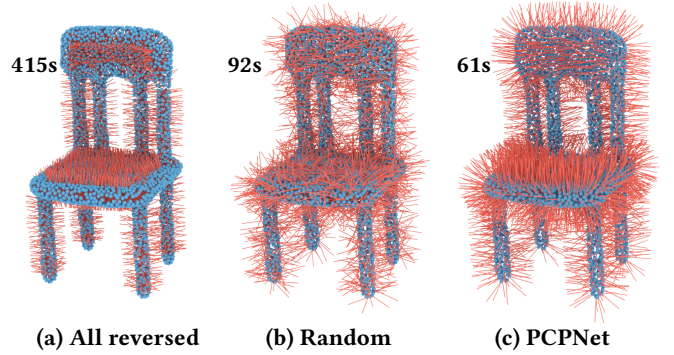
\begin{figure}[h]
    \centering
    \begin{overpic}[width=0.99\linewidth]{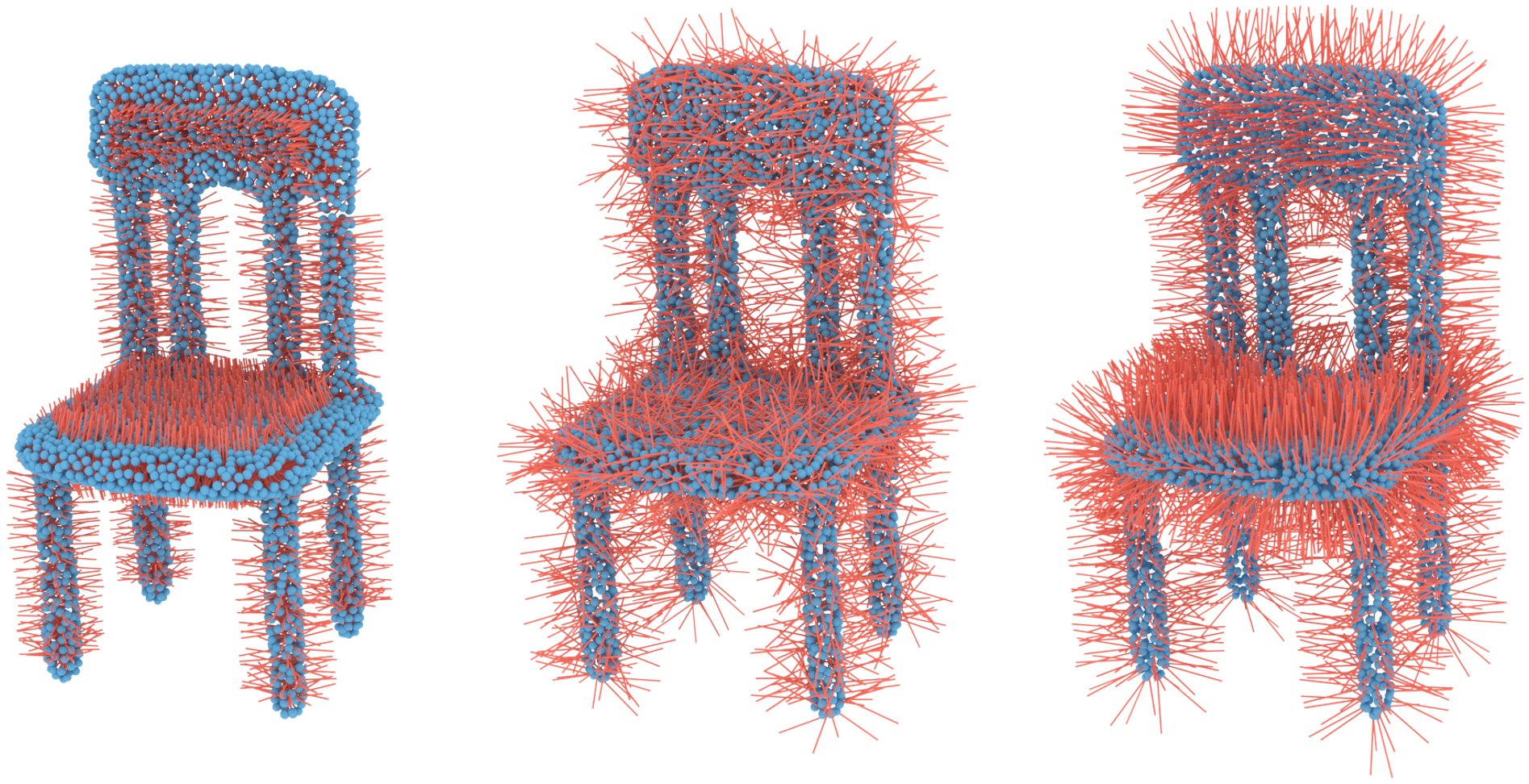}
    \put(2, -3){\textbf{(a)~All reversed}}
    \put(38, -3){\textbf{(b)~Random}}
    \put(74, -3){\textbf{(c)~PCPNet}}
    \put(-2, 40){\textbf{415s}}
    \put(32, 40){\textbf{92s}}
    \put(68, 40){\textbf{61s}}
    \end{overpic}
    % \caption{\XR{Different Initialization, the time is 415s, 92s and 61s.} 
    \caption{\ZY{
    Normal orientation results using different initialization strategies: (a) all reversed normal initialization, (b) random initialization, and (c) initialization by PCPNet~\cite{guerrero2018pcpnet}. Their timings are $415$s, $92$s, and $61$s, respectively.} }
    \label{fig:init}
\end{figure}

\ZY{As shown in Figure~\ref{fig:init}, our method exhibits high robustness across different initialization strategies, as demonstrated through three strategies: (1) reversed normal initialization, (2) random initialization, and (3) initialization using PCPNet~\cite{guerrero2018pcpnet}. Our method achieves more accurate normal orientation results for any of the initialization strategies.
An interesting observation is that
our method requires much less computational cost 
if initialized by PCPNet~\cite{guerrero2018pcpnet}. Note that the stop criteria for the three strategies are the same, i.e., when the difference of the objective value between two successive iterations are small enough. 
}

% \XR{We have tested the following initialization strategies: (1) random initialization; (2) completely reversed normal; and (3) PCPNet. See Figure~\ref{fig:init}. We found that a good initialization can potentially improve the run-time performance. But we stress that our method can still get faithful results even with a poor initialization. }
% @zhiyang 这里说明一下我们是统计让给他们收敛到同一个函数值的时间对比。
% copy that.

\vfill
\newpage

\section{Gradient Function}
\ZY{Herein, we provide the gradient of our objective function.} 
% \ZY{@Rui add ref here.} 
Note we parameterize each normal vector $\normal_i$ with $(u_i, v_i)$. 
\begin{equation}
\normal_i =  \left(\sin(u_i)\cos(v_i),\sin(u_i)\sin(v_i),\cos(u_i)\right)
\label{eq:ni}
\end{equation}

\ZY{We start by deriving the gradient of winding-number field $w(q)$ w.r.t. $u_i$ and $v_i$. Let $L=(p_i-q)$, then we have:}
%首先是winding num 公式的梯度

\begin{flalign}
\frac{\partial w(q)}{\partial v_i} &= \frac{a_iL\cdot n_i}{4\pi \left \| L \right \|^3}(-L_xsin(u_i)sin(v_i)+L_ysin(u_i)cos(v_i))&&\\\nonumber
\frac{\partial w(q)}{\partial u_i} &= \frac{a_iL\cdot n_i}{4\pi \left \| L \right \|^3}(L_xcos(u_i)cos(v_i)+L_ycos(u_i)sin(v_i)-L_zsin(u_i))&&
\end{flalign}

% \begin{equation}
% \frac{\partial w(q)}{\partial u_i}= \frac{a_iL\cdot n_i}{4\pi \left \| L \right \|^3}(L_xcos(u_i)cos(v_i)+L_ycos(u_i)sin(v_i)-L_zsin(u_i))

% \end{equation}
% \begin{equation}
% \frac{\partial w(q)}{\partial v_i}= \frac{a_iL\cdot n_i}{4\pi \left \| L \right \|^3}(-L_xsin(u_i)sin(v_i)+L_ysin(u_i)cos(v_i))
% \end{equation}

%然后这里是我们的三个优化项对于每个法向(u_i,v_i)的梯度。
% \XR{Then we give the gradient function of our optimization function.}
\ZY{Similarly, the gradient of each objective term w.r.t.  $u_i$ and $v_i$ is given as follows}

\begin{flalign}
f_{01}'(u_i) &= \sum_j^M(\frac{4}{\sqrt{\sigma}}(\frac{w_j-c}{\sqrt{\sigma}})^3-\frac{2}{\sqrt{\sigma}}(\frac{w_j-c}{\sqrt{\sigma}})-\frac{1}{D})\frac{\partial w_j}{\partial u_i}&&\\
f_{01}'(v_i) &= \sum_j^{M}(\frac{4}{\sqrt{\sigma}}(\frac{w_j-c}{\sqrt{\sigma}})^3-\frac{2}{\sqrt{\sigma}}(\frac{w_j-c}{\sqrt{\sigma}})-\frac{1}{D})\frac{\partial w_j}{\partial v_i}&&
\\
f_B'(u_i) &= -\sum_j^N\sum_k^{M_i}\frac{2\left \| w_k^j-\bar{w}^j \right \|}{M_j}(\frac{\partial w_k^j}{\partial u_i}-\sum_k^{M_j}\frac{1}{M_j}\frac{\partial w_k^j}{\partial u_i})&&\\
f_B'(v_i) &= -\sum_j^N\sum_k^{M_i}\frac{2\left \| w_k^j-\bar{w}^j \right \|}{M_j}(\frac{\partial w_k^j}{\partial u_i}-\sum_k^{M_j}\frac{1}{M_j}\frac{\partial w_k^j}{\partial v_i})&&
\end{flalign}
% \begin{equation}
% f_{01}'(u_i) = \sum_j^M(\frac{4}{\sqrt{\sigma}}(\frac{w_j-c}{\sqrt{\sigma}})^3-\frac{2}{\sqrt{\sigma}}(\frac{w_j-c}{\sqrt{\sigma}})-\frac{1}{D})\frac{\partial w_j}{\partial u_i}
% \end{equation}
% \begin{equation}
% f_{01}'(v_i) = \sum_j^{M}(\frac{4}{\sqrt{\sigma}}(\frac{w_j-c}{\sqrt{\sigma}})^3-\frac{2}{\sqrt{\sigma}}(\frac{w_j-c}{\sqrt{\sigma}})-\frac{1}{D})\frac{\partial w_j}{\partial v_i}
% \end{equation}
% \begin{equation}
% f_B'(u_i) = -\sum_j^N\sum_k^{M_i}\frac{2\left \| w_k^j-\bar{w}^j \right \|}{M_j}(\frac{\partial w_k^j}{\partial u_i}-\sum_k^{M_j}\frac{1}{M_j}\frac{\partial w_k^j}{\partial u_i})
% \end{equation}
% \begin{equation}
% f_B'(v_i) = -\sum_j^N\sum_k^{M_i}\frac{2\left \| w_k^j-\bar{w}^j \right \|}{M_j}(\frac{\partial w_k^j}{\partial u_i}-\sum_k^{M_j}\frac{1}{M_j}\frac{\partial w_k^j}{\partial v_i})
% \end{equation}

% \begin{flalign}
% f_A'(u_i) = &&\\\nonumber
% \sum_k^{M_i}\frac{-(L_xcos(u_i)cos(v_i)+L_ycos(u_i)sin(v_i)-L_zsin(u_i))w_k^i}{M_i}+ &&\\\nonumber
% \sum_j^N\sum_k^{M_i}\frac{-(n_j\cdot (q_k^j-p_j))}{M_j}\frac{\partial w_j}{\partial u_i}
% \end{flalign}
\begin{flalign}
f_A'(u_i) &=\sum_j^N\sum_k^{M_i}\frac{-(n_j\cdot (q_k^j-p_j))}{M_j}\frac{\partial w_j}{\partial u_i}&&\\\nonumber
            &+ \sum_k^{M_i}\frac{-(L_xcos(u_i)cos(v_i)+L_ycos(u_i)sin(v_i)-L_zsin(u_i))w_k^i}{M_i}&&
\end{flalign}

\begin{flalign}
f_A'(v_i) &= \sum_j^N\sum_k^{M_i}\frac{-(n_j\cdot (q_k^j-p_j))}{M_j}\frac{\partial w_j}{\partial v_i}&&\\\nonumber
&+\sum_k^{M_i}\frac{(L_xsin(u_i)sin(v_i)-L_ysin(u_i)cos(v_i))w_k^i}{M_i}&&
\end{flalign}

% \begin{equation}
% f_A'(v_i) = 
% \\
% \sum_k^{M_i}\frac{(L_xsin(u_i)sin(v_i)-L_ysin(u_i)cos(v_i))w_k^i}{M_i}+\sum_j^N\sum_k^{M_i}\frac{-(n_j\cdot (q_k^j-p_j))}{M_j}\frac{\partial w_j}{\partial v_i}
% \end{equation}

\FloatBarrier
\bibliographystyle{ACM-Reference-Format}
\bibliography{sample-base}

%%% -*-BibTeX-*-
%%% Do NOT edit. File created by BibTeX with style
%%% ACM-Reference-Format-Journals [18-Jan-2012].

\begin{thebibliography}{66}

%%% ====================================================================
%%% NOTE TO THE USER: you can override these defaults by providing
%%% customized versions of any of these macros before the \bibliography
%%% command.  Each of them MUST provide its own final punctuation,
%%% except for \shownote{}, \showDOI{}, and \showURL{}.  The latter two
%%% do not use final punctuation, in order to avoid confusing it with
%%% the Web address.
%%%
%%% To suppress output of a particular field, define its macro to expand
%%% to an empty string, or better, \unskip, like this:
%%%
%%% \newcommand{\showDOI}[1]{\unskip}   % LaTeX syntax
%%%
%%% \def \showDOI #1{\unskip}           % plain TeX syntax
%%%
%%% ====================================================================

\ifx \showCODEN    \undefined \def \showCODEN     #1{\unskip}     \fi
\ifx \showDOI      \undefined \def \showDOI       #1{#1}\fi
\ifx \showISBNx    \undefined \def \showISBNx     #1{\unskip}     \fi
\ifx \showISBNxiii \undefined \def \showISBNxiii  #1{\unskip}     \fi
\ifx \showISSN     \undefined \def \showISSN      #1{\unskip}     \fi
\ifx \showLCCN     \undefined \def \showLCCN      #1{\unskip}     \fi
\ifx \shownote     \undefined \def \shownote      #1{#1}          \fi
\ifx \showarticletitle \undefined \def \showarticletitle #1{#1}   \fi
\ifx \showURL      \undefined \def \showURL       {\relax}        \fi
% The following commands are used for tagged output and should be
% invisible to TeX
\providecommand\bibfield[2]{#2}
\providecommand\bibinfo[2]{#2}
\providecommand\natexlab[1]{#1}
\providecommand\showeprint[2][]{arXiv:#2}

\bibitem[Alliez et~al\mbox{.}(2007)]%
        {alliez2007voronoi}
\bibfield{author}{\bibinfo{person}{Pierre Alliez}, \bibinfo{person}{David
  Cohen-Steiner}, \bibinfo{person}{Yiying Tong}, {and} \bibinfo{person}{Mathieu
  Desbrun}.} \bibinfo{year}{2007}\natexlab{}.
\newblock \showarticletitle{Voronoi-based variational reconstruction of
  unoriented point sets}. In \bibinfo{booktitle}{\emph{Proc. of Symp. of
  Geometry Processing}}, Vol.~\bibinfo{volume}{7}. \bibinfo{pages}{39--48}.
\newblock


\bibitem[Amenta and Bern(1998)]%
        {amenta1998surface}
\bibfield{author}{\bibinfo{person}{Nina Amenta} {and} \bibinfo{person}{Marshall
  Bern}.} \bibinfo{year}{1998}\natexlab{}.
\newblock \showarticletitle{Surface reconstruction by Voronoi filtering}. In
  \bibinfo{booktitle}{\emph{Proceedings of the fourteenth annual symposium on
  Computational geometry}}. \bibinfo{pages}{39--48}.
\newblock


\bibitem[Amenta et~al\mbox{.}(2001)]%
        {amenta2001power}
\bibfield{author}{\bibinfo{person}{Nina Amenta}, \bibinfo{person}{Sunghee
  Choi}, {and} \bibinfo{person}{Ravi~Krishna Kolluri}.}
  \bibinfo{year}{2001}\natexlab{}.
\newblock \showarticletitle{The power crust}. In
  \bibinfo{booktitle}{\emph{Proceedings of the sixth ACM symposium on Solid
  modeling and applications}}. \bibinfo{pages}{249--266}.
\newblock


\bibitem[Aroudj et~al\mbox{.}(2017)]%
        {aroudj2017visibility}
\bibfield{author}{\bibinfo{person}{Samir Aroudj}, \bibinfo{person}{Patrick
  Seemann}, \bibinfo{person}{Fabian Langguth}, \bibinfo{person}{Stefan Guthe},
  {and} \bibinfo{person}{Michael Goesele}.} \bibinfo{year}{2017}\natexlab{}.
\newblock \showarticletitle{Visibility-consistent thin surface reconstruction
  using multi-scale kernels}.
\newblock \bibinfo{journal}{\emph{ACM Trans. on Graphics}}
  \bibinfo{volume}{36}, \bibinfo{number}{6} (\bibinfo{year}{2017}),
  \bibinfo{pages}{1--13}.
\newblock


\bibitem[Avron et~al\mbox{.}(2010)]%
        {avron2010l1}
\bibfield{author}{\bibinfo{person}{Haim Avron}, \bibinfo{person}{Andrei Sharf},
  \bibinfo{person}{Chen Greif}, {and} \bibinfo{person}{Daniel Cohen-Or}.}
  \bibinfo{year}{2010}\natexlab{}.
\newblock \showarticletitle{l1-sparse reconstruction of sharp point set
  surfaces}.
\newblock \bibinfo{journal}{\emph{ACM Trans. on Graphics}}
  \bibinfo{volume}{29}, \bibinfo{number}{5} (\bibinfo{year}{2010}),
  \bibinfo{pages}{1--12}.
\newblock


\bibitem[Barill et~al\mbox{.}(2018)]%
        {barill2018fast}
\bibfield{author}{\bibinfo{person}{Gavin Barill}, \bibinfo{person}{Neil~G
  Dickson}, \bibinfo{person}{Ryan Schmidt}, \bibinfo{person}{David~IW Levin},
  {and} \bibinfo{person}{Alec Jacobson}.} \bibinfo{year}{2018}\natexlab{}.
\newblock \showarticletitle{Fast winding numbers for soups and clouds}.
\newblock \bibinfo{journal}{\emph{ACM Trans. on Graphics}}
  \bibinfo{volume}{37}, \bibinfo{number}{4} (\bibinfo{year}{2018}),
  \bibinfo{pages}{1--12}.
\newblock


\bibitem[Ben-Shabat and Gould(2020)]%
        {ben2020deepfit}
\bibfield{author}{\bibinfo{person}{Yizhak Ben-Shabat} {and}
  \bibinfo{person}{Stephen Gould}.} \bibinfo{year}{2020}\natexlab{}.
\newblock \showarticletitle{Deepfit: 3d surface fitting via neural network
  weighted least squares}. In \bibinfo{booktitle}{\emph{{ECCV}}}. Springer,
  \bibinfo{pages}{20--34}.
\newblock


\bibitem[Ben-Shabat et~al\mbox{.}(2019)]%
        {ben2019nesti}
\bibfield{author}{\bibinfo{person}{Yizhak Ben-Shabat}, \bibinfo{person}{Michael
  Lindenbaum}, {and} \bibinfo{person}{Anath Fischer}.}
  \bibinfo{year}{2019}\natexlab{}.
\newblock \showarticletitle{Nesti-net: Normal estimation for unstructured 3d
  point clouds using convolutional neural networks}. In
  \bibinfo{booktitle}{\emph{{IEEE} {CVPR}}}. \bibinfo{pages}{10112--10120}.
\newblock


\bibitem[Boltcheva and L{\'e}vy(2017)]%
        {boltcheva2017surface}
\bibfield{author}{\bibinfo{person}{Dobrina Boltcheva} {and}
  \bibinfo{person}{Bruno L{\'e}vy}.} \bibinfo{year}{2017}\natexlab{}.
\newblock \showarticletitle{Surface reconstruction by computing restricted
  Voronoi cells in parallel}.
\newblock \bibinfo{journal}{\emph{Computer-Aided Design}}  \bibinfo{volume}{90}
  (\bibinfo{year}{2017}), \bibinfo{pages}{123--134}.
\newblock


\bibitem[Cazals and Pouget(2005)]%
        {cazals2005estimating}
\bibfield{author}{\bibinfo{person}{Fr{\'e}d{\'e}ric Cazals} {and}
  \bibinfo{person}{Marc Pouget}.} \bibinfo{year}{2005}\natexlab{}.
\newblock \showarticletitle{Estimating differential quantities using polynomial
  fitting of osculating jets}.
\newblock \bibinfo{journal}{\emph{Comp. Aided Geom. Design}}
  \bibinfo{volume}{22}, \bibinfo{number}{2} (\bibinfo{year}{2005}),
  \bibinfo{pages}{121--146}.
\newblock


\bibitem[Chang et~al\mbox{.}(2015)]%
        {chang2015shapenet}
\bibfield{author}{\bibinfo{person}{Angel~X Chang}, \bibinfo{person}{Thomas
  Funkhouser}, \bibinfo{person}{Leonidas Guibas}, \bibinfo{person}{Pat
  Hanrahan}, \bibinfo{person}{Qixing Huang}, \bibinfo{person}{Zimo Li},
  \bibinfo{person}{Silvio Savarese}, \bibinfo{person}{Manolis Savva},
  \bibinfo{person}{Shuran Song}, \bibinfo{person}{Hao Su}, {et~al\mbox{.}}}
  \bibinfo{year}{2015}\natexlab{}.
\newblock \showarticletitle{Shapenet: An information-rich 3d model repository}.
\newblock \bibinfo{journal}{\emph{arXiv preprint arXiv:1512.03012}}
  (\bibinfo{year}{2015}).
\newblock


\bibitem[Chi and Song(2021)]%
        {Chi_2021_ICCV}
\bibfield{author}{\bibinfo{person}{Cheng Chi} {and} \bibinfo{person}{Shuran
  Song}.} \bibinfo{year}{2021}\natexlab{}.
\newblock \showarticletitle{GarmentNets: Category-Level Pose Estimation for
  Garments via Canonical Space Shape Completion}. In
  \bibinfo{booktitle}{\emph{{IEEE} {ICCV}}}. \bibinfo{pages}{3324--3333}.
\newblock


\bibitem[Dey and Goswami(2004)]%
        {dey2004provable}
\bibfield{author}{\bibinfo{person}{Tamal~K Dey} {and} \bibinfo{person}{Samrat
  Goswami}.} \bibinfo{year}{2004}\natexlab{}.
\newblock \showarticletitle{Provable surface reconstruction from noisy
  samples}. In \bibinfo{booktitle}{\emph{Proceedings of the twentieth annual
  symposium on Computational Geometry}}. \bibinfo{pages}{330--339}.
\newblock


\bibitem[Dey et~al\mbox{.}(2005)]%
        {dey2005normal}
\bibfield{author}{\bibinfo{person}{Tamal~K Dey}, \bibinfo{person}{Gang Li},
  {and} \bibinfo{person}{Jian Sun}.} \bibinfo{year}{2005}\natexlab{}.
\newblock \showarticletitle{Normal estimation for point clouds: A comparison
  study for a Voronoi based method}. In \bibinfo{booktitle}{\emph{Proceedings
  Eurographics/IEEE VGTC Symposium Point-Based Graphics, 2005.}} IEEE,
  \bibinfo{pages}{39--46}.
\newblock


\bibitem[Dou et~al\mbox{.}(2022)]%
        {dou2022coverage}
\bibfield{author}{\bibinfo{person}{Zhiyang Dou}, \bibinfo{person}{Cheng Lin},
  \bibinfo{person}{Rui Xu}, \bibinfo{person}{Lei Yang},
  \bibinfo{person}{Shiqing Xin}, \bibinfo{person}{Taku Komura}, {and}
  \bibinfo{person}{Wenping Wang}.} \bibinfo{year}{2022}\natexlab{}.
\newblock \showarticletitle{Coverage Axis: Inner Point Selection for 3D Shape
  Skeletonization}. In \bibinfo{booktitle}{\emph{Computer Graphics Forum}},
  Vol.~\bibinfo{volume}{41}. Wiley Online Library, \bibinfo{pages}{419--432}.
\newblock


\bibitem[Grilli et~al\mbox{.}(2017)]%
        {grilli2017review}
\bibfield{author}{\bibinfo{person}{Eleonora Grilli}, \bibinfo{person}{Fabio
  Menna}, {and} \bibinfo{person}{Fabio Remondino}.}
  \bibinfo{year}{2017}\natexlab{}.
\newblock \showarticletitle{A review of point clouds segmentation and
  classification algorithms}.
\newblock \bibinfo{journal}{\emph{The International Archives of Photogrammetry,
  Remote Sensing and Spatial Information Sciences}}  \bibinfo{volume}{42}
  (\bibinfo{year}{2017}), \bibinfo{pages}{339}.
\newblock


\bibitem[Grimm and Smart(2011)]%
        {grimm2011shape}
\bibfield{author}{\bibinfo{person}{Cindy Grimm} {and}
  \bibinfo{person}{William~D Smart}.} \bibinfo{year}{2011}\natexlab{}.
\newblock \showarticletitle{Shape classification and normal estimation for
  non-uniformly sampled, noisy point data}.
\newblock \bibinfo{journal}{\emph{Computers \& Graphics}} \bibinfo{volume}{35},
  \bibinfo{number}{4} (\bibinfo{year}{2011}), \bibinfo{pages}{904--915}.
\newblock


\bibitem[Guerrero et~al\mbox{.}(2018)]%
        {guerrero2018pcpnet}
\bibfield{author}{\bibinfo{person}{Paul Guerrero}, \bibinfo{person}{Yanir
  Kleiman}, \bibinfo{person}{Maks Ovsjanikov}, {and} \bibinfo{person}{Niloy~J
  Mitra}.} \bibinfo{year}{2018}\natexlab{}.
\newblock \showarticletitle{Pcpnet learning local shape properties from raw
  point clouds}. In \bibinfo{booktitle}{\emph{Computer Graphics Forum}},
  Vol.~\bibinfo{volume}{37}. Wiley Online Library, \bibinfo{pages}{75--85}.
\newblock


\bibitem[Hardy et~al\mbox{.}(1952)]%
        {hardy1952inequalities}
\bibfield{author}{\bibinfo{person}{Godfrey~Harold Hardy},
  \bibinfo{person}{John~Edensor Littlewood}, \bibinfo{person}{George
  P{\'o}lya}, \bibinfo{person}{Gy{\"o}rgy P{\'o}lya}, {et~al\mbox{.}}}
  \bibinfo{year}{1952}\natexlab{}.
\newblock \bibinfo{booktitle}{\emph{Inequalities}}.
\newblock \bibinfo{publisher}{Cambridge university press}.
\newblock


\bibitem[Hashimoto and Saito(2019)]%
        {hashimoto2019normal}
\bibfield{author}{\bibinfo{person}{Taisuke Hashimoto} {and}
  \bibinfo{person}{Masaki Saito}.} \bibinfo{year}{2019}\natexlab{}.
\newblock \showarticletitle{Normal Estimation for Accurate 3D Mesh
  Reconstruction with Point Cloud Model Incorporating Spatial Structure.}. In
  \bibinfo{booktitle}{\emph{CVPR workshops}}, Vol.~\bibinfo{volume}{1}.
\newblock


\bibitem[Hoppe et~al\mbox{.}(1992)]%
        {hoppe1992surface}
\bibfield{author}{\bibinfo{person}{Hugues Hoppe}, \bibinfo{person}{Tony
  DeRose}, \bibinfo{person}{Tom Duchamp}, \bibinfo{person}{John McDonald},
  {and} \bibinfo{person}{Werner Stuetzle}.} \bibinfo{year}{1992}\natexlab{}.
\newblock \showarticletitle{Surface reconstruction from unorganized points}. In
  \bibinfo{booktitle}{\emph{Proc. ACM {SIGGRAPH}}}. \bibinfo{pages}{71--78}.
\newblock


\bibitem[Hou et~al\mbox{.}(2022)]%
        {hou2022iterative}
\bibfield{author}{\bibinfo{person}{Fei Hou}, \bibinfo{person}{Chiyu Wang},
  \bibinfo{person}{Wencheng Wang}, \bibinfo{person}{Hong Qin},
  \bibinfo{person}{Chen Qian}, {and} \bibinfo{person}{Ying He}.}
  \bibinfo{year}{2022}\natexlab{}.
\newblock \showarticletitle{Iterative Poisson surface reconstruction (iPSR) for
  unoriented points}.
\newblock \bibinfo{journal}{\emph{ACM Trans. on Graphics (Proc. {SIGGRAPH})}}
  (\bibinfo{year}{2022}).
\newblock


\bibitem[Hu et~al\mbox{.}(2020)]%
        {hu2020fast}
\bibfield{author}{\bibinfo{person}{Yixin Hu}, \bibinfo{person}{Teseo
  Schneider}, \bibinfo{person}{Bolun Wang}, \bibinfo{person}{Denis Zorin},
  {and} \bibinfo{person}{Daniele Panozzo}.} \bibinfo{year}{2020}\natexlab{}.
\newblock \showarticletitle{Fast tetrahedral meshing in the wild}.
\newblock \bibinfo{journal}{\emph{ACM Transactions on Graphics (TOG)}}
  \bibinfo{volume}{39}, \bibinfo{number}{4} (\bibinfo{year}{2020}),
  \bibinfo{pages}{117--1}.
\newblock


\bibitem[Hu et~al\mbox{.}(2018)]%
        {Hu2018TMW}
\bibfield{author}{\bibinfo{person}{Yixin Hu}, \bibinfo{person}{Qingnan Zhou},
  \bibinfo{person}{Xifeng Gao}, \bibinfo{person}{Alec Jacobson},
  \bibinfo{person}{Denis Zorin}, {and} \bibinfo{person}{Daniele Panozzo}.}
  \bibinfo{year}{2018}\natexlab{}.
\newblock \showarticletitle{Tetrahedral Meshing in the Wild}.
\newblock \bibinfo{journal}{\emph{ACM Trans. on Graphics}}
  (\bibinfo{year}{2018}).
\newblock


\bibitem[Huang et~al\mbox{.}(2019)]%
        {huang2019variational}
\bibfield{author}{\bibinfo{person}{Zhiyang Huang}, \bibinfo{person}{Nathan
  Carr}, {and} \bibinfo{person}{Tao Ju}.} \bibinfo{year}{2019}\natexlab{}.
\newblock \showarticletitle{Variational implicit point set surfaces}.
\newblock \bibinfo{journal}{\emph{ACM Transactions on Graphics (TOG)}}
  \bibinfo{volume}{38}, \bibinfo{number}{4} (\bibinfo{year}{2019}),
  \bibinfo{pages}{1--13}.
\newblock


\bibitem[Huang et~al\mbox{.}(2022)]%
        {huang2022surface}
\bibfield{author}{\bibinfo{person}{Zhangjin Huang}, \bibinfo{person}{Yuxin
  Wen}, \bibinfo{person}{Zihao Wang}, \bibinfo{person}{Jinjuan Ren}, {and}
  \bibinfo{person}{Kui Jia}.} \bibinfo{year}{2022}\natexlab{}.
\newblock \showarticletitle{Surface Reconstruction from Point Clouds: A Survey
  and a Benchmark}.
\newblock \bibinfo{journal}{\emph{arXiv preprint arXiv:2205.02413}}
  (\bibinfo{year}{2022}).
\newblock


\bibitem[Jacobson et~al\mbox{.}(2021)]%
        {gptoolbox}
\bibfield{author}{\bibinfo{person}{Alec Jacobson} {et~al\mbox{.}}}
  \bibinfo{year}{2021}\natexlab{}.
\newblock \bibinfo{title}{{gptoolbox}: Geometry Processing Toolbox}.
\newblock
\newblock
\newblock
\shownote{http://github.com/alecjacobson/gptoolbox}.


\bibitem[Jacobson et~al\mbox{.}(2013)]%
        {jacobson2013robust}
\bibfield{author}{\bibinfo{person}{Alec Jacobson}, \bibinfo{person}{Ladislav
  Kavan}, {and} \bibinfo{person}{Olga Sorkine-Hornung}.}
  \bibinfo{year}{2013}\natexlab{}.
\newblock \showarticletitle{Robust inside-outside segmentation using
  generalized winding numbers}.
\newblock \bibinfo{journal}{\emph{ACM Trans. on Graphics}}
  \bibinfo{volume}{32}, \bibinfo{number}{4} (\bibinfo{year}{2013}),
  \bibinfo{pages}{1--12}.
\newblock


\bibitem[Jakob et~al\mbox{.}(2019)]%
        {jakob2019parallel}
\bibfield{author}{\bibinfo{person}{Johannes Jakob}, \bibinfo{person}{Christoph
  Buchenau}, {and} \bibinfo{person}{Michael Guthe}.}
  \bibinfo{year}{2019}\natexlab{}.
\newblock \showarticletitle{Parallel globally consistent normal orientation of
  raw unorganized point clouds}. In \bibinfo{booktitle}{\emph{Computer Graphics
  Forum}}, Vol.~\bibinfo{volume}{38}. Wiley Online Library,
  \bibinfo{pages}{163--173}.
\newblock


\bibitem[Jelic and Marsiglio(2012)]%
        {jelic2012double}
\bibfield{author}{\bibinfo{person}{V Jelic} {and} \bibinfo{person}{F
  Marsiglio}.} \bibinfo{year}{2012}\natexlab{}.
\newblock \showarticletitle{The double-well potential in quantum mechanics: a
  simple, numerically exact formulation}.
\newblock \bibinfo{journal}{\emph{European Journal of Physics}}
  \bibinfo{volume}{33}, \bibinfo{number}{6} (\bibinfo{year}{2012}),
  \bibinfo{pages}{1651}.
\newblock


\bibitem[Kazhdan(2005)]%
        {kazhdan2005reconstruction}
\bibfield{author}{\bibinfo{person}{Michael Kazhdan}.}
  \bibinfo{year}{2005}\natexlab{}.
\newblock \showarticletitle{Reconstruction of solid models from oriented point
  sets}. In \bibinfo{booktitle}{\emph{Eurographics Symposium on Geometry
  Processing}}. \bibinfo{pages}{73--es}.
\newblock


\bibitem[Kazhdan et~al\mbox{.}(2006)]%
        {kazhdan2006poisson}
\bibfield{author}{\bibinfo{person}{Michael Kazhdan}, \bibinfo{person}{Matthew
  Bolitho}, {and} \bibinfo{person}{Hugues Hoppe}.}
  \bibinfo{year}{2006}\natexlab{}.
\newblock \showarticletitle{Poisson surface reconstruction}. In
  \bibinfo{booktitle}{\emph{Eurographics Symposium on Geometry Processing}},
  Vol.~\bibinfo{volume}{7}.
\newblock


\bibitem[Kazhdan and Hoppe(2013)]%
        {kazhdan2013screened}
\bibfield{author}{\bibinfo{person}{Michael Kazhdan} {and}
  \bibinfo{person}{Hugues Hoppe}.} \bibinfo{year}{2013}\natexlab{}.
\newblock \showarticletitle{Screened poisson surface reconstruction}.
\newblock \bibinfo{journal}{\emph{ACM Trans. on Graphics}}
  \bibinfo{volume}{32}, \bibinfo{number}{3} (\bibinfo{year}{2013}),
  \bibinfo{pages}{1--13}.
\newblock


\bibitem[Kolluri et~al\mbox{.}(2004)]%
        {kolluri2004spectral}
\bibfield{author}{\bibinfo{person}{Ravikrishna Kolluri},
  \bibinfo{person}{Jonathan~Richard Shewchuk}, {and} \bibinfo{person}{James~F
  O'Brien}.} \bibinfo{year}{2004}\natexlab{}.
\newblock \showarticletitle{Spectral surface reconstruction from noisy point
  clouds}. In \bibinfo{booktitle}{\emph{Proceedings of the 2004
  Eurographics/ACM SIGGRAPH symposium on Geometry processing}}.
  \bibinfo{pages}{11--21}.
\newblock


\bibitem[K{\"o}nig and Gumhold(2009)]%
        {konig2009consistent}
\bibfield{author}{\bibinfo{person}{S{\"o}ren K{\"o}nig} {and}
  \bibinfo{person}{Stefan Gumhold}.} \bibinfo{year}{2009}\natexlab{}.
\newblock \showarticletitle{Consistent Propagation of Normal Orientations in
  Point Clouds}. In \bibinfo{booktitle}{\emph{VMV}}. \bibinfo{pages}{83--92}.
\newblock


\bibitem[Lenssen et~al\mbox{.}(2020)]%
        {lenssen2020deep}
\bibfield{author}{\bibinfo{person}{Jan~Eric Lenssen},
  \bibinfo{person}{Christian Osendorfer}, {and} \bibinfo{person}{Jonathan
  Masci}.} \bibinfo{year}{2020}\natexlab{}.
\newblock \showarticletitle{Deep iterative surface normal estimation}. In
  \bibinfo{booktitle}{\emph{{IEEE} {CVPR}}}. \bibinfo{pages}{11247--11256}.
\newblock


\bibitem[Levin(1998)]%
        {levin1998approximation}
\bibfield{author}{\bibinfo{person}{David Levin}.}
  \bibinfo{year}{1998}\natexlab{}.
\newblock \showarticletitle{The approximation power of moving least-squares}.
\newblock \bibinfo{journal}{\emph{Mathematics of computation}}
  \bibinfo{volume}{67}, \bibinfo{number}{224} (\bibinfo{year}{1998}),
  \bibinfo{pages}{1517--1531}.
\newblock


\bibitem[Levin(2004)]%
        {levin2004mesh}
\bibfield{author}{\bibinfo{person}{David Levin}.}
  \bibinfo{year}{2004}\natexlab{}.
\newblock \showarticletitle{Mesh-independent surface interpolation}.
\newblock In \bibinfo{booktitle}{\emph{Geometric modeling for scientific
  visualization}}. \bibinfo{publisher}{Springer}, \bibinfo{pages}{37--49}.
\newblock


\bibitem[Li et~al\mbox{.}(2010)]%
        {li2010robust}
\bibfield{author}{\bibinfo{person}{Bao Li}, \bibinfo{person}{Ruwen Schnabel},
  \bibinfo{person}{Reinhard Klein}, \bibinfo{person}{Zhiquan Cheng},
  \bibinfo{person}{Gang Dang}, {and} \bibinfo{person}{Shiyao Jin}.}
  \bibinfo{year}{2010}\natexlab{}.
\newblock \showarticletitle{Robust normal estimation for point clouds with
  sharp features}.
\newblock \bibinfo{journal}{\emph{Computers \& Graphics}} \bibinfo{volume}{34},
  \bibinfo{number}{2} (\bibinfo{year}{2010}), \bibinfo{pages}{94--106}.
\newblock


\bibitem[Li et~al\mbox{.}(2022)]%
        {li2022neaf}
\bibfield{author}{\bibinfo{person}{Shujuan Li}, \bibinfo{person}{Junsheng
  Zhou}, \bibinfo{person}{Baorui Ma}, \bibinfo{person}{Yu-Shen Liu}, {and}
  \bibinfo{person}{Zhizhong Han}.} \bibinfo{year}{2022}\natexlab{}.
\newblock \showarticletitle{NeAF: Learning Neural Angle Fields for Point Normal
  Estimation}.
\newblock \bibinfo{journal}{\emph{arXiv preprint arXiv:2211.16869}}
  (\bibinfo{year}{2022}).
\newblock


\bibitem[Lin et~al\mbox{.}(2022)]%
        {PGR2022Siyou}
\bibfield{author}{\bibinfo{person}{Siyou Lin}, \bibinfo{person}{Dong Xiao},
  \bibinfo{person}{Zuoqiang Shi}, {and} \bibinfo{person}{Bin Wang}.}
  \bibinfo{year}{2022}\natexlab{}.
\newblock \showarticletitle{Surface Reconstruction from Point Clouds without
  Normals by Parametrizing the Gauss Formula}.
\newblock \bibinfo{journal}{\emph{ACM Trans. on Graphics}}
  \bibinfo{volume}{42}, \bibinfo{number}{2} (\bibinfo{year}{2022}),
  \bibinfo{numpages}{19}~pages.
\newblock
\showISSN{0730-0301}
\urldef\tempurl%
\url{https://doi.org/10.1145/3554730}
\showDOI{\tempurl}


\bibitem[Liu et~al\mbox{.}(2015)]%
        {liu2015quality}
\bibfield{author}{\bibinfo{person}{Xiuping Liu}, \bibinfo{person}{Jie Zhang},
  \bibinfo{person}{Junjie Cao}, \bibinfo{person}{Bo Li}, {and}
  \bibinfo{person}{Ligang Liu}.} \bibinfo{year}{2015}\natexlab{}.
\newblock \showarticletitle{Quality point cloud normal estimation by guided
  least squares representation}.
\newblock \bibinfo{journal}{\emph{Computers \& Graphics}}  \bibinfo{volume}{51}
  (\bibinfo{year}{2015}), \bibinfo{pages}{106--116}.
\newblock


\bibitem[Meister(1769)]%
        {meister1769generalia}
\bibfield{author}{\bibinfo{person}{Albrecht Ludwig~Friedrich Meister}.}
  \bibinfo{year}{1769}\natexlab{}.
\newblock \bibinfo{booktitle}{\emph{Generalia de genesi figurarum planarum et
  inde pendentibus earum affectionibus}}.
\newblock


\bibitem[M{\'e}rigot et~al\mbox{.}(2010)]%
        {merigot2010voronoi}
\bibfield{author}{\bibinfo{person}{Quentin M{\'e}rigot}, \bibinfo{person}{Maks
  Ovsjanikov}, {and} \bibinfo{person}{Leonidas~J Guibas}.}
  \bibinfo{year}{2010}\natexlab{}.
\newblock \showarticletitle{Voronoi-based curvature and feature estimation from
  point clouds}.
\newblock \bibinfo{journal}{\emph{IEEE Trans. on Vis. and Comp. Graphics}}
  \bibinfo{volume}{17}, \bibinfo{number}{6} (\bibinfo{year}{2010}),
  \bibinfo{pages}{743--756}.
\newblock


\bibitem[Metzer et~al\mbox{.}(2021)]%
        {dipole_propagation}
\bibfield{author}{\bibinfo{person}{Gal Metzer}, \bibinfo{person}{Rana Hanocka},
  \bibinfo{person}{Denis Zorin}, \bibinfo{person}{Raja Giryes},
  \bibinfo{person}{Daniele Panozzo}, {and} \bibinfo{person}{Daniel Cohen-Or}.}
  \bibinfo{year}{2021}\natexlab{}.
\newblock \showarticletitle{Orienting Point Clouds with Dipole Propagation}.
\newblock \bibinfo{journal}{\emph{ACM Trans. on Graphics}}
  \bibinfo{volume}{40}, \bibinfo{number}{4}, Article \bibinfo{articleno}{165}
  (\bibinfo{date}{jul} \bibinfo{year}{2021}), \bibinfo{numpages}{14}~pages.
\newblock
\showISSN{0730-0301}
\urldef\tempurl%
\url{https://doi.org/10.1145/3450626.3459835}
\showDOI{\tempurl}


\bibitem[Mitra and Nguyen(2003)]%
        {mitra2003estimating}
\bibfield{author}{\bibinfo{person}{Niloy~J Mitra} {and} \bibinfo{person}{An
  Nguyen}.} \bibinfo{year}{2003}\natexlab{}.
\newblock \showarticletitle{Estimating surface normals in noisy point cloud
  data}. In \bibinfo{booktitle}{\emph{special issue of International Journal of
  Computational Geometry and Applications}}. \bibinfo{pages}{322--328}.
\newblock


\bibitem[Nuvoli et~al\mbox{.}(2022)]%
        {nuvoli2022skinmixer}
\bibfield{author}{\bibinfo{person}{Stefano Nuvoli}, \bibinfo{person}{Nico
  Pietroni}, \bibinfo{person}{Paolo Cignoni}, \bibinfo{person}{Riccardo
  Scateni}, {and} \bibinfo{person}{Marco Tarini}.}
  \bibinfo{year}{2022}\natexlab{}.
\newblock \showarticletitle{SkinMixer: Blending 3D Animated Models}.
\newblock \bibinfo{journal}{\emph{ACM Transactions on Graphics (TOG)}}
  \bibinfo{volume}{41}, \bibinfo{number}{6} (\bibinfo{year}{2022}),
  \bibinfo{pages}{1--15}.
\newblock


\bibitem[OuYang and Feng(2005)]%
        {ouyang2005normal}
\bibfield{author}{\bibinfo{person}{Daoshan OuYang} {and}
  \bibinfo{person}{Hsi-Yung Feng}.} \bibinfo{year}{2005}\natexlab{}.
\newblock \showarticletitle{On the normal vector estimation for point cloud
  data from smooth surfaces}.
\newblock \bibinfo{journal}{\emph{Computer-Aided Design}} \bibinfo{volume}{37},
  \bibinfo{number}{10} (\bibinfo{year}{2005}), \bibinfo{pages}{1071--1079}.
\newblock


\bibitem[Pauly et~al\mbox{.}(2003)]%
        {pauly2003shape}
\bibfield{author}{\bibinfo{person}{Mark Pauly}, \bibinfo{person}{Richard
  Keiser}, \bibinfo{person}{Leif~P Kobbelt}, {and} \bibinfo{person}{Markus
  Gross}.} \bibinfo{year}{2003}\natexlab{}.
\newblock \showarticletitle{Shape modeling with point-sampled geometry}.
\newblock \bibinfo{journal}{\emph{ACM Trans. on Graphics}}
  \bibinfo{volume}{22}, \bibinfo{number}{3} (\bibinfo{year}{2003}),
  \bibinfo{pages}{641--650}.
\newblock


\bibitem[Pomerleau et~al\mbox{.}(2015)]%
        {pomerleau2015review}
\bibfield{author}{\bibinfo{person}{Fran{\c{c}}ois Pomerleau},
  \bibinfo{person}{Francis Colas}, \bibinfo{person}{Roland Siegwart},
  {et~al\mbox{.}}} \bibinfo{year}{2015}\natexlab{}.
\newblock \showarticletitle{A review of point cloud registration algorithms for
  mobile robotics}.
\newblock \bibinfo{journal}{\emph{Foundations and Trends{\textregistered} in
  Robotics}} \bibinfo{volume}{4}, \bibinfo{number}{1} (\bibinfo{year}{2015}),
  \bibinfo{pages}{1--104}.
\newblock


\bibitem[Rusu and Cousins(2011)]%
        {Rusu_ICRA2011_PCL}
\bibfield{author}{\bibinfo{person}{Radu~Bogdan Rusu} {and}
  \bibinfo{person}{Steve Cousins}.} \bibinfo{year}{2011}\natexlab{}.
\newblock \showarticletitle{{3D is here: Point Cloud Library (PCL)}}. In
  \bibinfo{booktitle}{\emph{{IEEE International Conference on Robotics and
  Automation (ICRA)}}}. \bibinfo{publisher}{IEEE}.
\newblock


\bibitem[Sell{\'a}n et~al\mbox{.}(2021)]%
        {sellan2021swept}
\bibfield{author}{\bibinfo{person}{Silvia Sell{\'a}n}, \bibinfo{person}{Noam
  Aigerman}, {and} \bibinfo{person}{Alec Jacobson}.}
  \bibinfo{year}{2021}\natexlab{}.
\newblock \showarticletitle{Swept volumes via spacetime numerical
  continuation}.
\newblock \bibinfo{journal}{\emph{ACM Transactions on Graphics (TOG)}}
  \bibinfo{volume}{40}, \bibinfo{number}{4} (\bibinfo{year}{2021}),
  \bibinfo{pages}{1--11}.
\newblock


\bibitem[Sell{\'a}n and Jacobson(2022)]%
        {sellan2022stochastic}
\bibfield{author}{\bibinfo{person}{Silvia Sell{\'a}n} {and}
  \bibinfo{person}{Alec Jacobson}.} \bibinfo{year}{2022}\natexlab{}.
\newblock \showarticletitle{Stochastic Poisson Surface Reconstruction}.
\newblock \bibinfo{journal}{\emph{ACM Transactions on Graphics (TOG)}}
  \bibinfo{volume}{41}, \bibinfo{number}{6} (\bibinfo{year}{2022}),
  \bibinfo{pages}{1--12}.
\newblock


\bibitem[Sun et~al\mbox{.}(2015)]%
        {sun2015denoising}
\bibfield{author}{\bibinfo{person}{Yujing Sun}, \bibinfo{person}{Scott
  Schaefer}, {and} \bibinfo{person}{Wenping Wang}.}
  \bibinfo{year}{2015}\natexlab{}.
\newblock \showarticletitle{Denoising point sets via L0 minimization}.
\newblock \bibinfo{journal}{\emph{Comp. Aided Geom. Design}}
  \bibinfo{volume}{35} (\bibinfo{year}{2015}), \bibinfo{pages}{2--15}.
\newblock


\bibitem[Wang et~al\mbox{.}(2012)]%
        {wang2012variational}
\bibfield{author}{\bibinfo{person}{Jun Wang}, \bibinfo{person}{Zhouwang Yang},
  {and} \bibinfo{person}{Falai Chen}.} \bibinfo{year}{2012}\natexlab{}.
\newblock \showarticletitle{A variational model for normal computation of point
  clouds}.
\newblock \bibinfo{journal}{\emph{The Visual Computer}} \bibinfo{volume}{28},
  \bibinfo{number}{2} (\bibinfo{year}{2012}), \bibinfo{pages}{163--174}.
\newblock


\bibitem[Wang et~al\mbox{.}(2022a)]%
        {wang2022computing}
\bibfield{author}{\bibinfo{person}{Ningna Wang}, \bibinfo{person}{Bin Wang},
  \bibinfo{person}{Wenping Wang}, {and} \bibinfo{person}{Xiaohu Guo}.}
  \bibinfo{year}{2022}\natexlab{a}.
\newblock \showarticletitle{Computing Medial Axis Transform with Feature
  Preservation via Restricted Power Diagram}.
\newblock \bibinfo{journal}{\emph{ACM Transactions on Graphics (TOG)}}
  \bibinfo{volume}{41}, \bibinfo{number}{6} (\bibinfo{year}{2022}),
  \bibinfo{pages}{1--18}.
\newblock


\bibitem[Wang et~al\mbox{.}(2022b)]%
        {wang2022restricted}
\bibfield{author}{\bibinfo{person}{Pengfei Wang}, \bibinfo{person}{Zixiong
  Wang}, \bibinfo{person}{Shiqing Xin}, \bibinfo{person}{Xifeng Gao},
  \bibinfo{person}{Wenping Wang}, {and} \bibinfo{person}{Changhe Tu}.}
  \bibinfo{year}{2022}\natexlab{b}.
\newblock \showarticletitle{Restricted Delaunay Triangulation for Explicit
  Surface Reconstruction}.
\newblock \bibinfo{journal}{\emph{ACM Transactions on Graphics (TOG)}}
  (\bibinfo{year}{2022}).
\newblock


\bibitem[Wang et~al\mbox{.}(2021)]%
        {wang2021neural}
\bibfield{author}{\bibinfo{person}{Zixiong Wang}, \bibinfo{person}{Pengfei
  Wang}, \bibinfo{person}{Qiujie Dong}, \bibinfo{person}{Junjie Gao},
  \bibinfo{person}{Shuangmin Chen}, \bibinfo{person}{Shiqing Xin}, {and}
  \bibinfo{person}{Changhe Tu}.} \bibinfo{year}{2021}\natexlab{}.
\newblock \showarticletitle{Neural-IMLS: Learning Implicit Moving Least-Squares
  for Surface Reconstruction from Unoriented Point clouds}.
\newblock \bibinfo{journal}{\emph{arXiv preprint arXiv:2109.04398}}
  (\bibinfo{year}{2021}).
\newblock


\bibitem[Xie et~al\mbox{.}(2004)]%
        {xie2004surface}
\bibfield{author}{\bibinfo{person}{Hui Xie}, \bibinfo{person}{Kevin~T
  McDonnell}, {and} \bibinfo{person}{Hong Qin}.}
  \bibinfo{year}{2004}\natexlab{}.
\newblock \showarticletitle{Surface reconstruction of noisy and defective data
  sets}. In \bibinfo{booktitle}{\emph{IEEE visualization 2004}}. IEEE,
  \bibinfo{pages}{259--266}.
\newblock


\bibitem[Xu et~al\mbox{.}(2022)]%
        {xu2022rfeps}
\bibfield{author}{\bibinfo{person}{Rui Xu}, \bibinfo{person}{Zixiong Wang},
  \bibinfo{person}{Zhiyang Dou}, \bibinfo{person}{Chen Zong},
  \bibinfo{person}{Shiqing Xin}, \bibinfo{person}{Mingyan Jiang},
  \bibinfo{person}{Tao Ju}, {and} \bibinfo{person}{Changhe Tu}.}
  \bibinfo{year}{2022}\natexlab{}.
\newblock \showarticletitle{RFEPS: Reconstructing Feature-Line Equipped
  Polygonal Surface}.
\newblock \bibinfo{journal}{\emph{ACM Trans. on Graphics (Proc. {SIGGRAPH
  Asia})}} \bibinfo{volume}{41}, \bibinfo{number}{6} (\bibinfo{year}{2022}),
  \bibinfo{pages}{1--15}.
\newblock


\bibitem[Yoon et~al\mbox{.}(2007)]%
        {yoon2007surface}
\bibfield{author}{\bibinfo{person}{Mincheol Yoon}, \bibinfo{person}{Yunjin
  Lee}, \bibinfo{person}{Seungyong Lee}, \bibinfo{person}{Ioannis
  Ivrissimtzis}, {and} \bibinfo{person}{Hans-Peter Seidel}.}
  \bibinfo{year}{2007}\natexlab{}.
\newblock \showarticletitle{Surface and normal ensembles for surface
  reconstruction}.
\newblock \bibinfo{journal}{\emph{Computer-Aided Design}} \bibinfo{volume}{39},
  \bibinfo{number}{5} (\bibinfo{year}{2007}), \bibinfo{pages}{408--420}.
\newblock


\bibitem[Zapata-Impata et~al\mbox{.}(2019)]%
        {zapata2019fast}
\bibfield{author}{\bibinfo{person}{Brayan~S Zapata-Impata},
  \bibinfo{person}{Pablo Gil}, \bibinfo{person}{Jorge Pomares}, {and}
  \bibinfo{person}{Fernando Torres}.} \bibinfo{year}{2019}\natexlab{}.
\newblock \showarticletitle{Fast geometry-based computation of grasping points
  on three-dimensional point clouds}.
\newblock \bibinfo{journal}{\emph{International Journal of Advanced Robotic
  Systems}} \bibinfo{volume}{16}, \bibinfo{number}{1} (\bibinfo{year}{2019}),
  \bibinfo{pages}{1729881419831846}.
\newblock


\bibitem[Zhang et~al\mbox{.}(2018)]%
        {zhang2018multi}
\bibfield{author}{\bibinfo{person}{Jie Zhang}, \bibinfo{person}{Junjie Cao},
  \bibinfo{person}{Xiuping Liu}, \bibinfo{person}{He Chen}, \bibinfo{person}{Bo
  Li}, {and} \bibinfo{person}{Ligang Liu}.} \bibinfo{year}{2018}\natexlab{}.
\newblock \showarticletitle{Multi-normal estimation via pair consistency
  voting}.
\newblock \bibinfo{journal}{\emph{IEEE Trans. on Vis. and Comp. Graphics}}
  \bibinfo{volume}{25}, \bibinfo{number}{4} (\bibinfo{year}{2018}),
  \bibinfo{pages}{1693--1706}.
\newblock


\bibitem[Zhang et~al\mbox{.}(2013)]%
        {zhang2013point}
\bibfield{author}{\bibinfo{person}{Jie Zhang}, \bibinfo{person}{Junjie Cao},
  \bibinfo{person}{Xiuping Liu}, \bibinfo{person}{Jun Wang},
  \bibinfo{person}{Jian Liu}, {and} \bibinfo{person}{Xiquan Shi}.}
  \bibinfo{year}{2013}\natexlab{}.
\newblock \showarticletitle{Point cloud normal estimation via low-rank subspace
  clustering}.
\newblock \bibinfo{journal}{\emph{Computers \& Graphics}} \bibinfo{volume}{37},
  \bibinfo{number}{6} (\bibinfo{year}{2013}), \bibinfo{pages}{697--706}.
\newblock


\bibitem[Zhou et~al\mbox{.}(2020)]%
        {zhou2020normal}
\bibfield{author}{\bibinfo{person}{Jun Zhou}, \bibinfo{person}{Hua Huang},
  \bibinfo{person}{Bin Liu}, {and} \bibinfo{person}{Xiuping Liu}.}
  \bibinfo{year}{2020}\natexlab{}.
\newblock \showarticletitle{Normal estimation for 3d point clouds via local
  plane constraint and multi-scale selection}.
\newblock \bibinfo{journal}{\emph{Computer-Aided Design}}
  \bibinfo{volume}{129} (\bibinfo{year}{2020}), \bibinfo{pages}{102916}.
\newblock


\bibitem[Zhu et~al\mbox{.}(2021)]%
        {zhu2021adafit}
\bibfield{author}{\bibinfo{person}{Runsong Zhu}, \bibinfo{person}{Yuan Liu},
  \bibinfo{person}{Zhen Dong}, \bibinfo{person}{Yuan Wang},
  \bibinfo{person}{Tengping Jiang}, \bibinfo{person}{Wenping Wang}, {and}
  \bibinfo{person}{Bisheng Yang}.} \bibinfo{year}{2021}\natexlab{}.
\newblock \showarticletitle{AdaFit: Rethinking Learning-based Normal Estimation
  on Point Clouds}. In \bibinfo{booktitle}{\emph{{IEEE} {ICCV}}}.
  \bibinfo{pages}{6118--6127}.
\newblock


\end{thebibliography}

\end{document}